\documentclass [10pt,twocolumn]{IEEEtran}
\usepackage{algorithm,algorithmic,amsbsy,amsmath,amssymb,epsfig,bbm,mathrsfs,multirow,amsthm}
\usepackage[T1]{fontenc}
\usepackage[latin9]{inputenc}
\usepackage{amsmath}
\usepackage{fixltx2e}
\usepackage{algorithm}
\usepackage{array,multirow,graphicx}
\usepackage{color}
\usepackage{cite}
\usepackage{float}
\usepackage{makecell}
\usepackage[x11names,dvipsnames,table]{xcolor} %for use in color links
\usepackage{colortbl}
\usepackage{multirow}
\usepackage{subcaption}
\usepackage{lipsum}
\definecolor{mygray}{gray}{0.6}
\definecolor{myblue}{rgb}{0.8,0.85,1}
\usepackage{array}
\newcolumntype{L}[1]{>{\raggedright\let\newline\\\arraybackslash\hspace{0pt}}m{#1}}
\newcolumntype{C}[1]{>{\centering\let\newline\\\arraybackslash\hspace{0pt}}m{#1}}
\newcolumntype{R}[1]{>{\raggedleft\let\newline\\\arraybackslash\hspace{0pt}}m{#1}}
\newcommand{\noun}[1]{\textsc{#1}}
\usepackage{array, tabularx, boldline}
\usepackage{eurosym}
\usepackage{amstext} % for \text
\DeclareRobustCommand{\officialeuro}{%
  \ifmmode\expandafter\text\fi
  {\fontencoding{U}\fontfamily{eurosym}\selectfont e}}

\usepackage{array, tabularx, boldline}
\usepackage{graphicx}
\usepackage{cellspace}
\graphicspath{{figures/}}

\newtheorem{assumption} {Assumption}

\begin{document}
%\[\title{Pricing Models in Internet of Things: A Survey}
\title{\huge Applications of Deep Reinforcement Learning in Communications and Networking: A Survey}
\author{Nguyen Cong Luong, Dinh Thai Hoang, \textit{Member, IEEE}, Shimin Gong, \textit{Member, IEEE}, Dusit Niyato, \textit{Fellow, IEEE}, Ping Wang, \textit{Senior Member, IEEE}, Ying-Chang Liang, \textit{Fellow, IEEE}, Dong In Kim, \textit{Senior Member, IEEE}
\thanks{N.~C.~Luong and D.~Niyato are with School of Computer Science and Engineering, Nanyang Technological University, Singapore. E-mails: clnguyen@ntu.edu.sg, dniyato@ntu.edu.sg.}
\thanks{D.~T.~Hoang is with the Faculty of Engineering and Information Technology,
University  of  Technology  Sydney, Australia. E-mail: hoang.dinh@uts.edu.au.}
\thanks{S.~Gong is with the Shenzhen Institute of  Advanced
Technology, Chinese Academy of Sciences, Shenzhen 518055, China. E-mail: sm.gong@siat.ac.cn.}
%\thanks{L.~B.~Le is with the Institut National de la Recherche Scientifique (INRS), Universit\'e du Qu\'ebec, Montr\'eal, QC J3X 1P7, Canada. E-mail: long.le@emt.inrs.ca.} %\vspace{-6mm}
\thanks{P.~Wang is with Department of Electrical Engineering \& Computer Science, York University, Canada. E-mail: pingw@yorku.ca.}
\thanks{Y.-C.~Liang is with Center for Intelligent Networking and Communications (CINC), with University of Electronic Science and Technology of China, Chengdu, China. E-mail: liangyc@ieee.org.} %\vspace{-6mm}
\thanks{D.~I.~Kim is with School of Information and Communication Engineering, Sungkyunkwan University, Korea. Email: dikim@skku.ac.kr.}
}

\maketitle
%====================================================================
\begin{abstract}
This paper presents a comprehensive literature review
on applications of deep reinforcement learning in communications and networking. Modern networks, e.g., Internet of Things (IoT) and Unmanned Aerial Vehicle (UAV) networks, become more decentralized and autonomous. In such networks, network entities need to make decisions locally to maximize the network performance under uncertainty of network environment. Reinforcement learning has been efficiently used to enable the network entities to obtain the optimal policy including, e.g., decisions or actions, given their states when the state and action spaces are small. However, in complex and large-scale networks, the state and action spaces are usually large, and the reinforcement learning may not be able to find the optimal policy in reasonable time. Therefore, deep reinforcement learning, a combination of reinforcement learning with deep learning, has been developed to overcome the shortcomings. In this survey, we first give a tutorial of deep reinforcement learning from fundamental concepts to advanced models. Then, we review deep reinforcement learning approaches proposed to address emerging issues in communications and networking. The issues include dynamic network access, data rate control, wireless caching, data offloading, network security, and connectivity preservation which are all important to next generation networks such as 5G and beyond. Furthermore, we present applications of deep reinforcement learning for traffic routing, resource sharing, and data collection. Finally, we highlight important challenges, open issues, and future research directions of applying deep reinforcement learning.

\end{abstract}
%main

%=====================
\section{Introduction}
%=====================
Reinforcement learning~\cite{sutton1998reinforcement} is one of the most important research directions of machine learning which has significant impacts to the development of Artificial Intelligence (AI) over the last 20 years. Reinforcement learning is a learning process in which an agent can periodically make decisions, observe the results, and then automatically adjust its strategy to achieve the optimal policy. However, this learning process, even though proved to converge, takes a lot of time to reach the best policy as it has to explore and gain knowledge of an entire system, making it unsuitable and inapplicable to large-scale networks. Consequently, applications of reinforcement learning are very limited in practice. Recently, deep learning~\cite{goodfellow2016deep} has been introduced as a new breakthrough technique. It can overcome the limitations of reinforcement learning, and thus open a new era for the development of reinforcement learning, namely \emph{Deep Reinforcement Learning} (DRL). The DRL embraces the advantage of Deep Neural Networks (DNNs) to train the learning process, thereby improving the learning speed and the performance of reinforcement learning algorithms. As a result, DRL has been adopted in a numerous applications of reinforcement learning in practice such as robotics, computer vision, speech recognition, and natural language processing~\cite{goodfellow2016deep}. One of the most famous applications of DRL is AlphaGo~\cite{alphago}, the first computer program which can beat a human professional without handicaps on a full-sized 19$\times$19 board.
% In 2017, MIT Technology Review also listed DRL as one of 10 breakthrough technology of the year~\cite{MITTechnologyReview}.  Other areas that have found various applications of DRL are  recently received a lot of attentions are networking and communications in which DRL can be used to effectively address various problems and challenges.

In the areas of communications and networking, DRL has been recently used as an emerging tool to effectively address various problems and challenges. In particular, modern networks such as Internet of Things (IoT), Heterogeneous Networks (HetNets), and Unmanned Aerial Vehicle (UAV) network become more decentralized, ad-hoc, and autonomous in nature. Network entities such as IoT devices, mobile users, and UAVs need to make local and autonomous decisions, e.g., spectrum access, data rate selection, transmit power control, and base station association, to achieve the goals of different networks including, e.g., throughput maximization and energy consumption minimization. Under uncertain and stochastic environments, most of the decision-making problems can be modeled by a so-called \textit{Markov Decision Process} (MDP)~\cite{puterman2014markov}. Dynamic programming \cite{bertsekas2005dynamic}, \cite{bellman2013dynamic} and other algorithms such as value iteration, as well as reinforcement learning techniques can be adopted to solve the MDP. However, the modern networks are large-scale and complicated, and thus the computational complexity of the techniques rapidly becomes unmanageable. As a result, DRL has been developing to be an alternative solution to overcome the challenge. In general, the DRL approaches provide the following advantages:
\begin{itemize}
\item DRL can obtain the solution of sophisticated  network optimizations. Thus, it enables network controllers, e.g., base stations, in modern networks to solve non-convex and complex problems, e.g., joint user association, computation, and transmission schedule, to achieve the optimal solutions without complete and accurate network information.
\item DRL allows network entities to learn and build knowledge about the communication and networking environment. Thus, by using DRL, the network entities, e.g., a mobile user, can learn optimal policies, e.g., base station selection, channel selection, handover decision, caching and offloading decisions, without knowing channel model and mobility pattern.
\item DRL provides autonomous decision-making. With the DRL approaches, network entities can make observation and obtain the best policy locally with minimum or without information exchange among each other. This not only reduces communication overheads but also improves security and robustness of the networks.
\item DRL improves significantly the learning speed, especially in the problems with large state and action spaces. Thus, in large-scale networks, e.g., IoT systems with thousands of devices, DRL allows network controller or IoT gateways to control dynamically user association, spectrum access, and transmit power for a massive number of IoT devices and mobile users.
%\item By providing adaptive decisions which allows clients in dynamic streaming systems to locally decide video segments with appropriate bitrates to maximize their Quality of Experience (QoE).
%\item DRL can represent a large number of action and states. Thus, DRL allows  transmitters in complex systems, e.g., space communication systems, to configure several transmit parameters, e.g., symbol rate, to achieve a number of performance objectives., e.g., throughput maximization.
%\item DRL can learn optimal policies based on past observations. Thus, DRL enables legitimate transmitters, e.g., base station and mobile users, to quickly select appropriately defending strategies, e.g., transmit power control, channel selection, and mobility strategy decision, to prevent attacks, e.g., jamming, without knowing the attack model.
%\item Neural networks can extract fingerprints from signals of transmitters. Thus, by using DRL, legitimate receivers can use the fingerprints to detect cyber-physical attacks with a high accuracy.
\item Several other problems in communications and networking such as cyber-physical attacks, interference management, and data offloading can be modeled as games, e.g., the non-cooperative game. DRL has been recently used as an efficient tool to solve the games, e.g., finding the Nash equilibrium, without the complete information.

\end{itemize}

%we provide a tutorial of deep reinforcement learning, and we review deep reinforcement learning approaches proposed to address emerging issues

\begin{figure}[t!]
    \begin{subfigure}[t]{0.5\linewidth}
        \centering
        \includegraphics[width=1.\linewidth]{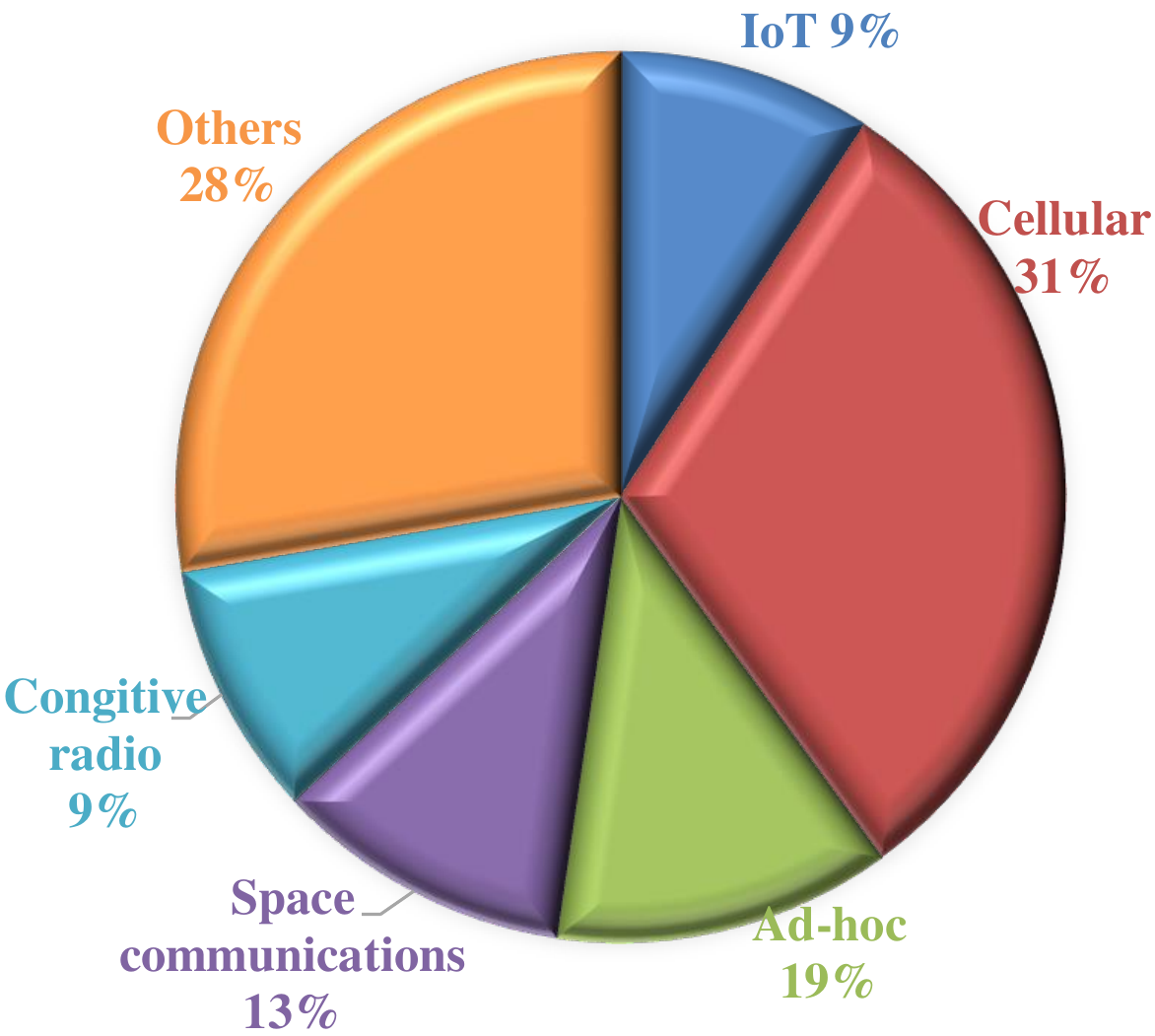}
        \caption{}
    \end{subfigure}%
 \hfill
    \begin{subfigure}[t]{0.5\linewidth}
        \centering
        \includegraphics[width=1.06\linewidth]{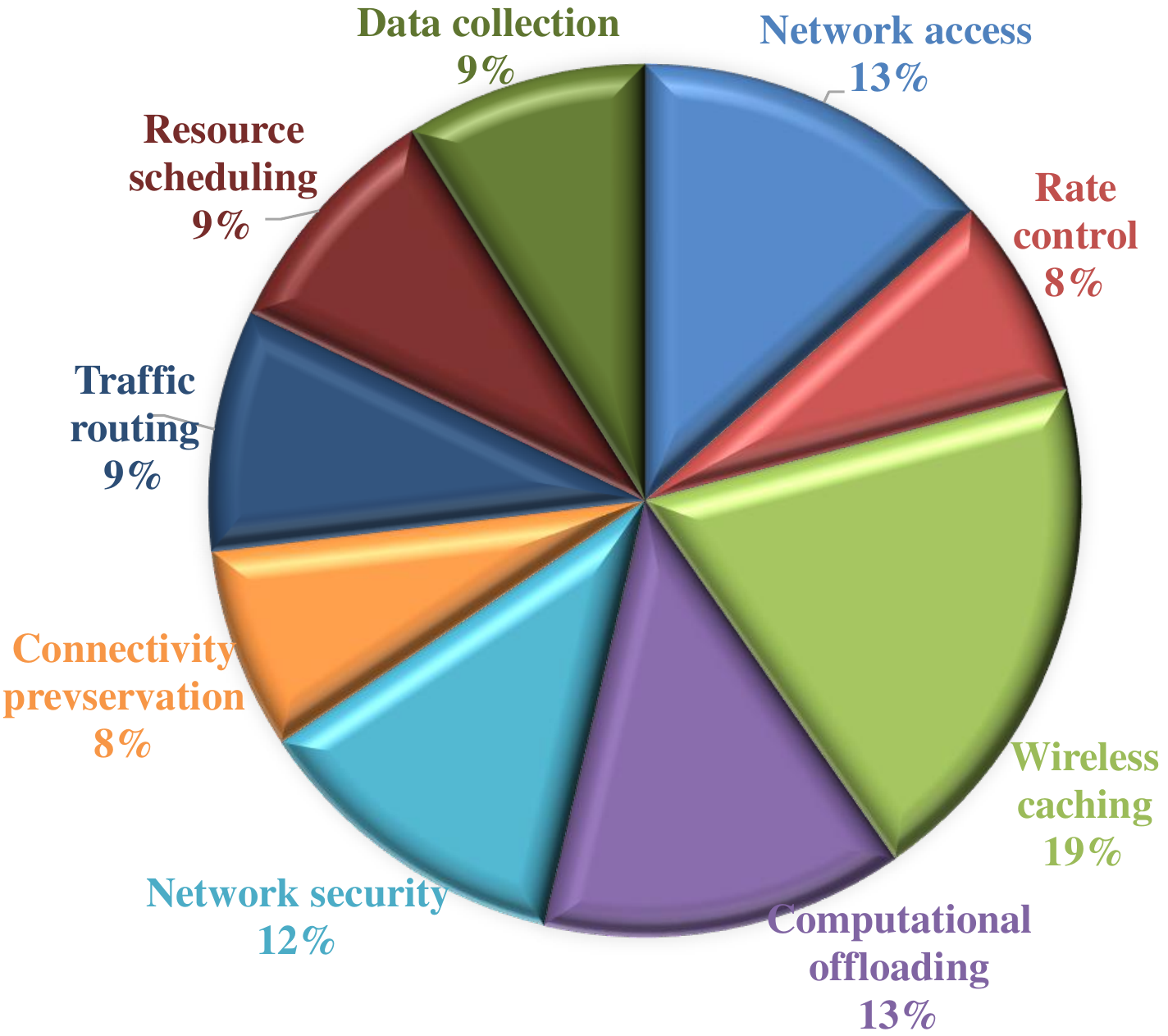}
        \caption{}
    \end{subfigure}
   \caption{\small Percentages of related work for (a) different networks and (b) different issues in the networks.}
    \label{Summary_Figure}
\end{figure}

\begin{figure*}[t!]
\centering
\includegraphics[width=15.7 cm, height=6.3 cm]{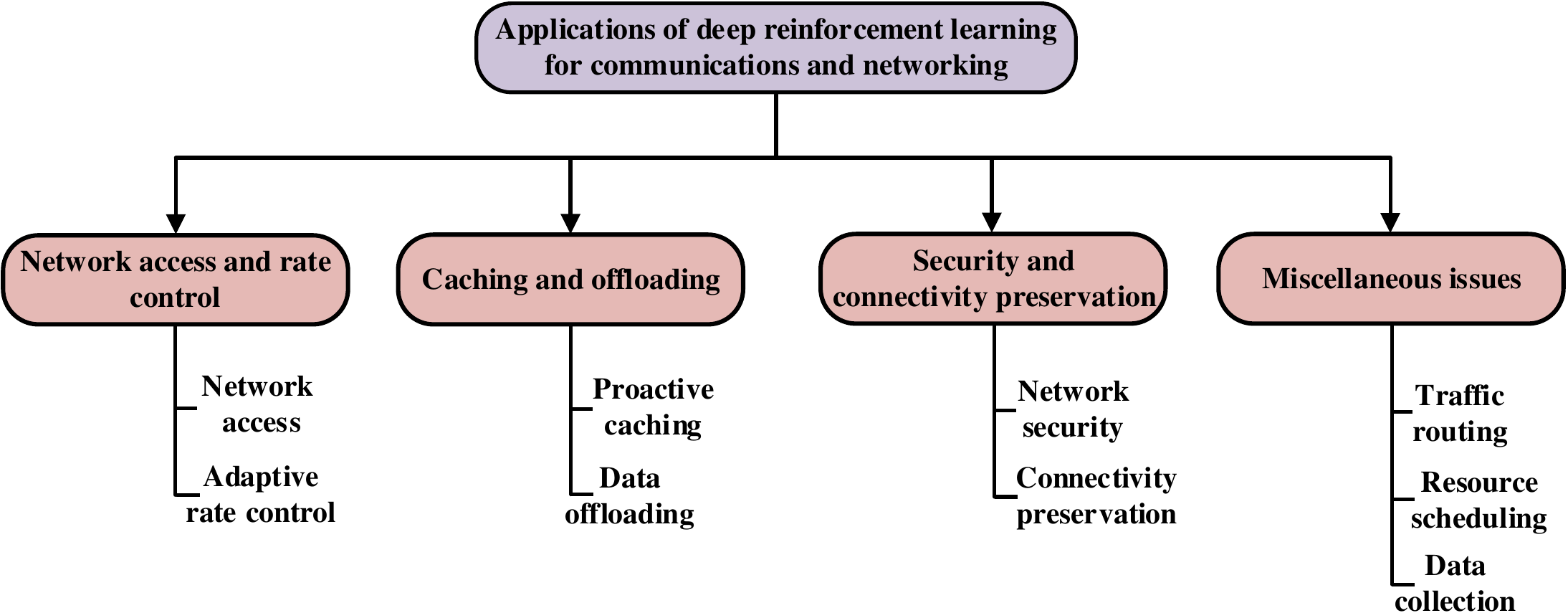}
 \caption{\small A taxonomy of the applications of deep reinforcement learning for communications and networking.}
 \label{Application_DRL}
\end{figure*}

Although there are some surveys related to DRL, they do
not focus on communications and networking. For example, the surveys of applications of DRL for computer vision and natural language processing can be found in \cite{li2017deep} and\cite{arulkumaran2017brief}. Also, there are surveys related to the use of only ``deep learning'' for networking. For example, the survey of machine learning for wireless networks is given in \cite{chen2017machine}, but it does not focus on the DRL approaches. To the best of our knowledge, there is no survey specifically discussing the applications of DRL in communications and networking. This motivates us to deliver the survey with the tutorial of DRL and the comprehensive literature review on the applications of DRL to address issues in communications and networking. For convenience, the related works in this survey are classified based on issues in communications and networking as shown in Fig.~\ref{Application_DRL}. The major issues include network access, data rate control, wireless caching, data offloading, network security, connectivity preservation, traffic routing, and data collection. Also, the percentages of DRL related works for different networks and different issues in the networks are shown in Figs.~\ref{Summary_Figure}(a) and~\ref{Summary_Figure}(b), respectively. From the figures, we observe that the majority of the related works are for the cellular networks. Also, the related works to the wireless caching and offloading have received more attention than the other issues.

\begin{table}[h!]
\scriptsize
  \caption{\small List of abbreviations}
  \label{tab:table_abb}\centering
  \begin{tabularx}{8.7cm}{|Sl|X|}
    \hline
  \cellcolor{mygray} \textbf{Abbreviation} &   \cellcolor{mygray} \textbf{Description} \\
            \hline
ANN/APF&Artificial Neural Network/Artificial Potential Field\\
          \hline
A3C&Asynchronous Advantage Actor-
Critic\\
          \hline
CRN&Cognitive Radio Network\\
          \hline
CNN&Convolutional Neural Network \\
          \hline
DRL/DQL&Deep Reinforcement Learning/Deep Q-Learning \\
          \hline
DNN&Deep Neural Network \\
          \hline
DQN/DDQN/DRQN  & Deep Q-Network/Double DQN/Deep Recurrent Q-Learning\\
%          \hline
%DRQN&Deep Recurrent Q-Learning \\
          \hline
DASH&Dynamic Adaptive Streaming over HTTP \\
          \hline
DoS&Denial-of-Service\\
            \hline
ESN&Echo State Network\\
          \hline
FNN/RNN&Feedforward Neural Network/Recurrent Neural
Network \\
          \hline
FSMC&Finite-State Markov Channel\\
          \hline
HVFT&High Volume Flexible Time\\
          \hline
ITS&Intelligent Transportation System\\
          \hline
LSM/LSTM&Liquid State Machine/Long Short-Term Memory\\
         \hline
  MEC&Mobile Edge Computing\\
                   \hline
MDP/POMDP&Markov Decision Process/Partially Observable MDP \\
          \hline
NFSP&Neural Fictitious Self-Play \\
          \hline
NFV&Network Function Virtualization\\
          \hline
RDPG&Recurrent Deterministic Policy
Gradient \\
          \hline
RCNN&Recursive Convolutional Neural Network\\
          \hline
RRH/BBU&Remote Radio Head/BaseBand Unit\\
\hline
RSSI &Received Signal Strength Indicators\\
          \hline
SPD&Sequential Prisoner's Dilemma \\
          \hline
SBS/BS&Small Base Station/Base Station \\
%          \hline
%SON&Self-Organizing Network\\
          \hline
SDN&Software-Defined Network\\
          \hline
SU/PU&Secondary User/Primary User\\
          \hline
UDN/UAN&Ultra-Density Network/Underwater Acoustic Network\\
          \hline
UAV&Unmanned Aerial Vehicle\\
          \hline
VANET/V2V&Vehicular Ad hoc Network/Vehicle-to-Vehicle\\
  \hline
  \end{tabularx}
\end{table}

The rest of this paper is organized as follows. Section~\ref{sec:ODRL} presents the introduction of reinforcement learning and discusses DRL techniques as well as their extensions.
Section~\ref{spectrum_rate_control} reviews the applications of DRL for dynamic network access and adaptive data rate control. Section~\ref{sec:caching_offloading} discusses the applications of DRL for wireless caching and data offloading. Section~\ref{sec:network_connectivity} presents DRL related works for network security and connectivity preservation. Section~\ref{sec:misc_issues} considers how to use DRL to deal with other issues in communications and networking. Important challenges,
open issues, and future research directions are outlined
in Section~\ref{sec:challenge_issue_future}. Section~\ref{sec:conclusion} concludes the paper. The list of abbreviations commonly appeared in this paper is given in Table~\ref{tab:table_abb}. Note that DRL consists of two different algorithms which are Deep Q-Learning (DQL) and policy gradients~\cite{lowe2017multi}. In particular, DQL is mostly used for the DRL related works. Therefore, in the rest of the paper, we use ``DRL'' and ``DQL'' interchangeably to refer to the DRL algorithms.

%==================================================================================
%==================================================================================
\section{Deep Reinforcement Learning: An Overview} \label{sec:ODRL}

In this section, we first present fundamental knowledge of Markov decision processes, reinforcement learning, and deep learning techniques which are important branches of machine learning theory. We then discuss DRL technique that can  capitalize on the capability of the deep learning to improve efficiency and performance in terms of the learning rate for reinforcement learning algorithms. Afterward, advanced DRL models and their extensions are reviewed.

%==============================================================
\subsection{Markov Decision Processes}

MDP~\cite{puterman2014markov} is a discrete time stochastic control process. MDP provides a mathematical framework for modeling decision-making problems in which outcomes are partly random and under control of a decision maker or an agent. MDPs are useful for studying optimization problems which can be solved by dynamic programming and reinforcement learning techniques. Typically, an MDP is defined by a tuple $(\mathcal{S}, \mathcal{A}, p, r )$ where $\mathcal{S}$ is a finite set of states, $\mathcal{A}$ is a finite set of actions, $p$ is a transition probability from state $s$ to state $s'$ after action $a$ is executed, and $r$ is the immediate reward obtained after action $a$ is performed. We denote $\pi$ as a ``policy'' which is a mapping from a state to an action. The goal of an MDP is to find an optimal policy to maximize the reward function. An MDP can be finite or infinite time horizon. For an infinite time horizon MDP, we aim to find an optimal policy $\pi^*$ to maximize the expected total reward defined by $\overset{\infty}{\underset{t=0}{\sum}} \gamma	 r_{t}(s_t,a_t)$, where $a_t=\pi^*(s_t)$, and $\gamma \in [0,1]$ is the discount factor.

%==============================================================
\subsubsection{Partially Observable Markov Decision Process}

In MDPs, we assume that the system state is fully observable by the agent. However, in many cases, the agent only can observe a part of the system state, and thus Partially Observable Markov Decision Processes (POMDPs)~\cite{monahan1982state} can be used to model the decision-making problems. A typical POMDP model is defined by a 6-tuple $(\mathcal{S}, \mathcal{A}, p, r , \Omega, \mathcal{O})$, where $\mathcal{S}, \mathcal{A}, p, r$ are defined the same as in the MDP model, $\Omega$ and $\mathcal{O}$ are defined as the set of observations and observation probabilities, respectively. At each time step, the system is at state $s \in \mathcal{S}$. Then, the agent takes an action $a \in \mathcal{A}$ and the system transits to state $s' \in \mathcal{S}$. At the same time, the agent has a new observation $o \in \Omega$ with probability $\mathcal{O} (o|s,a,s')$. Finally, the agent receives an immediate reward $r$ that is equal to $r(s,a)$ in the MDP. Similar to the MDP model, the agent in POMDP also aims to find the optimal policy $\pi^*$ in order to maximize its expected long-term discounted reward $\overset{\infty}{\underset{t=0}{\sum}} \gamma r_{t}(s_t,\pi^*(s_t))$.

%==============================================================
\subsubsection{Markov Games}

In game theory, a Markov game, or a stochastic game~\cite{Shapley1953Stochastic}, is a dynamic game with probabilistic transitions played by multiple players, i.e., agents. %The game is played in a sequence of stages. At the beginning of the game, the agents select actions and each agent receives a payoff that depends on the current system state and the chosen actions of all agents. The game then transits to a new state whose distribution depends on the previous state and the actions chosen by the agents. The procedure is repeated at the new state and play continues for a finite or infinite number of stages. The total payoff to a player is often taken to be the discounted sum of the stage payoffs or the limit inferior of the averages of the stage payoffs.
A typical Markov game model is defined by a tuple $(\mathcal{I}, \mathcal{S}, \{\mathcal{A}^i\}_{i\in\mathcal{I}}, p, \{r^i\}_{i\in\mathcal{I}})$, where
\begin{itemize}
	\item $\mathcal{I} \triangleq \{1,\ldots,i,\ldots,I\}$ is a set of agents,
	\item $\mathcal{S} \triangleq \{ \mathcal{S}^1, \ldots, \mathcal{S}^i, \ldots, \mathcal{S}^I \}$ is the global state space of the all agents with $\mathcal{S}^i$ being the state space of agent $i$,
	\item $\{\mathcal{A}^i\}_{i\in\mathcal{I}}$ are sets of action spaces of the agents with $\mathcal{A}^i$ being the action space of agent $i$,
	\item $p \triangleq \mathcal{S} \times \mathcal{A}^1 \times \cdots \times \mathcal{A}^I \rightarrow [0,1] $ is the transition probability function of the system.
	\item $\{r^i\}_{i\in\mathcal{I}}$ are payoff functions of the agents with \\ $r^i \triangleq \mathcal{S} \times \mathcal{A}^1 \times \cdots \times \mathcal{A}^I \rightarrow \mathbb{R}$, i.e., the payoff of agent $i$ obtained after all actions of the agents are executed.
\end{itemize}

In a Markov game, the agents start at some initial state $s_0 \in \mathcal{S}$. After observing the current state, all the agents simultaneously select their actions $a=\{a^1,\ldots,a^I\}$ and they will receive their corresponding rewards together with their own new observations. At the same time, the system will transit to a new state $s' \in \mathcal{S}$ with probability $p(s'|s,a)$. The procedure is repeated at the new state and continues for a finite or infinite number of stages. In this game, all the agents try to find their optimal policies to maximize their own expected long-term average rewards, i.e., $\overset{\infty}{\underset{t=0}{\sum}} \gamma_i	 r^i_{t}(s_t,\pi_i^*(s_t))$, $\forall i$. The set of all optimal policies of this game, i.e., $\{\pi^*_1,\ldots,\pi^*_I\}$ is known to be the equilibrium of this game. If there is a finite number of players and the sets of states and actions are finite, then the Markov game always has a Nash equilibrium~\cite{Hu2003Nash} under a finite number of stages. The same is true for Markov games with infinite stages, but the total payoff of agents is the discounted sum~\cite{Hu2003Nash}.

%===============================================
%===============================================
\subsection{Reinforcement Learning}

\begin{figure*}[!htb]
	\begin{center}
		$\begin{array}{ccc}
		\epsfxsize=2.1 in \epsffile{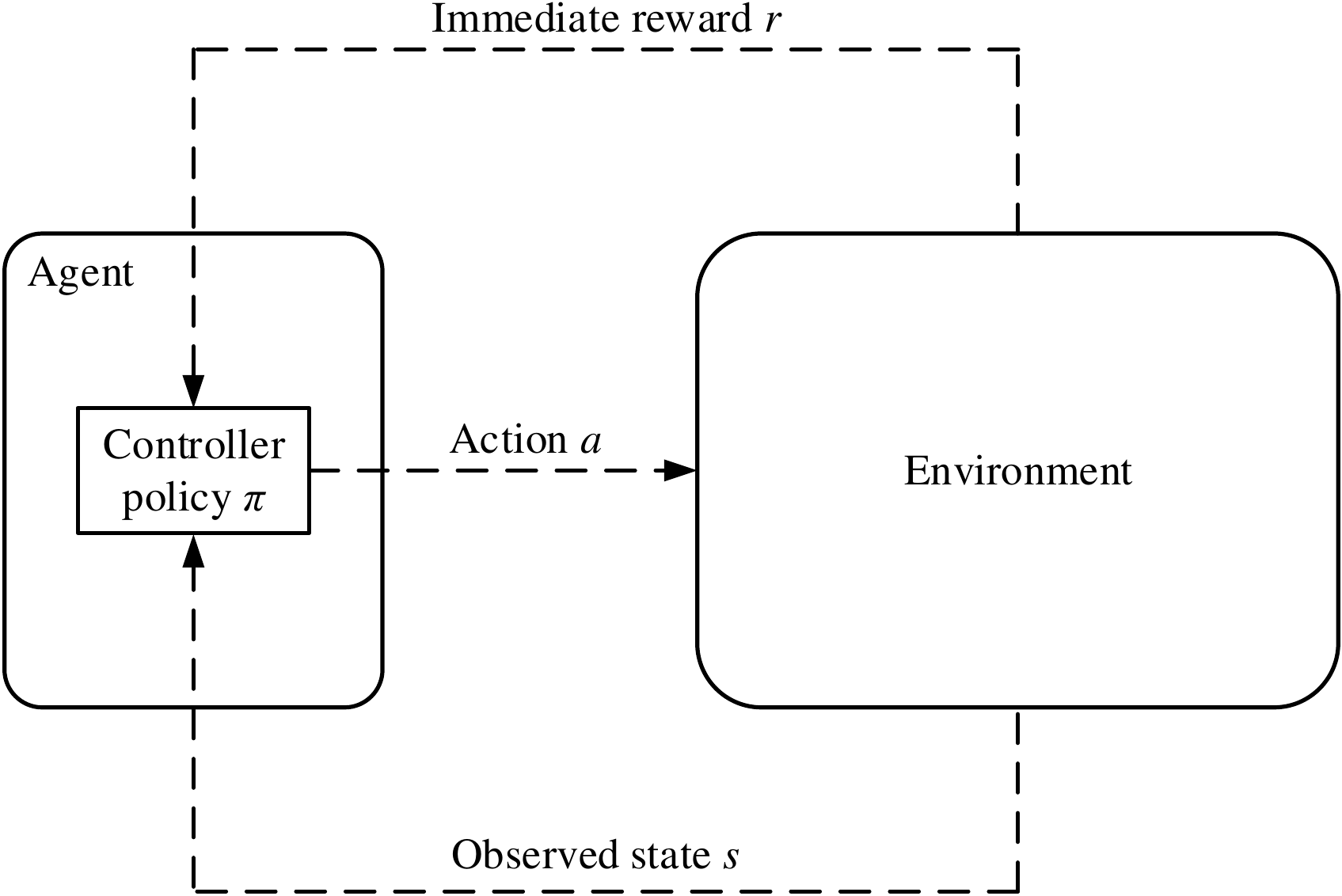} &
		\epsfxsize=2.1 in \epsffile{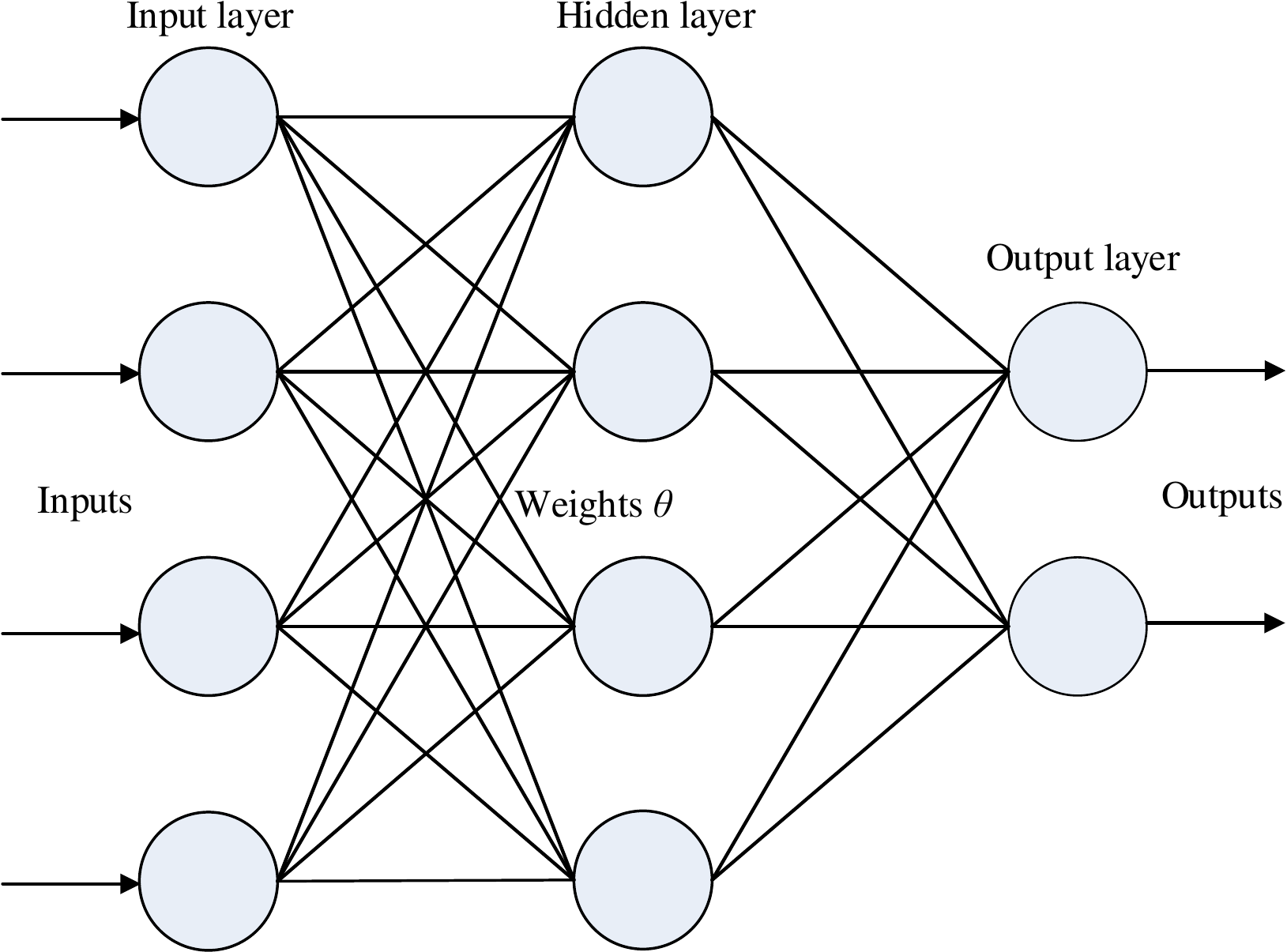} &
		\epsfxsize=2.8 in \epsffile{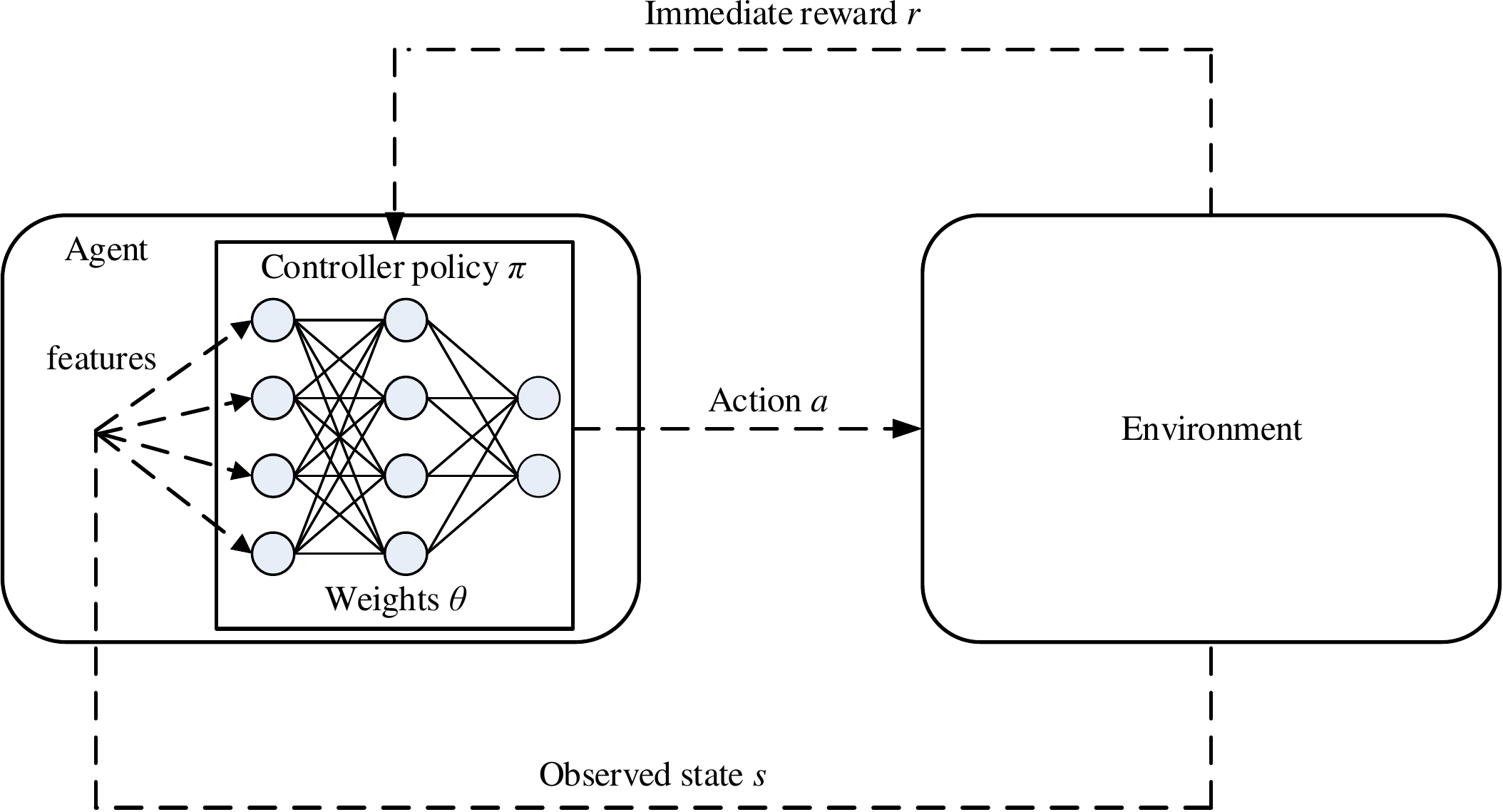} \\
		\text{(a)} & \text{(b)} & \text{(c)}
		\end{array}$
		\caption{(a) Reinforcement learning, (b) Artificial neural network, and (c) Deep Q-learning.}
		\label{fig:RL_DNN_DQL}
	\end{center}
\end{figure*}

Reinforcement learning, an important branch of machine learning, is an effective tool and widely used in the literature to address MDPs~\cite{sutton1998reinforcement}. In a reinforcement learning process, an agent can learn its optimal policy through interaction with its environment. In particular, the agent first observes its current state, and then takes an action, and receives its immediate reward together with its new state as illustrated in Fig.~\ref{fig:RL_DNN_DQL}(a). The observed information, i.e., the immediate reward and new state, is used to adjust the agent's policy, and this process will be repeated until the agent's policy approaches to the optimal policy. In reinforcement learning, $Q$-learning is the most effective method and widely used in the literature. In the following, we will discuss the $Q$-learning algorithm and its extensions for advanced MDP models.

%\begin{figure}[h]
%	\centering
%	%\includegraphics[width=\linewidth]{Application_pricing_model_change}
%	\includegraphics[width=\linewidth]{Figure/RL}
%	\caption{\small Reinforcement learning.}
%	\label{fig:RL}
%\end{figure}

%==============================================================
\subsubsection{$Q$-Learning Algorithm}

%Dusit: You may consider to add one figure to show relation between state, action, agent, and environment for reinforcement learning. This can be different and can be used to compare with Fig. 5.

In an MDP, we aim to find an optimal policy $\pi^*: \mathcal{S} \rightarrow \mathcal{A}$ for the agent to minimize the overall cost for the system. Accordingly, we first define value function $\mathcal{V}^\pi: \mathcal{S} \rightarrow \mathbb{R}$ that represents the expected value obtained by following policy $\pi$ from each state $s \in \mathcal{S}$. The value function $\mathcal{V}$ for policy $\pi$ quantifies the goodness of the policy through an infinite horizon and discounted MDP that can be expressed as follows:
\begin{equation}
\begin{aligned}
\mathcal{V}^\pi(s) & =  \mathbb{E}_{\pi} \Big[	\sum_{t=0}^{\infty} \gamma r_{t}(s_t,a_t) |s_0=s 	\Big]	 \\
& = \mathbb{E}_{\pi}\Big[ r_{t}(s_t,a_t) +  \gamma \mathcal{V}^\pi(s_{t+1}) |s_0=s 	 \Big] .
\end{aligned}		
\end{equation}

Since we aim to find the optimal policy $\pi^*$, an optimal action at each state can be found through the optimal value function expressed by $\mathcal{V}^{*}(s) = \underset{a_t}{\max} \Big\{ \mathbb{E}_{\pi}\big[r_{t}(s_t,a_t) +  \gamma \mathcal{V}^\pi(s_{t+1}) \big]	\Big\}$	.

If we denote $\mathcal{Q}^{*}(s,a) \triangleq r_{t}(s_t,a_t) +  \gamma \mathbb{E}_{\pi} \big[ \mathcal{V}^\pi(s_{t+1})	 \big]$ as the optimal $Q$-function for all state-action pairs, then the optimal value function can be written by $\mathcal{V}^{*}(s) = \underset{a}{\max} \big\{	\mathcal{Q}^{*}(s,a) \big\}$. Now, the problem is reduced to find optimal values of $Q$-function, i.e., $\mathcal{Q}^{*}(s,a)$, for all state-action pairs, and this can be done through iterative processes. In particular, the $Q$-function is updated according to the following rule:
\begin{equation}
\begin{aligned}
\label{eq:Q-update}
\mathcal{Q}_{t+1}(s,a)  = & \mathcal{Q}_{t}(s,a) + \\
& \alpha_t \Big[ r_t(s,a) +   \gamma \max_{a'}	\mathcal{Q}_{t}(s,a') 	- 	 \mathcal{Q}_{t}(s,a)		\Big]	.
\end{aligned}
\end{equation}
The core idea behind this update is to find the Temporal Difference (TD) between the predicted $\mathcal{Q}$-value, i.e., $r_{t}(s,a) +   \gamma \underset{a'}{\max}	 \mathcal{Q}_{t}(s,a')$ and its current value, i.e., $\mathcal{Q}_{t}(s,a)$. In~(\ref{eq:Q-update}), the learning rate $\alpha_t$ is used to determine the impact of new information to the existing $\mathcal{Q}$-value. The learning rate can be chosen to be a constant, or it can be adjusted dynamically during the learning process. However, it must satisfy Assumption~\ref{ass:step-size} to guarantee the convergence for the $Q$-learning algorithm.

\begin{assumption}
\label{ass:step-size}
The step size $\alpha_{t}$ is deterministic, nonnegative and satisfies the following conditions: $\alpha_t \in [0,1]$, $\overset{\infty}{\underset{t=0}{\sum}} \alpha_{t} = \infty$, and $\phantom{5} \overset{\infty}{\underset{t=0}{\sum}} ( \alpha_{t} )^{2} < \infty$ .	
\end{assumption}
The step size adaptation $\alpha_t=\frac{1}{t}$ is one of the most common examples used in reinforcement learning. More discussions for selecting an appropriate step size can be found in~\cite{Dabney2014Thesis}. The details of the $Q$-learning algorithm are then provided in Algorithm~\ref{alg:Q-learning}.

%******************************************************
\begin{algorithm}
	\caption{The $Q$-learning algorithm}
	\label{alg:Q-learning}
	\begin{algorithmic}[0]
		%-------------- Initialize -----------------
		\STATE \textbf{Input:} For each state-action pair $(s,a)$, initialize the table entry $\mathcal{Q}(s,a)$ arbitrarily, e.g., to zero.
		Observe the current state $s$, initialize a value for the learning rate $\alpha$ and the discount factor $\gamma$.
		%--------------- for loop -----------------------
		\FOR{$t:=1$ to $T$}
		\STATE From the current state-action pair $(s,a)$, execute action $a$ and obtain the immediate reward $r$ and a new state $s'$.
		\STATE Select an action $a'$ based on the state $s'$ and then update the table entry for $\mathcal{Q}(s,a)$ as follows:
		\begin{equation}
		\begin{aligned}
		\mathcal{Q}_{t+1}(s,a) & \leftarrow  \mathcal{Q}_{t}(s,a) + \alpha_t \Big[ r_t(s,a) +   \\
		& \gamma \max_{a'}	\mathcal{Q}_{t}(s',a') 	- 	\mathcal{Q}_{t}(s,a)		 \Big]
		\end{aligned}
		\label{eq:Q-learning_update}
		\end{equation}
		\STATE Replace $s \leftarrow s'$.
		\ENDFOR
		\STATE \textbf{Output:} $\pi^*(s)=\arg \max_{a} \mathcal{Q}^*(s,a)$.
	\end{algorithmic}
\end{algorithm}
%******************************************************

Once either all $\mathcal{Q}$-values converge or a certain number of iterations is reached, the algorithm will terminate. The algorithm then yields the optimal policy indicating an action to be taken at each state such that $\mathcal{Q}^*(s,a)$ is maximized for all states in the state space, i.e., $\pi^*(s)=\arg \underset{a}{\max} \mathcal{Q}^*(s,a)$. Under the assumption of the step size (i.e., Assumption~\ref{ass:step-size}), it is proved in~\cite{watkins1992q} that the $Q$-learning algorithm converges to the optimum action-values with probability one.

%\begin{remark}
%\label{rem:epsilon}
%In practice, selecting action $a$ can be done through using a popular method, i.e., $\epsilon$-greedy strategy. Under this strategy, the agent chooses a random action with probability $\epsilon$, and otherwise the agent selects an action that maximizes $\mathcal{Q}(s,a)$. The random action is necessary in exploring the whole state space. Thus, in $\mathcal{Q}$-learning algorithm, we need to balance between the exploration, i.e., $\epsilon$, and the exploitation, i.e., $1-\epsilon$.
%\end{remark}

%==============================================================
\subsubsection{SARSA: An Online Q-Learning Algorithm}

Although the $Q$-learning algorithm can find the optimal policy for the agent without requiring knowledge about the environment, this algorithm works in an offline fashion. In particular, Algorithm~\ref{alg:Q-learning} can obtain the optimal policy only after all $\mathcal{Q}$-values converge. Therefore, this section presents an alternative online learning algorithm, i.e., the SARSA algorithm, which allows the agent to approach the optimal policy in an online fashion.

Different from the $Q$-learning algorithm, the SARSA algorithm is an online algorithm which allows the agent to choose optimal actions at each time step in a real-time fashion without waiting until the algorithm converges. In the $Q$-learning algorithm, the policy is updated according to the maximum reward of available actions regardless of which policy is applied, i.e., an off-policy method. In contrast, the SARSA algorithm interacts with the environment and updates the policy directly from the actions taken, i.e., an on-policy method. Note that the SARSA algorithm updates $\mathcal{Q}$-values from the quintuple $\mathcal{Q}(s,a,r,s',a')$.

%==============================================================
\subsubsection{Q-Learning for Markov Games}

To apply Q-learning algorithm to the Markov game context, we first define the $Q$-function for agent $i$ by $\mathcal{Q}_i(s,a^i,\textbf{a}^{-i})$, where $\textbf{a}^{-i} \triangleq \{a^1,\ldots,a^{i-1},a^{i+1},\ldots,a^I\}$ denotes the set of actions of all agents except $i$. Then, the Nash $Q$-function of agent $i$ is defined by:
\begin{equation}
\begin{aligned}
 \mathcal{Q}^*_i(s,a^i,\textbf{a}^{-i}) & =  r^i(s,a^i,\textbf{a}^{-i}) + \\
& \beta \sum_{s' \in \mathcal{S}}  p(s'|s,a^i,\textbf{a}^{-i}) \mathcal{V}^i(s',\pi_1^*,\ldots,\pi_I^*),
\end{aligned}
\end{equation}
where $(\pi_1^*,\ldots,\pi_I^*)$ is the joint Nash equilibrium strategy, $r^i(s,a^i,\textbf{a}^{-i})$ is agent $i$'s immediate reward in state $s$ under the joint action $(a^i,\textbf{a}^{-i})$, and $\mathcal{V}^i(s',\pi_1^*,\ldots,\pi_I^*)$ is the total discounted reward over an infinite time horizon starting from state $s'$ given that all the agents follow the equilibrium strategies.

In~\cite{Hu2003Nash}, the authors propose a multi-agent Q-learning algorithm for general-sum Markov games which allows the agents to perform updates based on assuming Nash equilibrium behavior over the current Q-values. In particular, agent $i$ will learn its $Q$-values by forming an arbitrary guess from starting time of the game. At each time step $t$, agent $i$ observes the current state and takes an action $a^i$. Then, it observes its immediate reward $r^i$, actions taken by others $\textbf{a}^{-i}$, others' immediate rewards, and the new system state $s'$. After that, agent $i$ calculates a Nash equilibrium $(\pi_1(s'),\ldots,\pi_I(s'))$ for the state game $(\mathcal{Q}_1^t(s'),\ldots,\mathcal{Q}_I^t(s'))$, and updates its $Q$-values according to:
\begin{equation}
\mathcal{Q}_i^{t+1} (s,a^i,\textbf{a}^{-i}) = (1-\alpha_t) \mathcal{Q}_i^{t} (s,a^i,\textbf{a}^{-i}) + \alpha_t [r_t^i + \gamma \mathscr{N}_t^i(s') ] ,
\label{eq:Markov_Q_learing}
\end{equation}
where $\alpha_t \in (0,1)$ is the learning rate and $\mathscr{N}_t^i(s') \triangleq \mathcal{Q}_i^{t}(s') \times \pi_1(s') \times \cdots \times \pi_I(s')$.

In order to calculate the Nash equilibrium, agent $i$ needs to know $(\mathcal{Q}_1^t(s'),\ldots,\mathcal{Q}_I^t(s'))$. However, the information about other agents' $\mathcal{Q}$-values is not given, and thus agent $i$ must learn this information too. To do so, agent $i$ will set estimations about others' $\mathcal{Q}$-values at the beginning of the game, e.g., $\mathcal{Q}_0^j (s,a^i,\textbf{a}^{-i}) = 0, \forall j, s$. As the game proceeds, agent $i$ observes other agents' immediate rewards and previous actions. That information can then be used to update agent $i$'s conjectures on other agents' $Q$-functions. Agent $i$ updates its beliefs about agent $j$'s $Q$-function, according to the same updating rule in~(\ref{eq:Markov_Q_learing}). Then, the authors prove that under some highly restrictive assumptions on the form of the state games during learning, the proposed multi-agent $Q$-learning algorithm is guaranteed to be converged.

%===============================================
%===============================================
\subsection{Deep  Learning}

Deep learning~\cite{goodfellow2016deep} is composed of a set of algorithms and techniques that attempt to find important features of data and to model its high-level abstractions. The main goal of deep learning is to avoid manual description of a data structure (like hand-written features) by automatic learning from the data. Its name refers to the fact that typically any neural network with two or more hidden layers is called DNN. Most deep learning models are based on an Artificial Neural Network (ANN), even though they can also include propositional formulas or latent variables organized layer-wise in deep generative models such as the nodes in Deep Belief Networks and Deep Boltzmann Machines.

An ANN is a computational nonlinear model based on the neural structure of the brain that is able to learn to perform tasks such as classification, prediction, decision-making, and visualization. An ANN consists of artificial neurons and is organized into three interconnected layers: input, hidden, and output as illustrated in Fig.~\ref{fig:RL_DNN_DQL}(b). The input layer contains input neurons that send information to the hidden layer. The hidden layer sends data to the output layer. Every neuron has weighted inputs (synapses), an activation function (defines the output given an input), and one output. Synapses are the adjustable parameters that convert a neural network to a parameterized system.

%\begin{figure}[h]
%	\centering
%	\includegraphics[width=\linewidth]{Figure/ArtificialNeuralNetwork}
%	\caption{\small Artificial Neural Network.}
%	\label{fig:ANN}
%\end{figure}

During the training phase, ANNs use backpropagation as an effective learning algorithm to compute quickly a gradient descent with respect to the weights. Backpropagation is a special case of automatic differentiation. In the context of learning, backpropagation is commonly used by the gradient descent optimization algorithm to adjust the weights of neurons by calculating the gradient of the loss function. This technique is also sometimes called backward propagation of errors, because the error is calculated at the output and distributed back through the network layers.

\begin{figure}[h]
	\centering
	\includegraphics[width=\linewidth]{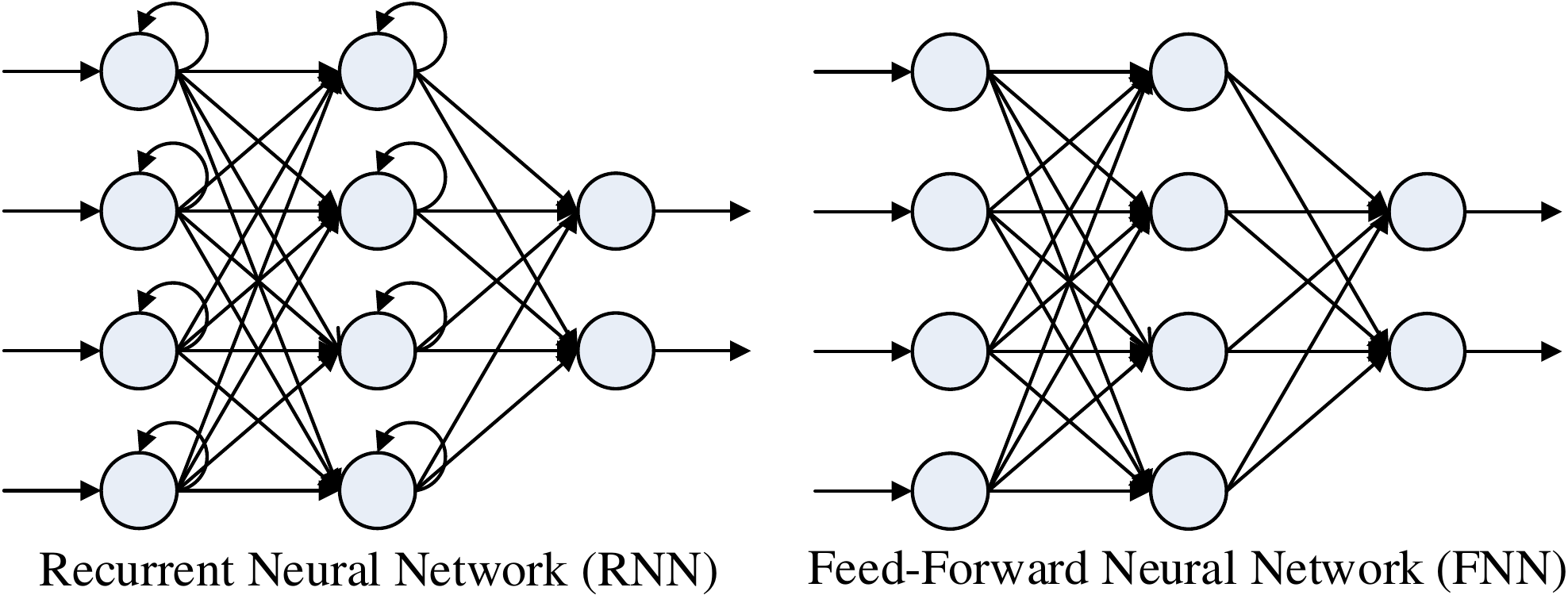}
	\caption{\small RNN vs CNN.}
	\label{fig:RNNvsCNN}
\end{figure}

A DNN is defined as an ANN with multiple hidden layers. There are two typical DNN models, i.e., Feedforward Neural Network (FNN) and Recurrent Neural Network (RNN). In the FNN, the information moves in only one direction, i.e., from the input nodes, through the hidden nodes and to the output nodes, and there are no cycles or loops in the network as shown in Fig.~\ref{fig:RNNvsCNN}. In FNNs, Convolutional Neural Network (CNN) is the most well known model with a wide range of applications especially in image and speech recognition. The CNN contains one or more convolutional layers, pooling or fully connected, and uses a variation of multilayer perceptrons discussed above. Convolutional layers use a convolution operation to the input passing the result to the next layer. This operation allows the network to be deeper with much fewer parameters.

Unlike FNNs, the RNN is a variant of a recursive artificial neural network in which connections between neurons make directed cycles. It means that an output depends not only on its immediate inputs, but also on the previous further step's neuron state. The RNNs are designed to utilize sequential data, when the current step has some relation with the previous steps. This makes the RNNs ideal for applications with a time component, e.g., time-series data, and natural language processing. However, all RNNs have feedback loops in the recurrent layer. This lets RNNs maintain information in memory over time. Nevertheless, it can be difficult to train standard RNNs to solve problems that require learning long-term temporal dependencies. The reason is that the gradient of the loss function decays exponentially with time, which is called the vanishing gradient problem. Thus, Long Short-Term Memory (LSTM) is often used in RNNs to address this issue. The LSTM is designed to model temporal sequences and their long-range dependencies are more accurate than conventional RNNs. The LSTM does not use an activation function within its recurrent components, the stored values are not modified, and the gradient does not tend to vanish during training. Usually, LSTM units are implemented in ``blocks'' with several units. These blocks have three or four ``gates'', e.g., input gate, forget gate, output gate, that control information flow drawing on the logistic function.

%===============================================
%===============================================
\subsection{Deep $Q$-Learning}

The $Q$-learning algorithm can efficiently obtain an optimal policy when the state space and action space are small. However, in practice, with complicated system models, these spaces are usually large. As a result, the $Q$-learning algorithm may not be able to find the optimal policy. Thus, Deep $Q$-Learning (DQL) algorithm is introduced to overcome this shortcoming. Intuitively, the DQL algorithm implements a Deep $Q$-Network (DQN), i.e., a DNN, instead of the $Q$-table to derive an approximate value of $Q^{*} (s,a)$ as shown in Fig.~\ref{fig:RL_DNN_DQL}(c).

%\begin{figure}[h]
%	\centering
%	%\includegraphics[width=\linewidth]{Application_pricing_model_change}
%	\includegraphics[width=\linewidth]{Figure/DeepQLearning}
%	\caption{\small Deep Q-learning.}
%	\label{fig:DQL}
%\end{figure}

As stated in~\cite{mnih2015human1}, the average reward obtained by reinforcement learning algorithms may not be stable or even diverge when a nonlinear function approximator is used. This stems from the fact that a small change of $\mathcal{Q}$-values may greatly affect the policy. Thus, the data distribution and the correlations between the $\mathcal{Q}$-values and the target values $R+\gamma \max_{a'} \mathcal{Q}(s',a')$ are varied. To address this issue, two mechanisms, i.e., experience replay and target $Q$-network, can be used.
\begin{itemize}
	\item \emph{Experience replay mechanism}: The algorithm first initializes a replay memory $\mathbf{D}$, i.e., the memory pool, with transitions $(s_t, a_t, r_t, s_{t+1})$, i.e., experiences, generated randomly, e.g., through using $\epsilon$-greedy policy. Then, the algorithm randomly selects samples, i.e., minibatches, of transitions from $\mathbf{D}$ to train the DNN. The Q-values obtained by the trained DNN will be used to obtain new experiences, i.e., transitions, and these experiences will be then stored in the memory pool $\mathbf{D}$. This mechanism allows the DNN trained more efficiently by using both old and new experiences. In addition, by using the experience replay, the transitions are more independent and identically distributed, and thus the correlations between observations can be removed.
	\item \textit{Fixed target $Q$-network:} In the training process, the $\mathcal{Q}$-value will be shifted. Thus, the value estimations can be out of control if a constantly shifting set of values is used to update the $Q$-network. This leads to the destabilization of the algorithm. To address this issue, the target $Q$-network is used to update frequently but slowly the primary $Q$-networks' values. In this way, the correlations between the target and estimated $\mathcal{Q}$-values are significantly reduced, thereby stabilizing the algorithm.	
\end{itemize}

The DQL algorithm with experience replay and fixed target $Q$-network is presented in Algorithm~\ref{alg:DQL_ER_FTN}. DQL inherits and promotes advantages of both reinforcement and deep learning techniques, and thus it has a wide range of applications in practice such as game development~\cite{alphago}, transportation~\cite{lin2018efficient}, and robotics~\cite{gu2017deep}.

%++++++++++++++++++++++++++++++++++++++++++++++++++++++++++++++++++++
\begin{algorithm}[!]
	\caption{The DQL Algorithm with Experience Replay and Fixed Target $Q$-Network}
	\label{alg:DQL_ER_FTN}
	\begin{algorithmic}[1]
		\STATE Initialize replay memory $\mathbf{D}$.
		\STATE Initialize the $Q$-network $\mathbf{Q}$ with random weights $\boldsymbol{\theta}$.
		\STATE Initialize the target $Q$-network $\hat{\mathbf{Q}}$ with random weights $\boldsymbol{\theta'}$.
		\FOR{\textit{episode=1 to T}}
		\STATE With probability $\epsilon$ select a random action $a_t$, otherwise select $a_t=\arg \max \mathcal{Q}^*(s_t, a_t, \boldsymbol{\theta})$.
		\STATE Perform action $a_t$ and observe immediate reward $r_t$ and next state $s_{t+1}$.
		\STATE Store transition $(s_t, a_t, r_t, s_{t+1})$ in $\mathbf{D}$.
		\STATE Select randomly samples c$(s_j, a_j, r_j, s_{j+1})$ from $\mathbf{D}$.
		\STATE The weights of the neural network then are optimized by using stochastic gradient descent with respect to the network parameter $\boldsymbol{\theta}$ to minimize the loss:
		\begin{equation}
		\label{eq:DQL}
		\Big[ r_j + \gamma\max_{a_{j+1}} \hat{\mathcal{Q}}(s_{j+1},a_{j+1};\boldsymbol{\theta'}) - \mathcal{Q}(s_j,a_j;\boldsymbol{\theta}) \Big]^2 .
		\end{equation}
		\STATE Reset $\hat{\mathbf{Q}} = \mathbf{Q}$ after every a fixed number of steps.
		\ENDFOR
	\end{algorithmic}
\end{algorithm}
%++++++++++++++++++++++++++++++++++++++++++++++++++++++++++++++++++++

%===============================================
%===============================================
\subsection{Advanced Deep $Q$-Learning Models }

%----------------------------------
%----------------------------------
\subsubsection{Double Deep  $Q$-Learning}

In some stochastic environments, the $Q$-learning algorithm performs poorly due to the large over-estimations of action values~\cite{thrun1993issues}. These over-estimations result from a positive bias that is introduced because $Q$-learning uses the maximum action value as an approximation for the maximum expected action value as shown in Eq.~(\ref{eq:Q-learning_update}). The reason is that the same samples are used to decide which action is the best, i.e., with highest expected reward, and the same samples are also used to estimate that action-value. Thus, to overcome the over-estimation problem of the $Q$-learning algorithm, the authors in~\cite{hasselt2010double} introduce a solution using two $Q$-value functions, i.e., $\mathcal{Q}_{1}$ and $\mathcal{Q}_{2}$, to simultaneously select and evaluate action values through the loss function as follows: $\Big[ r_j + \gamma \mathcal{Q}_{2} \Big(s_{j+1}, \arg \underset{a_{j+1}}{\max} \mathcal{Q}_{1} \big(s_{j+1}, a_{j+1};\boldsymbol{\theta}_{1} \big); \boldsymbol{\theta}_{2} \Big) - \mathcal{Q}_{1}(s_j,a_j;\boldsymbol{\theta}_{1}) \Big]^2$.

Note that the selection of an action, in the $\arg \max$, is still due to the online weights $\boldsymbol{\theta}_1$. This means that, as in $Q$-learning, we are still estimating the value of the greedy policy according to the current values, as defined by $\boldsymbol{\theta}_1$. However, the second set of weights $\boldsymbol{\theta}_2$ is used to evaluate fairly the value of this policy. This second set of weights can be updated symmetrically by switching the roles of $\boldsymbol{\theta}_1$ and $\boldsymbol{\theta}_2$. Inspired by this idea, the authors in~\cite{hasselt2010double} then develop Double Deep $Q$-Learning (DDQL) model~\cite{hasselt2016deep} using a Double Deep Q-Network (DDQN) with the loss function updated as follows:

\begin{equation}
\begin{aligned}
\Big[ r_j & + \gamma \hat{\mathcal{Q}} \Big(s_{j+1}, \arg \max_{a_{j+1}} \mathcal{Q} \big(s_{j+1}, a_{j+1};\boldsymbol{\theta} \big); \boldsymbol{\theta'} \Big) \\
& - \mathcal{Q}(s_j,a_j;\boldsymbol{\theta}) \Big]^2.
\label{eq:DDQL_update}
\end{aligned}
\end{equation}

Unlike double $Q$-learning, the weights of the second network $\boldsymbol{\theta_2}$ are replaced with the weights of the target networks $\boldsymbol{\theta'}$ for the evaluation of the current greedy policy as shown in Eq.~(\ref{eq:DDQL_update}). The update to the target network stays unchanged from DQN, and remains a periodic copy of the online network. Due to the effectiveness of DDQL, there are some applications of DDQL introduced recently to address dynamic spectrum access problems in multichannel wireless networks~\cite{naparstek2017deep} and resource allocation in heterogeneous networks~\cite{nan2018deep}.

%----------------------------------
%----------------------------------
\subsubsection{Deep  $Q$-Learning with Prioritized Experience Replay}

Experience replay mechanism allows the reinforcement learning agent to remember and reuse experiences, i.e., transitions, from the past. In particular, transitions are uniformly sampled from the replay memory $\mathbf{D}$. However, this approach simply replays transitions at the same frequency as that the agent was originally experienced, regardless of their significance. Therefore, the authors in~\cite{schaul2015prioritized1} develop a framework for prioritizing experiences, so as to replay important transitions more frequently, and therefore learn more efficiently. Ideally, we want to sample more frequently those transitions from which there is much to learn. As a proxy for learning potential, the proposed Prioritized Experience Replay (PER)~\cite{schaul2015prioritized1} samples transitions with probability $p_t$ relative to the last encountered absolute error defined as follows:
\begin{equation}
p_t \varpropto \Big| r_j + \gamma\max_{a'} \hat{\mathcal{Q}}(s_{j+1},a';\boldsymbol{\theta'}) - \mathcal{Q}(s_j,a_j;\boldsymbol{\theta)}	\Big|^\omega,
\end{equation}
where $\omega$ is a hyper-parameter that determines the shape of the distribution. New transitions are inserted into the replay buffer with maximum priority, providing a bias towards recent transitions. Note that stochastic transitions may also be favoured, even when there is little left to learn about them. Through real experiments on many Atari games, the authors demonstrate that DQL with PER outperforms DQL with uniform replay on 41 out of 49 games. However, this solution is only appropriate to implement when we can find and define the important experiences in the replay memory $\mathbf{D}$.

%----------------------------------
%----------------------------------
\subsubsection{Dueling Deep $Q$-Learning}

The $Q$-values, i.e., $\mathcal{Q}(s,a)$, used in the $Q$-learning algorithm, i.e., Algorithm~\ref{alg:Q-learning}, are to express how good it is to take a certain action at a given state. The value of an action $a$ at a given state $s$ can actually be decomposed into two fundamental values. The first value is the state-value function, i.e., $\mathscr{V}(s)$, to estimate the importance of being in a particular state $s$. The second value is the action-value function, i.e., $\mathscr{A}(a)$, to estimate the importance of selecting an action $a$ compared with other actions. As a result, the $Q$-value function can be expressed by two fundamental value functions as follows: $\mathcal{Q}(s,a) = \mathscr{V}(s) + \mathscr{A}(a)$.

Stemming from the fact that in many MDPs, it is unnecessary to estimate both values, i.e., action and state values of Q-function $\mathcal{Q}(s,a)$, at the same time. For example, in many racing games, moving left or right matters if and only if the agent meets the obstacles or enemies. Inspired by this idea, the authors in~\cite{wang2016dueling} introduce an idea of using two streams, i.e., two sequences, of fully connected layers instead of using a single sequence with fully connected layers for the DQN. The two streams are constructed such that they are able to provide separate estimations on the action and state value functions, i.e., $\mathscr{V}(s)$ and $\mathscr{A}(a)$. Finally, the two streams are combined to generate a single output $\mathcal{Q}(s,a)$ as follows:
\begin{equation}
\begin{aligned}
\mathcal{Q} (s,a;\boldsymbol{\alpha}, \boldsymbol{\beta}) =  \mathscr{V}(s;\boldsymbol{\beta}) + \Big( \mathscr{A} (s,a;\boldsymbol{\alpha}) - \frac{\sum_{a'} \mathscr{A} (s,a';\boldsymbol{\alpha})}{|\mathcal{A}|}		\Big) ,
\end{aligned}
\end{equation}
where $\boldsymbol{\beta}$ and $\boldsymbol{\alpha}$ are the parameters of the two streams $\mathscr{V}(s;\boldsymbol{\beta})$ and $\mathscr{A} (s,a';\boldsymbol{\alpha})$, respectively. Here, $|\mathcal{A}|$ is the total number of actions in the action space $\mathcal{A}$. Then, the loss function is derived in the similar way to~(\ref{eq:DQL}) as follows: \\
$\Big[ r_j + \gamma \underset{a_{j+1}}{\max} \hat{\mathcal{Q}}(s_{j+1},a_{j+1};\boldsymbol{\alpha'}, \boldsymbol{\beta'}) - \mathcal{Q}(s_j,a_j;\boldsymbol{\alpha}, \boldsymbol{\beta}) \Big]^2$. Through the simulation, the authors show that the proposed dueling DQN can outperform DDQN~\cite{hasselt2016deep} in 50 out of 57 learned Atari games. However, the proposed dueling architecture only clearly benefits for MDPs with large action spaces. For small state spaces, the performance of dueling DQL is even not as good as that of double DQL as shown in simulation results in~\cite{wang2016dueling}.

%----------------------------------
%----------------------------------
\subsubsection{Asynchronous Multi-step Deep Q-Learning}

Most of the $Q$-learning methods such as DQL and dueling DQL rely on the experience replay method. However, such kind of method has several drawbacks. For example, it uses more memory and computation resources per real interaction, and it requires off-policy learning algorithms that can update from data generated by an older policy. This limits the applications of DQL. Therefore, the authors in~\cite{mnih2016asynchronous} introduce a method using multiple agents to train the DNN in parallel. In particular, the authors propose a training procedure which utilizes asynchronous gradient decent updates from multiple agents at once. Instead of training one single agent that interacts with its environment, multiple agents are interacting with their own version of the environment simultaneously. After a certain amount of timesteps, accumulated gradient updates from an agent are applied to a global model, i.e., the DNN. These updates are asynchronous and lock free. In addition, to tradeoff between bias and variance in the policy gradient, the authors adopt $n$-step updates method~\cite{sutton1998reinforcement} to update the reward function. In particular, the truncated $n$-step reward function can be defined by $r_t^{(n)} = \underset{k=0}{\overset{n-1}{\sum}} \gamma^{(k)} r_{t+k+1}$. Thus, the alternative loss for each agent will be derived by:
\begin{equation}
\Big[ r_j^{(n)} + \gamma_j^{(n)} \max_{a'} \hat{\mathcal{Q}}(s_{j+n},a';\boldsymbol{\theta'}) - \mathcal{Q}(s_j,a_j;\boldsymbol{\theta}) \Big]^2 .
\end{equation}

The effects of training speed and quality of the proposed asynchronous DQL with multi-step learning are analyzed for various reinforcement learning methods, e.g., 1-step $Q$-learning, 1-step SARSA, and n-step $Q$-learning. They show that asynchronous updates have a stabilizing effect on policy and value updates. Also, the proposed method outperforms the current state-of-the-art algorithms on the Atari games while training for half of the time on a single multi-core CPU instead of a GPU. As a result, some recent applications of asynchronous DQL have been developed for handover control problems in wireless systems~\cite{wang2018handover}

%----------------------------------
%----------------------------------
\subsubsection{Distributional  Deep Q-learning}

All aforementioned methods use the Bellman equation to approximate the expected value of future rewards. However, if the environment is stochastic in nature and the future rewards follow multimodal distribution, choosing actions based on expected value may not lead to the optimal outcome. For example, we know that the expected transmission time of a packet in a wireless network is 20 minutes. However, this information may not be so meaningful because it may overestimate the transmission time most of the time. For example, the expected transmission time is calculated based on the normal transmissions (without collisions) and the interference transmissions (with collisions). Although the interference transmissions are very rare to happen, but it takes a lot of time. Then, the estimation about the expected transmission is overestimated most of the time. This makes estimations not useful for the DQL algorithms.

Thus, the authors in~\cite{bellemare2017distributional} introduce a solution using distributional reinforcement learning to update $Q$-value function based on its distribution rather than its expectation. In particular, let $\mathcal{Z}(s,a)$ be the return obtained by starting from state $s$, executing action $a$, and following the current policy, then $\mathcal{Q}(s,a) = \mathbb{E}[\mathcal{Z}(s,a)]$. Here, $\mathcal{Z}$ represents the distribution of future rewards, which is no longer a scalar quantity like $Q$-values. Then we obtain the distributional version of Bellman equation as follows: $\mathcal{Z}(s,a) = r + \gamma \mathcal{Z}(s',a')$. For example, if we use the DQN and extract an experience $(s,a,r,s')$ from the replay buffer, then the sample of the target distribution is $\mathcal{Z}(s,a) = r + \gamma \mathcal{Z}(s',a^*)$ with $a^* = \arg \underset{a'}{\max} \mathcal{Q}(s,a')$. Although the proposed distributional deep Q-learning is demonstrated to outperform the conventional DQL~\cite{mnih2015human1} on many Atari 2600 Games (45 out of 57 games), its performance relies much on the distribution function $\mathcal{Z}$. If $\mathcal{Z}$ is well defined, the performance of distributional deep $Q$-learning is much more significant than that of the DQL. Otherwise, its performance is even worse than that of the DQL.

%----------------------------------
%----------------------------------
\subsubsection{ Deep $Q$-learning with Noisy Nets}

In~\cite{fortunato2018noisy1}, the authors introduce Noisy Net, a type of neural network whose bias and weights are iteratively perturbed during training by a parametric function of the noise. This network basically adds the Gaussian noise to the last (fully-connected) layers of the network. The parameters of this noise can be adjusted by the model during training, which allows the agent to decide when and in what proportion it wants to introduce the uncertainty to its weights. In particular, to implement the noisy network, we first replace the $\epsilon$-greedy policy by a randomized action-value function. Then, the fully connected layers of the value network are parameterized as a noisy network, where the parameters are drawn from the noisy network parameter distribution after every replay step. For replay, the current noisy network parameter sample is held fixed across the batch. Since the DQL takes one step of optimization for every action step, the noisy network parameters are re-sampled before every action. After that, the loss function can be updated as follows:
\begin{equation}
\mathcal{L} = \mathbb{E} \Big[ \mathbb{E}_{(s,a,r,s') \thicksim \mathbf{D}}	\big[ r + \gamma \max_{a' \in \mathcal{A}} \hat{\mathcal{Q}} (s',a',\epsilon'; \boldsymbol{\theta'}) - \mathcal{Q}(s,a,\epsilon;\boldsymbol{\theta}) \big]	\Big] ,
\end{equation}
where the outer and inner expectations are with respect to distributions of the noise variables $\epsilon$ and $\epsilon'$ for the noisy value functions $\hat{\mathcal{Q}} (s',a',\epsilon'; \boldsymbol{\theta'})$ and $\mathcal{Q}(s,a,\epsilon;\boldsymbol{\theta})$, respectively.

Through experimental results, the authors demonstrate that by adding the Gaussian noise layer to the DNN, the performance of conventional DQL~\cite{mnih2015human1}, dueling DQL~\cite{wang2016dueling}, and asynchronous DQL~\cite{mnih2016asynchronous} can be significantly improved for a wide range of Atari games. However, the impact of noise to the performance of the deep DQL algorithms is still under debating in the literature, and thus analysis on the impact of noise layer requires further investigations.

\begin{table*}[!]
	\centering
	\caption{Performance comparison among DQL algorithms}
	\label{tab:DQL_Comparisons}
	\begin{tabular}{||c||c||c||c||c||} \hline
		\textbf{DQL Algorithms} & \textbf{No Operations} & \textbf{Human Starts} & \textbf{Publish} & \textbf{Developer} \\ \hline \hline
		DQL & 79\% & 68\% & Nature 2015~\cite{mnih2015human1} & Google DeepMind \\ \hline
		DDQL & 117\% & 110\%  & AAAI 2016~\cite{hasselt2016deep} & Google DeepMind \\ \hline
		Prioritized DDQL & 140\% & 128\% & ICLR 2015~\cite{schaul2015prioritized1} & Google DeepMind \\ \hline
		Dueling DDQL & 151\% & 117\% & ICML 2016~\cite{wang2016dueling} & Google DeepMind\\ \hline
		Asynchronous DQL & - & 116\% & ICML 2016~\cite{mnih2016asynchronous} & Google DeepMind \\ \hline
		Distributional DQL & 164\% & 125\% & ICML 2017~\cite{bellemare2017distributional} & Google DeepMind \\ \hline	
		Noisy Nets DQL & 118\% & 102\% & ICLR 2018~\cite{fortunato2018noisy1} &  Google DeepMind \\ \hline
		\textbf{Rainbow} & \textbf{223\%} & \textbf{153\%} & AAAI 2018~\cite{hessel2018rainbow} & Google DeepMind \\ \hline
	\end{tabular}
\end{table*}

%----------------------------------
%----------------------------------
\subsubsection{Rainbow Deep $Q$-learning}

In~\cite{hessel2018rainbow}, the authors propose a solution which integrates all advantages of seven aforementioned solutions (including DQL) into a single learning agent, called Rainbow DQL. In particular, this algorithm first defines the loss function based on the asynchronous multi-step and distributional DQL. Then, the authors combine the multi-step distributional loss with double $Q$-learning by using the greedy action in $s_{t+n}$ selected according to the $Q$-network as the bootstrap action $a^*_{t+n}$, and evaluate the action by using the target network.
%After that, the algorithm combines the multi-step distributional loss with double $Q$-learning by using the greedy action in $s_{t+n}$ selected according to the neural network as the bootstrap action $a^*_{t+n}$, and evaluating the action using the target network.

In standard proportional prioritized replay~\cite{schaul2015prioritized1} technique, the absolute TD-error is used to prioritize the transitions. Here, TD-error at a time slot is the error in the estimate made at the time slot. However, in the proposed Rainbow DQL algorithm, all distributional Rainbow variants prioritize transitions by the Kullbeck-Leibler (KL) loss because this loss may be more robust to noisy stochastic environment. Alternatively, the dueling architecture of DNNs is presented in~\cite{wang2016dueling}. Finally, the Noisy Net layer~\cite{hessel2018rainbow} is used to replace all linear layers in order to reduce the number of independent noise variables. Through simulation, the authors show that this is the most advanced technique which outperforms almost all current DQL algorithms in the literature over 57 Atari 2600 games.

In Table~\ref{tab:DQL_Comparisons}, we summarize the DQL algorithms and their performance under the parameter settings used in~\cite{hessel2018rainbow}. As observed in Table~\ref{tab:DQL_Comparisons}, all of the DQL algorithms have been developed by Google DeepMind based on the original work in~\cite{mnih2015human1}. So far, through experimental results on Atari 2600 games, the Rainbow DQL presents very impressive results over all other DQL algorithms. However, more experiments need to be further conducted in different domains to confirm the real efficiency of the Rainbow DQL algorithm.

%===============================================
%===============================================
\subsection{Deep Q-Learning for Extensions of MDPs}

%----------------------------------
%----------------------------------
\subsubsection{Deep Deterministic Policy Gradient Q-Learning for Continuous Action}
\label{DQN_continous_action}

Although DQL algorithm can solve problems with high-dimensional state spaces, it can only handle discrete and low-dimensional action spaces. However, systems in many applications have continuous, i.e., real values, and high dimensional action spaces. The DQL algorithms cannot be straightforwardly applied to continuous actions since they rely on choosing the best action that maximizes the $Q$-value function. In particular, a full search in a continuous action space to find the optimal action is often infeasible.

In~\cite{lillicrap2016continuous}, the authors introduce a model-free off-policy actor-critic algorithm using deep function approximators that can learn policies in high-dimensional, continuous action spaces. The key idea is based on the deterministic policy gradient (DPG) algorithm proposed in~\cite{Silver2014Deterministic}. In particular, the DPG algorithm maintains a parameterized actor function $\mu(s;\boldsymbol{\theta}^{\mu})$ with parameter vector $\boldsymbol{\theta}$ which specifies the current policy by deterministically mapping states to a specific action. The critic $Q(s,a)$ is learned by using the Bellman equation as in $Q$-learning. The actor is updated by applying the chain rule to the expected return from the start distribution $J$ with respect to the actor parameters as follows:
\begin{equation}
\begin{aligned}
& \nabla_{\boldsymbol{\theta}^{\mu}} J  \thickapprox \mathbb{E}_{s_t \thicksim \rho^{\beta}} \big[ \nabla_{\boldsymbol{\theta}^{\mu}} Q(s,a;\boldsymbol{\theta}^{Q}) |_{s=s_t, a=\mu(s_t|\boldsymbol{\theta}^{\mu})} \big] \\
& \thickapprox \mathbb{E}_{s_t \thicksim \rho^{\beta}} \Big[ \nabla_{a} Q(s,a;\boldsymbol{\theta}^{Q}) |_{s=s_t, a=\mu(s_t)} \nabla_{\boldsymbol{\theta}{\mu}} \mu (s;\boldsymbol{\theta}^{\mu}) |_{s=s_t} \Big] .
\end{aligned}
\end{equation}

Based on this update rule, the authors then introduce Deep DPG (DDPG) algorithm which can learn competitive policies by using low-dimensional observations (e.g. cartesian coordinates or joint angles) under the same hyper-parameters and network structure. The detail of the DDPG algorithm is presented in \ref{alg:DDPG}. The algorithm makes a copy of the actor and critic networks $Q'(s,a;\boldsymbol{\theta}^{Q'})$ and $\mu'(s;\boldsymbol{\theta}^{\mu'})$, respectively, to calculate the target values. The weights of these target networks are then updated with slowly tracking on the learned networks, i.e., $\boldsymbol{\theta}' \leftarrow \tau \boldsymbol{\theta} + (1-\tau) \boldsymbol{\theta}'$ with $\tau \ll 1$. This means that the target values are constrained to change slowly, greatly improving the stability of learning. Note that a major challenge of learning in continuous action spaces is exploration. Therefore, in Algorithm~\ref{alg:DDPG}, an exploration policy $\mu'$ is constructed by adding noise sampled from a noise process $\mathcal{N}$ to the actor policy.

%++++++++++++++++++++++++++++++++++++++++++++++++++++++++++++++++++++
\begin{algorithm}[!]
	\caption{DDPG algorithm}
	\label{alg:DDPG}
	\begin{algorithmic}[1]
		\STATE Randomly initialize critic network $Q(s,a;\boldsymbol{\theta}^{Q})$ and actor $\mu(s;\boldsymbol{\theta}^{\mu})$ with weights $\boldsymbol{\theta}^{Q}$ and $\boldsymbol{\theta}^{\mu}$, respectively.
		\STATE Initialize target network $Q'$ and $\mu'$ with weights \\
		$\boldsymbol{\theta}^{Q'} \leftarrow \boldsymbol{\theta}^{Q}$, and $\boldsymbol{\theta}^{\mu'} \leftarrow \boldsymbol{\theta}^{\mu}$, respectively.
		\STATE Initialize replay memory $\mathbf{D}$.
		\FOR{\textit{episode=1 to M}}
		\STATE Initialize a random process $N$ for action exploration
		\STATE Receive initial observation state $s_1$
		\FOR{\textit{t=1 to T}}
		\STATE Select action $a_t = \mu(s_t;\boldsymbol{\theta}^{\mu}) + \mathcal{N}_t$ according to the current policy and exploration noise.
		\STATE Execute action $a_t$ and observe reward $r_t$ and new state $s_{t+1}$.
		\STATE Store transition $(s_t, a_t, r_t, s_{t+1})$ in $\mathbf{D}$.
		\STATE Sample a random mini-batch of $N$ transitions $(s_i, a_i, r_i, s_{i+1})$ from $\mathbf{D}$.
		\STATE Set $y_i = r_i + \gamma \mathcal{Q}^{'} \big(s_{i+1}, \mu^{'} (s_{i+1};\boldsymbol{\theta}^{\mu'}) ; \boldsymbol{\theta}^{Q'} \big) $.
		\STATE Update critic by minimizing the loss: \\
		$L= \frac{1}{N} \sum_{i} (y_i - \mathcal{Q} \big(s_i,a_i;\boldsymbol{\theta}^{Q}) \big)^2 $
		\STATE Update the actor policy by using the sampled policy gradient: $\nabla_{\boldsymbol{\theta}^{\mu}} J \thickapprox \frac{1}{N} \sum_i \nabla_a \mathcal{Q}(s,a;\boldsymbol{\theta}^Q)|_{s=s_i,a=\mu(s_i)}$
		$\nabla_{\boldsymbol{\theta}^{\mu}} \mu(s|\boldsymbol{\theta}^{\mu})|_{s=s_i} $
		\STATE Update the target networks:\\
		 $\boldsymbol{\theta}^{Q'} \leftarrow \tau \boldsymbol{\theta}^{Q} + (1-\tau) \boldsymbol{\theta}^{Q'}$\\
		 $\boldsymbol{\theta}^{\mu'} \leftarrow \tau \boldsymbol{\theta}^{\mu} + (1-\tau) \boldsymbol{\theta}^{\mu'}$
		\ENDFOR
		\ENDFOR
	\end{algorithmic}
\end{algorithm}
%++++++++++++++++++++++++++++++++++++++++++++++++++++++++++++++++++++

%----------------------------------
%----------------------------------
\subsubsection{Deep Recurrent Q-Learning for POMDPs}

To tackle problems with partially observable environments by deep reinforcement learning, the authors in~\cite{Hausknecht2015Deep} propose a framework called Deep Recurrent Q-Learning (DRQN) in which an LSTM layer was used to replace the first post-convolutional fully-connected layer of the conventional DQN. The recurrent structure is able to integrate an arbitrarily long history to better estimate the current state instead of utilizing a fixed-length history as in DQNs. Thus, DRQNs estimate the function $\mathcal{Q}(o_t,h_{t-1};\boldsymbol{\theta})$ instead of $\mathcal{Q}(s_t,a_t);\boldsymbol{\theta})$, where $\boldsymbol{\theta}$ denotes the parameters of entire network, $h_{t-1}$ denotes the output of the LSTM layer at the previous step, i.e., $h_t=LSTM(h_{t-1},o_t)$. DRQN matches DQN's performance on standard MDP problems and outperforms DQN in partially observable domains. Regarding the training process, DRQN only considers the convolutional features of the observation history instead of explicitly incorporating the actions. Through the experiments, the authors demonstrate that DRQN is capable of handling partial observability, and recurrency confers benefits when the quality of observations changes during evaluation time.

%----------------------------------
%----------------------------------
\subsubsection{Deep SARSA Learning}

In~\cite{Zhao2016Deep}, the authors introduce a DQL technique based on SARSA learning to help the agent determine optimal policies in an online fashion. As shown in Algorithm~\ref{alg:DSARSA}, given the current state $s$, a CNN is used to obtain the current state-action value $\mathcal{Q}(s,a)$. Then, the current action $a$ is selected by the $\epsilon$-greedy algorithm. After that, the immediate reward $r$ and the next state $s'$ can be observed. In order to estimate the current $\mathcal{Q}(s,a)$, the next state-action value $\mathcal{Q}(s',a')$ is obtained. Here, when the next state $s'$ is used as the input of the CNN, $\mathcal{Q}(s',a')$ can be obtained as the output. Then, a label vector related to $\mathcal{Q}(s,a)$ is defined as $\mathcal{Q}(s',a')$ which represents the target vector. The two vectors only have one different component, i.e., $r+\gamma \mathcal{Q}(s',a') \rightarrow \mathcal{Q}(s,a)$. It should be noted that during the training phase, the next action $a'$ for estimating the current state-action value is never greedy. On the contrary, there is a small probability that a random action is chosen for exploration.

%++++++++++++++++++++++++++++++++++++++++++++++++++++++++++++++++++++
\begin{algorithm}[!]
	\caption{Deep SARSA learning algorithm}
	\label{alg:DSARSA}
	\begin{algorithmic}[1]
		\STATE Initialize data stack $\mathbf{D}$ with size of $N$
		\STATE Initialize parameters $\boldsymbol{\theta}$ of the CNN
		\FOR{\textit{episode=1 to M}}
		\STATE Initialize state $s_1$ and pre-process state $\phi_1 = \phi(s_1)$
		\STATE Select $a_1$ by the $\epsilon$-greedy method
		\FOR{\textit{t=1 to T}}
		\STATE Take action $a_t$, observe $r_t$ and next state $s_{t+1}$
		\STATE $\phi_{t+1} = \phi(s_{t+1})$
		\STATE Store data $(\phi_t,a_t,r_t,\phi_{t+1})$ into stack $\mathbf{D}$
		\STATE Sample data from stack $\mathbf{D}$
		\STATE Select action $a'$ by the $\epsilon$-greedy method
		\IF{episode terminates at step $j+1$} \STATE {Set $y_j = r_j$}
		\ELSE \STATE{set $y_j=r_j + \mathcal{Q}(\phi_{t+1},a';\boldsymbol{\theta})$} \ENDIF
		\STATE Minimize the loss function: $(y_j - \mathcal{Q}(\phi_{t},a';\boldsymbol{\theta}))^2$
		\STATE Update $a_t \leftarrow a'$
		\ENDFOR
		\ENDFOR
	\end{algorithmic}
\end{algorithm}
%++++++++++++++++++++++++++++++++++++++++++++++++++++++++++++++++++++

%----------------------------------
%----------------------------------
\subsubsection{Deep $Q$-Learning for Markov Games}

In~\cite{Wang2018Towards}, the authors introduce the general notion of sequential prisoner's dilemma (SPD) to model real world prisoner's dilemma (PD) problems. Since SPD is more complicated than PD, existing approaches addressing learning in matrix PD games cannot be directly applied in SPD. Thus, the authors propose a multi-agent DRL approach for mutual cooperation in SDP games. The deep multi-agent reinforcement learning towards mutual cooperation consists of two phases, i.e., offline and online phases. The offline phase generates policies with varying cooperation degrees. Since the number of policies with different cooperation degrees is infinite, it is computationally infeasible to train all the policies from scratch. To address this issue, the algorithm first trains representative policies using actor-critic until it converges, i.e., cooperation and defection baseline policy. Second, the algorithm synthesizes the full range of policies from the above baseline policies. Another task is to detect effectively the cooperation degree of the opponent. The algorithm divides this task into two steps. First, the algorithm trains an LSTM-based cooperation degree detection network offline, which will be then used for real-time detection during the online phase. In the online phase, the agent plays against the opponents by reciprocating with a policy of a slightly higher cooperation degree than that of the opponent. On one hand, intuitively the algorithm is cooperation-oriented and seeks for mutual cooperation whenever possible. On the other hand, the algorithm is also robust against selfish exploitation and resorts to defection strategy to avoid being exploited whenever necessary.

Unlike~\cite{Wang2018Towards} which considers a repeated normal form game with complete information, in~\cite{Heinrich2016Deep}, the authors introduce an application of DRL for extensive form games with imperfect information. In particular, the authors in~\cite{Heinrich2016Deep} introduce Neural Fictitious Self-Play (NFSP), a DRL method for learning approximate Nash equilibria of imperfect-information games. NFSP combines FSP with neural network function approximation. An NFSP agent has two neural networks. The first network is trained by reinforcement learning from memorized experience of play against fellow agents. This network learns an approximate best response to the historical behaviour of other agents. The second network is trained by supervised learning from memorized experience of the agent's own behaviour. This network learns a model that averages over the agent's own historical strategies. The agent behaves according to a mixture of its average strategy and best response strategy.

In the NSFP, all players of the game are controlled by separate NFSP agents that learn from simultaneous play against each other, i.e., self-play. An NFSP agent interacts with its fellow agents and memorizes its experience of game transitions and its own best response behaviour in two memories, $\mathcal{M}_{RL}$ and $\mathcal{M}_{SL}$. NFSP treats these memories as two distinct datasets suitable for DRL and supervised classification, respectively. The agent trains a neural network, $Q(s,a;\boldsymbol{\theta}^Q)$, to predict action values from data in $\mathcal{M}_{RL}$ using off-policy reinforcement learning. The resulting network defines the agent's approximate best response strategy, $\beta=\epsilon$-greedy($Q$), which selects a random action with probability $\epsilon$ and otherwise chooses the action that maximizes the predicted action values. The agent trains a separate neural network $\Pi(s,a;\boldsymbol{\theta}^{\Pi})$ to imitate its own past best response behavior by using supervised classification on the data in $\mathcal{M}_{SL}$. NFSP also makes use of two technical innovations in order to ensure the stability of the resulting algorithm as well as to enable simultaneous self-play learning. Through experimental results, the authors show that the NFSP can converge to approximate Nash equilibria in a small poker game.

\textbf{Summary:} In this section, we have presented the basics of reinforcement learning, deep learning, and DQL. Furthermore, we have discussed various advanced DQL techniques and their extensions. Different DQL techniques can be used to solve different problems in different network scenarios. In the next sections, we review DQL related works for various problems in communications and networking.

%===============================================
%=====================
\section{Network Access and Rate Control}
\label{spectrum_rate_control}
%=====================
Modern networks such as IoT become more decentralized and ad-hoc in nature. In such networks, entities such as sensors and mobile users need to make independent decisions, e.g., channel and base station selections, to achieve their own goals, e.g., throughput maximization. However, this is challenging due to the dynamic and the uncertainty of network status. Learning algorithms such as DQL allow to learn and build knowledge about the networks that are used to enable the network entities to make their optimal decisions. In this section, we review the applications of DQL for the following issues:
\begin{itemize}
\item \textit{Dynamic spectrum access:} Dynamic spectrum access allows users to locally select channels to maximize their throughput. However, the users may not have full observations of the system, e.g., channel states. Thus, DQL can be used as an effective tool for dynamic spectrum access.
\item \textit{Joint user association and spectrum access:} User association is implemented to determine which user to be assigned to which Base Station (BS). The joint user association and spectrum access problems are studied in \cite{fooladivanda2013joint} and \cite{lin2015optimizing}. However, the problems are typically combinatorial and non-convex which require nearly complete and accurate network information to obtain the optimal strategy. DQL is able to provide distributed solutions which can be effectively used for the problems.
\item \textit{Adaptive rate control:} This refers to bitrate/data rate control in dynamic and unpredictable environments such as Dynamic Adaptive Streaming over HTTP (DASH). Such a system allows clients or users to independently choose video segments with different bitrates to download. The client's objective is to maximize its Quality of Experience (QoE). DQL can be adopted to effectively solve the problem instead of dynamic programming which has high complexity and demands complete information.
\end{itemize}
%=======================
\subsection{Network Access}
\label{spectrum_rate_control_spectrum}
%========================
This section discusses how to use DQL to solve the spectrum access and user association in networks.
%========================
%=======================
\subsubsection{Dynamic Spectrum Access}
\label{spectrum_rate_control_spectrum_channel}
%========================
%========================
%\cite{wang2017deep}: Spectrum access (channel selection state)/wireless sensor network\\  The channel state varies over time slots according to a 2-state Markov chain. The actions of the sensor include choosing one of $N$ channels   Since the user is only able to sense the selected channel and no full observation of the system is available, the problem can be formulated as a partially observable Markov decision process (POMDP)
The authors in~\cite{wang2017deep} propose a dynamic channel access scheme of a sensor based on the DQL for IoT. At each time slot, the sensor selects one of $M$ channels for transmitting its packet. The channel state is either in low interference, i.e., successful transmission, or in high interference, i.e., transmission failure. Since the sensor only knows the channel state after selecting the channel, the sensor's optimization decision problem can be formulated as a POMDP. In particular, the action of sensor is to select one of $M$ channels. The sensor receives a positive reward ``+1'' if the selected channel is in low interference, and a negative reward ``-1'' otherwise. The objective is to find an optimal policy which maximizes the sensor's the expected accumulated discounted reward over time slots. A DQN\footnote{Remind that DQN is the core of the DQL algorithms.} using FNN with experience replay~\cite{volodymyr2013playing} is then adopted to find the optimal policy. The input of the DQN is a state of the sensor which is the combination of actions and observations, i.e., the rewards, in the past time slots. The output includes Q-values corresponding to the actions. To balance the exploration of the current best Q-value with the exploration of the better one, the $\epsilon$-greedy policy is adopted for the action selection mechanism. The simulation results based on real data from \cite{Govindantutornet} show that the proposed scheme can achieve the average accumulated reward close to the myopic policy~\cite{zhao2008myopic} without a full knowledge of the system.
%To balance the exploration and exploitation, the $\epsilon$-greedy policy is applied for the action selection : \footnote{http://anrg.usc.edu/www/tutornet/}

\cite{wang2017deep} can be considered to be a pioneer work using the DQL for the channel access. However, the DQL keeps following the learned policy over time slots and stops learning a suitable policy. Actual IoT environments are dynamic, and the DQN in the DQL needs to be re-trained. An adaptive DQL scheme is proposed in~\cite{wang2018deeptrans} which evaluates the accumulated reward of the current policy for every period. When the reward is reduced by a given threshold, the DQN is re-trained to find a new good policy. The simulation results~\cite{wang2018deeptrans} show that when the states of the channels change, the adaptive DQL scheme can detect the change and start re-learning to obtain the high reward.
% Realistic IoT systems often have multiple sensors, and the co-channel interference and packet forwarding problems need to be studied.

%\cite{zhu2017new}: Channel selection for packet forwarding (transmission schedule) /IoT\\ Note that the sensor is equipped buffers, and it can delay forwarding the packets as the channels are in high interference. However, this may cause packet loss in the buffers due to incoming packets.
 \begin{figure}[t!]
\centering
\includegraphics[width=6.5 cm, height=4 cm]{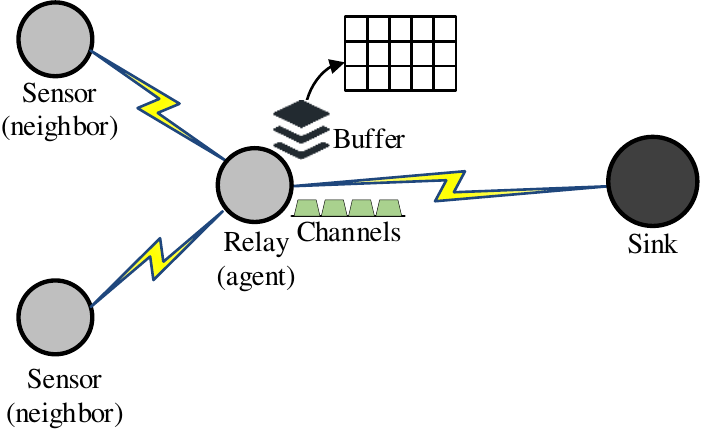}
 \caption{\small Joint channel selection and packet forwarding in IoT.}
 \label{channel_selection_buffer_one_agent}
\end{figure}
The models in \cite{wang2017deep} and~\cite{wang2018deeptrans} are constrained to only one sensor. Consider a multi-sensor scenario, the authors in \cite{zhu2017new} address the joint channel selection and packet forwarding using the DQL. The model is shown in Fig.~\ref{channel_selection_buffer_one_agent} in which one sensor as a relay forwards packets received from its neighboring sensors to the sink. The sensor is equipped with a buffer to store the received packets. At each time slot, the sensor selects a set of channels for the packet forwarding so as to maximize its utility, i.e., the ratio of the number of transmitted packets to the transmit power. Similar to \cite{wang2017deep}, the sensor's problem can be formulated as an MDP. The action is to select a set of channels, the number of packets transmitted on the channels, and a modulation mode. To avoid packet loss, the state is defined as the combination of the buffer state and channel state. The MDP is then solved by the DQL in which the input is the state and the output is the action selection. The DQL uses the stacked autoencoder to reduce the massive calculation and storage in the Q-learning phase. The sensor's utility function is proved to be bounded which can guarantee the convergence of the algorithm. As shown in the simulation results, the proposed scheme can converge after a certain number of iterations. Also, the proposed scheme significantly improves the system utility compared with the random action selection scheme. However, as the packet arrival rate increases, the system utility of the proposed scheme decreases since the sensor needs to consume more power to transmit all packets.

%\cite{chu2018reinforcement}: multi-access control/IoT\\
 Consuming more power leads to poor sensor's performance due to its energy constraint, i.e., a shorter IoT system lifetime. The channel access problem in the energy harvesting-enabled IoT system is investigated in \cite{chu2018reinforcement}. The model consists of one BS and energy harvesting-based sensors. The BS as a controller allocates channels to the sensors. However, the uncertainty of ambient energy availability at the sensors may make the channel allocation inefficient. For example, the channel allocated to the sensor with low available energy may not be fully utilized since the sensor cannot communicate later.

 Therefore, the BS's problem is to predict the sensors' battery states and select sensors for the channel access so as to maximize the total rate. Since the sensors are distributed randomly over a geographical area, the complete statistical knowledge of the system dynamics, e.g., the battery states and channel states, may not be available. Thus, the DQL is used to solve the problem of the BS, i.e., the agent. The DQL uses a DQN consisting of two LSTM-based neural network layers. The first layer generates the predicted battery states of sensors, and the second layer uses the predicted states along with Channel State Information (CSI) to determine the channel access policy. The state space consists of (i) channel access scheduling history, (ii) the history of predicted battery information, (iii) the history of the true battery information, and (iv) the current CSI of the sensors. The action space contains all sets of sensors to be selected for the channel access, and the reward is the difference between the total rate and the prediction error. As shown in the simulation results, the proposed scheme outperforms the myopic policy~\cite{zhao2008myopic} in terms of total rate. Moreover, the battery prediction error obtained from the proposed scheme is close to zero. %The stable stage of the battery prediction loss is earlier than that of the total rate.

%The output includes the predicted battery states of the sensors. The predicted battery states obtained from the first LSTM layer and the current channel state information of all sensors are the input of the second LSTM layer.

%\cite{ye2017deep}: Channel and power selection/V2V communications\\ deep Q-learning with experience replay
The above schemes, e.g.,~\cite{wang2017deep} and~\cite{chu2018reinforcement}, focus on the rate maximization. In IoT systems such as Vehicle-to-Vehicle (V2V) communications, latency also needs to be considered due to the mobility of V2V transmitters/receivers and vital applications in the
traffic safety. One of the problems of each V2V transmitter is to select a channel and a transmit power level to maximize its capacity under a latency constraint. Given the decentralized network, a DQN is adopted to make optimal decisions as proposed in \cite{ye2017deep}. The model consists of V2V transmitters, i.e., agents, which share a set of channels. The actions of each V2V transmitter include choosing channels and transmit power levels. The reward is a function of the V2V transmitter's capacity and latency. The state observed by the V2V transmitter consists of (i) the instantaneous CSI of the corresponding V2V link, (ii) the interference to the V2V link in the previous time slot, (iii) the channels selected by the V2V transmitter' neighbors in the previous time slot, and (iv) the remaining time to meet the latency constraint. The state is also an input of the DQN. The output includes Q-values corresponding to the actions. As shown in the simulation results, by dynamically adjusting the power and channel selection when V2V links are likely to violate the latency constraint, the proposed scheme has more V2V transmitters meeting the latency constraint compared with the random channel allocation.

%\cite{challita2017proactive}: unlicensed channels access/ LTE-U systems\\ a proactive approach for allocating spectrum resources to SBSs in the unlicensed bands. In this problem, the constraint of channel access proportions for the SBS and for the WLAN on the channels during time slots is introduced.
To reduce spectrum cost, the above IoT systems often use unlicensed channels. However, this may cause the interference to existing networks, e.g., WLANs. The authors in \cite{challita2017proactive} propose to use the DQN to jointly address the dynamic channel access and interference management. The model consists of Small Base Stations (SBSs) which share unlicensed channels in an LTE network. At each time slot, the SBS selects one of channels for transmitting its packet. However, there may be WLAN traffics on the selected channel, and thus the SBS accesses the selected channel with a probability. The actions of the SBS include pairs of channel selection and channel access probability. The problem of the SBS is to determine an action vector so as to maximize its total throughput, i.e., its utility, over all channels and time slots. The resource allocation problem can be formulated as a non-cooperative game, and the DQN using LSTM can be adopted to solve the game. The input of the DQN is the history traffic of the SBSs and the WLAN on the channels. The output includes predicted action vectors of the SBSs. The utility function of each SBS is proved to be convex, and thus the DQN-based algorithm converges to a Nash equilibrium of the game. The simulation results based on real traffic data from \cite{balazinska2003ibm} show that the proposed scheme can improve the average throughput up to 28\% compared with the standard Q-learning~\cite{watkins1992q}. Moreover, deploying more SBSs in the LTE network does not allow more airtime fraction for the network. This implies that the proposed scheme can avoid causing performance degradation to the WLAN. However, the proposed scheme requires synchronization between the SBSs and the WLAN which is challenging in real networks.

% \cite{naparstek2017deep}: Spectrum access (current state is mapped to spectrum access actions)/general wireless network\\
% The aforementioned schemes mostly use the experience replay to learn from pass observations. In a multi-user system, such learning is undesirable due to the interactions among the users. . Thus, it is necessary to collect a predefined number of episodes at each iteration and create target values for all the episodes.
In the same cellular network context, the authors in \cite{naparstek2017deep} address the dynamic spectrum access problem for multiple users sharing $K$ channels. At a time slot, the user selects a channel with a certain attempt probability or chooses not to transmit at all. The state is the history of the user's actions and its local observations, and the user's strategy is mapping from the history to an attempt probability. The problem of the user is to find a vector of the strategies, i.e., the policy, over time slots to maximize its expected accumulated discounted data rate of the user.

The above problem is solved by training a DQN. The input of the DQN includes past actions and the corresponding observations. The output includes estimated Q-values of the actions. To avoid the overestimation in the Q-learning, the DDQN~\cite{hasselt2010double} is used. Moreover, the dueling DQN~\cite{wang2015dueling} is employed to improve the estimated Q-value. The DQN is then offline trained at a base station. Similar to \cite{challita2017proactive}, the multichannel random access is modeled as a non-cooperative game. As proved in~\cite{naparstek2017deep}, the game has a subgame perfect equilibrium. Note that some users can keep increasing their attempt probability to increase their rates. This makes the equilibrium point inefficient, and thus the strategy space of the users is restricted to avoid the situation. The simulation results show that the proposed scheme can achieve twice the channel throughput compared with the slotted-Aloha~\cite{li2010multiagent}. The reason is that in the proposed scheme, each user only learns from its local observation without an online coordination or carrier sensing. However, the proposed scheme requires the central unit which may raise the message exchanges as the training is frequently updated.

% in which the first neural network is for selecting actions and the second one is for estimating the Q-value associated with the selected actionat which no user has an incentive to unilaterally deviate its strategy we develop a mechanism that restricts the strategy space for all users when training the DQN, referred to as common training

%\cite{yu2017deep}: Multiple access (channel states, actions are transmit or wait)/HetNets\\
%\cite{chen2017liquid}: spectrum access and user association/UAVs\\

%\cite{liu2018deep}: Dynamic channel allocation (deciding channel allocation to terminal users)/satellite systems\\

In the aforementioned models, the number of users is fixed in all time slots, and the arrival of new users is not considered. The authors in~\cite{liu2018deep} address the channel allocation to new arrival users in a multibeam satellite system. The multibeam satellite system generates a geographical footprint subdivided into multiple beams which provide services to ground User Terminals (UTs). The system has a set of channels. If there exist available channels, the system allocates a channel to the new arrived UT, i.e., the new service is satisfied. Otherwise, the service is blocked. The system's problem is to find a channel allocation decision to minimize the total service blocking probability of the new UT over time slots without causing the interference to the current UTs.

The system's problem can be viewed as a temporal correlated sequential decision-making optimization problem which is effectively solved by the DQN. Here, the satellite system is the agent. The action is an index indicating which channel is allocated to the new arrived UT. The reward is positive when the new service is satisfied and is negative when the service is blocked. The state includes the set of current UTs, the current channel allocation matrix, and the new arrived UT. Note that the state has the spatial correlation feature due to the co-channel interference, and thus it can be represented in an image-like fashion, i.e., an image tensor. Therefore, the DQN adopts the CNN to extract useful features of the state. The simulation results show that the proposed DQN algorithm converges after a certain number of training steps. Also, by allocating available channels to the new arrived UTs, the proposed scheme can improve the system traffic up to 24.4\% compared with the fixed channel allocation scheme. However, as the number of current UTs increases, the number of available channels is low or even zero. Therefore, the dynamic channel allocation decisions of the proposed scheme become meaningless, and the performance difference between the two schemes becomes insignificant. For the future work, a joint channel and power allocation algorithm based on the DQL can be investigated.

%=====================
\subsubsection{Joint User Association and Spectrum Access}
\label{spectrum_rate_control_spectrum_user_association_channel}
%=====================
The joint user association and spectrum access problems are typically non-convex. DQL is able to provide distributed solutions, and thus it can be effectively used to solve the problems without requiring complete and accurate network information.
%The joint user association and spectrum access problems are studied in \cite{fooladivanda2013joint} and \cite{lin2015optimizing}. However, the approaches often require nearly complete and accurate network information to obtain the optimal strategy. Such complete information may not be available in large-scale networks, e.g., dense HetNets.

%\cite{nan2018deep}: Joint base station selection and channel selection (user association and resource allocation)/HetNets\\
The authors in~\cite{nan2018deep} consider a HetNet which consists of multiple users and BSs including macro base stations and femto base stations. The BSs share a set of orthogonal channels, and the users are randomly located in the network. The problem of each user is to select one BS and a channel to maximize its data rate while guaranteeing that the Signal-to-Interference-plus-Noise Ratio (SINR) of the user is higher than a minimum Qualtiy of Service (QoS) requirement. The DQL is adopted to solve the problem in which each user is an agent, and its state is a vector including QoS states of all users, i.e., the global state. Here, the QoS state of the user refers to whether its SINR exceeds the minimum QoS requirement or not. At each time slot, the user takes an action. If the QoS is satisfied, the user receives utility as its immediate reward. Otherwise, it receives a negative reward, i.e., an action selection cost. Note that the cumulative reward of one user depends on actions of other users, then the user's problem can be defined as an MDP. Similar to \cite{naparstek2017deep}, the DDQN and the dueling DQN are used to learn the optimal policy, i.e., the joint BS and channel selections, for the user to maximize its cumulative reward. The simulation results from~\cite{nan2018deep} show that the proposed scheme outperforms the Q-learning implemented in~\cite{watkins1992q} in terms of convergence speed and system capacity.
% In particular, the fast convergence speed enables the proposed scheme to meet the real-time QoS requirements. However, the proposed scheme requires the user to report its local state at each time slot. For this, the user needs to monitor Received Signal Strength Indicators (RSSIs) from its neighboring BSs, and then it temporarily connects to the BS with the maximum RSSI and randomly selects a channel. However, given a high density of the BSs, the RSSIs from different BSs may not be different, and it is challenging for the user to determine its local state. The CNN and Recurrent Neural Network (RNN) can be adopted to extract potential spatial features, i.e., location information of the BSs, and sequential features from the raw wireless signal as proposed in~\cite{cao2018aif}. These features can be utilized to facilitate the state determination of users.

% One problem is that how the user reports its initially local state when it is not associated to any BS
%However, the proposed scheme requires each user to report its current state to the BS. To know its current state, each UE can measure received power from neighboring BSs and selects the BS with the maximum received signal reference power.  Then, each UE reports its own current state to its current associated BS cumulative reward of one UE is inevitably influenced by other UEs' actions in the cellular network

The scheme proposed in~\cite{nan2018deep} is considered to be the first work using the DQL for the joint user association and spectrum access problem. Inspired by this work, the authors in~\cite{chen2017liquid} propose to use the DQL for a joint user association, spectrum access, and content caching problem. The network model is an LTE network which consists of UAVs serving ground users. The UAVs are equipped with storage units and can act as cached-enabled LTE-BSs. The UAVs are able to access both licensed and unlicensed bands in the network. The UAVs are controlled by a cloud-based server, and the transmissions from the cloud to the UAVs are implemented by using the licensed cellular band. The problem of each UAV is to determine (i) its optimal user association, (ii) the bandwidth allocation indicators on the licensed band, (iii) the time slot indicators on the unlicensed band, and (iv) a set of popular contents that the users can request to maximize the number of users with stable queue, i.e., users satisfied with content transmission delay.

The UAV's problem is combinatorial and non-convex, and the DQL can be used to solve it. The UAVs do not know the users' content requests, and thus the Liquid State Machine approach (LSM)~\cite{maass2011liquid} is adopted to predict the content request distribution of the users and to perform resource allocation. In particular, predicting the content request distribution is implemented at the cloud based on an LSM-based prediction algorithm. Then, given the request distributions, each UAV as an agent uses an LSM-based learning algorithm to find its optimal users association. Specifically, the input of the LSM-based learning algorithm consists of actions, i.e., UAV-user association schemes, that other UAVs take, and the output includes the expected numbers of users with stable queues corresponding to actions that the UAV can take. After the user association is done, the optimal content caching is determined based on the results of \cite[Theorem 2]{chen2017caching}, and the optimal spectrum allocation is done by using linear programming. Based on the Gordon's Theorem~\cite{szita2006reinforcement}, the proposed DQL is proved to converge with probability one. The simulation results using content request data from \cite{index_youku} show that the proposed DQL can converge in around 400 iterations. Compared with the Q-learning, the proposed DQN improves the convergence time up to 33\% . Moreover, the proposed DQL significantly improves the number of users with stable queues up to 50\% compared with the Q-learning without cache. In fact, energy efficiency is also important for the UAVs, and thus applying the DQL for a joint user association, spectrum access, and power allocation problem needs to be investigated.

%\cite{andrea2018heterogeneous}: channel access/M2M communications (paper not downloaded)\\

%=====================
\subsection{Adaptive Rate Control}
%=====================
%\cite{gadaleta2017d}: bitrate adaptation/HTTP standards\\
 \begin{figure}[t!]
\centering
\includegraphics[width=8.1 cm, height=3.3 cm]{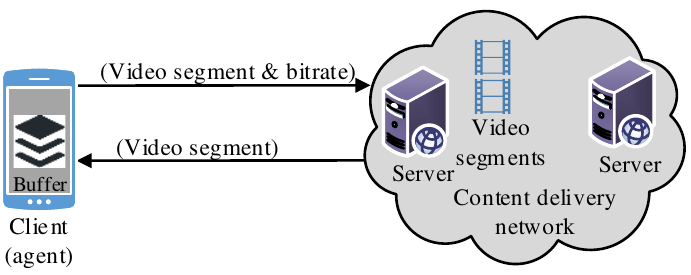}
 \caption{\small A dynamic adaptive streaming system based on HTTP standard.}
 \label{HTTP_video_stream}
\end{figure}
Dynamic Adaptive Streaming over HTTP (DASH) becomes the dominant standard for video streaming~\cite{stockhammer2011dynamic}. DASH is able to leverage existing content delivery network infrastructure and is compatible with a multitude of client-side applications. A general DASH system is shown in Fig.~\ref{HTTP_video_stream} in which the videos are stored in servers as multiple segments, i.e., chunks. Each segment is encoded at different compression levels to generate \textit{representations} with different bitrates, i.e., different video visual quality. At each time slot, the client chooses a representation, i.e., a segment with a certain bitrate, to download. The client's problem is to find an optimal policy which maximizes its QoE such as maximizing average bitrate and minimizing rebuffering, i.e., the time which the video playout freezes.

%The authors in \cite{gadaleta2017d} proposed to use the DQN for the birate adaptation in a dynamic streaming over HTTP standard.

As presented in \cite{gadaleta2017d}, the above problem can be modeled as an MDP in which the agent is the client and the action is choosing a representation to download. To maximize the QoE, the reward is defined as a function of (i) visual quality of the video, (ii) video quality stability, (iii) rebuffering event, and (iv) buffer state. Given the reward formulation, the state of the client should include (i) the video quality of the last downloaded segment, (ii) the current buffer state, (iii) the rebuffering time, and (iv) the channel capacities experienced during downloading of segments in the past time slots. The MDP can be solved by using dynamic programming, but the computational complexity rapidly becomes unmanageable as the size of the problem increases. Thus, the authors in \cite{gadaleta2017d} adopt the DQL to solve the problem. Similar to \cite{chu2018reinforcement}, the LSTM networks are used in which the input is the state of the client, and the output includes Q-values corresponding to the client's possible actions. To improve the performance of the standard LSTM, peephole connections are added into the LSTM networks. The simulation results based on dataset from \cite{klaue2003evalvid} show that the proposed DQL algorithm can converge much faster than Q-learning. Moreover, the proposed DQL improves the video quality and reduces the rebuffering since it is able to dynamically manage the buffer by considering the buffer state and channel capacity.
%\cite{mao2017neural}: adaptive bitrate/video streaming systems\\

The network model and the optimization problem in \cite{gadaleta2017d} are also found in~\cite{mao2017neural}. However, different from \cite{gadaleta2017d}, the authors in~\cite{mao2017neural} adopt the Asynchronous  Advantage  Actor-
Critic (A3C) method~\cite{mnih2016asynchronous} for the DQL to further enhance and speed up the training. As presented in Section~\ref{DQN_continous_action}, A3C includes two neural networks, namely, \textit{actor network} and \textit{critic network}. The actor network is to choose bitrates for the client, and the critic network helps train the actor network. For the actor network, the input is the client's state, and the output is a policy, i.e., a probability distribution over possible actions given states that the client can take. Here, the action is choosing the next representation, i.e., the next segment with a certain bitrate, to download. For the critic network, the input is the client's state, and the output is the expected total reward when following the policy obtained from the actor network. The simulation results based on the mobile dataset from \cite{riiser2013commute} show that the proposed DQL can improve the average QoE up to 25\% compared with the bitrate control scheme~\cite{yin2015control}. Also, by having sufficient buffer to
handle the network's throughput fluctuations, the proposed DQL reduces the rebuffering around 32.8\% compared with the baseline scheme.

In practice, the DQL algorithm proposed in~\cite{mao2017neural} can be easily deployed in a multi-client network since A3C is able to support parallel training for multiple agents. Accordingly, each client, i.e., an agent, is configured to observe its reward. Then, the client sends a tuple including its state, action, and reward to a server. The server uses the actor-critic algorithm to update its actor network model. The server then pushes the newest model to the agent. This update process can happen asynchronously among all agents which improves quality and speeds up the training. Although the parallel training scheme may incur a Round-Trip Time (RTT) between the clients and the server, the simulation results in~\cite{mao2017neural} show that the RTT between the clients and the server reduces the average QoE by only 3.5\%. The performance degradation is small, and thus the proposed DQL can be implemented in real network systems.

In~\cite{gadaleta2017d} and~\cite{mao2017neural}, the input of the DQL, i.e., the client's state, includes the video quality of the last downloaded video segment. The video segment is raw which may cause ``state explosion'' to the state space~\cite{huang2018qarc}. To reduce the state space and to improve the QoE, the authors in \cite{huang2018qarc} propose to use a video quality prediction network. The prediction network extracts useful features from the raw video segments using CNN and RNN. Then, the output of the prediction network, i.e., the predicted video quality, is used as one of the inputs of the DQL which is proposed in~\cite{mao2017neural}. The simulation results based on the broadband dataset from~\cite{broadbanddataset} show that the proposed DQL can improve the average QoE up to 25\% compared with the Google Hangout, i.e., a communication platform developed by Google. Moreover, the proposed DQL can reduce the average latency of video transmission around 45\% due to the small state space.

%\cite{chinchali2018cellular}: traffic rate (scheduling)/IoT\\

Apart from the DASH systems, the DQL can be effectively used for the rate control in High Volume Flexible Time (HVFT) applications. HVFT applications use cellular networks to deliver IoT traffic. The HVFT applications have a large volume of traffic, and the traffic scheduling, e.g., data rate control, in the HVFT applications is necessary. One common approach is to assign static priority classes per traffic type, and then traffic scheduling is based on its priority class. However, such an approach does not evolve to accommodate new traffic classes. Thus, learning methods such as DQL should be used to provide adaptive rate control mechanisms as proposed in~\cite{chinchali2018cellular}. The network model is a single cell including one BS as a central controller and multiple mobile users. The problem at the BS is to find a proper policy, i.e., data rate for the users, to maximize the amount of transmitted HVFT traffic while minimizing performance degradation to existing data traffics. It is shown in~\cite{chinchali2018cellular} that the problem can be formulated as an MDP. The agent is the BS, and the state includes the current network state and the useful features extracted from network states in the past time slots. The network state at a time slot includes (i) the congestion metric, i.e., the cell's traffic load, at the time slot, (ii) the total number of network connections, and (iii) the cell efficiency, i.e., the cell quality. The action that the BS takes is a combination of the traffic rate for the users. To achieve the BS' objective, the reward is defined as a function of (i) the sum of HVFT traffic, (ii) traffic loss to existing applications due to the presence of the HVFT traffic, and (iii) the amount of bytes served below desired minimum throughput. The DQL using the actor and critic networks with LSTM is then adopted. By using the real network data collected in Melbourne, the simulation results show that the proposed DQL increases the HVFT traffic up to 2 times compared with the heuristic control scheme. However, how the proposed scheme reduces the traffic loss is not shown.

In the aforementioned approaches, the maximum number of objectives is constrained, e.g., to 3 in \cite{zhang2018cache}. The authors in~\cite{ferreira2018multi} show that the DQL can be used for the rate control to achieve multiple objectives in complex communication systems. The network model is a future space communication system which is expected to operate in unpredictable environments, e.g., orbital dynamics, atmospheric and space weather, and dynamic channels. In the system, the transmitter needs to be configured with several transmit parameters, e.g., symbol rate and encoding rate, to achieve multiple conflict objectives, e.g., low Bit Error Rate (BER), throughput improvement, power and spectral efficiency. The adaptive coding and modulation schemes, i.e., \cite{tarchi2013adaptive}, can be used. However, the methods allow to achieve only limited number objectives. Learning algorithms such as the DQL can be thus used. The agent is the transmitter in the system. The action is a combination of  (i) symbol rate, (ii) energy per symbol, (iii) modulation mode, (iv) number of bits per symbol, and (v) encoding rate. The objective is to maximize the system performance. Thus, the reward is defined as a \textit{fitness} function of performance parameters including (i) BER estimated at the receiver, (ii) throughput, (iii) spectral efficiency, (iv) power consumption, and (v) transmit power efficiency. The state is the system performance measured by the transmitter, and thus the state is the reward. To achieve multiple objectives, the DQL is implemented by using a set of multiple neural networks in parallel. The input of the DQL is the current state and the channel conditions, and the output is the predicted action. The neural networks are trained by using the Levenberg-Marquardt backpropagation
algorithm~\cite{hagan1994training}. The simulation results show that the proposed DQL can achieve the fitness score, i.e., the weighted sum of different objectives, close to the ideal, i.e., the exhaustive search approach. This implies that the DQL is able to select near-optimal actions and learn the relationship between rewards and actions given dynamic channel conditions.

\textbf{Summary:} This section reviews applications of DQL for the dynamic network access and adaptive rate control. The reviewed approaches are summarized along with the references in Table~\ref{DQN_network_access_rate_control_summary_table}. We observe that the problems are mostly modeled as an MDP. Moreover, DQL approaches for the IoT and DASH systems receive more attentions than other networks. Future networks, e.g., 5G networks, involve multiple network entities with multiple conflicting objectives, e.g., provider's revenue versus users' utility maximization. This poses a number of challenges to the traditional resource management mechanisms that deserve in-depth investigation. In the next section, we review
the adoption of DQL for the emerging services, i.e., offloading and caching.

\begin{table*}
\caption{\label{tab: }A summary of approaches using DQL for network access and adaptive rate control.}
\label{DQN_network_access_rate_control_summary_table}
\begin{centering}
\begin{tabular}{|>{\centering\arraybackslash}m{0.7cm}|>{\centering}m{0.5cm}|>{\centering\arraybackslash}m{0.8cm}|>{\centering}m{1.8cm}|>{\centering}m{1.4cm}|>{\centering}m{3.5cm}|>{\centering}m{2.2cm}|>{\centering}m{2.5cm}|>{\centering}m{1.4cm}|}
\hline
 \cellcolor{mygray}  \textbf{\noun{Issues}} &  \cellcolor{mygray} \textbf{\noun{Ref.}}&  \cellcolor{mygray} \textbf{\noun{Model}}&  \cellcolor{mygray} \textbf{\noun{Learning algorithms}}&  \cellcolor{mygray} \textbf{\noun{Agent}}  &  \cellcolor{mygray} \textbf{\noun{States}} &  \cellcolor{mygray}\textbf{\noun{Actions}} & \cellcolor{mygray} \textbf{\noun{Rewards}} &  \cellcolor{mygray} \textbf{\noun{Networks}} \tabularnewline
\hline
\hline
\parbox[t]{2mm}{\multirow{9}{*}{\rotatebox[origin=c]{90}{ \hspace{-5 cm} Network access}}}
&\cite{wang2017deep}& POMDP &DQN using FNN& Sensor &Past channel selections and observations&Channel selection & Score +1 or -1 & IoT \tabularnewline \cline{2-9}
&\cite{zhu2017new}& MDP &DQN using FNN & Sensor&Current buffer state and channel state&Channel, packets, and modulation mode selection &Ratio of number of transmitted packets to transmit power & IoT \tabularnewline \cline{2-9}
&\cite{chu2018reinforcement}& MDP &DQN with LSTM & Base station&Channel access history, predicted and true battery information history, and current CSI &Sensor selection for channel access& Total rate and prediction error & IoT \tabularnewline \cline{2-9}
&\cite{ye2017deep}& MDP &DQN with LSTM &V2V transmitter &Current CSI, past interference, past channel selections, and remaining time to meet the latency constraints&Channel and transmit power selection & Capacity and latency & IoT \tabularnewline \cline{2-9}
& \cite{challita2017proactive}& Game &DQN with LSTM &Small base station & Traffic history of small base stations and the WLAN& Channel selection and channel access probability & Throughput& LTE network \tabularnewline \cline{2-9}
&\cite{naparstek2017deep}&Game &DDQN and dueling DQN & Mobile user &  Past channel selections and observations&Channel selection & Data rate& CRN \tabularnewline \cline{2-9}
&\cite{liu2018deep}& MDP &DQN with CNN & Satellite system &Current user terminals, channel allocation matrix, and the new arrival user&Channel selection & Score +1 or -1& Satellite system \tabularnewline \cline{2-9}
&\cite{nan2018deep}& MDP &DDQN and dueling DQN & Mobile user &QoS states&Base station and channel selection&Utility& HetNet \tabularnewline \cline{2-9}
&\cite{chen2017liquid}& Game &DQN with LSM& UAV&Content request distribution  &Base station selection&Users with stable queues& LTE network \tabularnewline \cline{2-9}
\hline
\parbox[t]{2mm}{\multirow{9}{*}{\rotatebox[origin=c]{90}{ \hspace{-3 cm} Rate control}}}
%\multirow{1}{*}{Rate control} &&  & & & & \tabularnewline
 &\cite{gadaleta2017d}&MDP & DQN with LSTM and peephole connections &Client & Last segment quality, current buffer state, rebuffering time, and channel capacities&Bitrate selection for segment & Video quality, rebuffering even, and buffer state& DASH system\tabularnewline \cline{2-9}
  &\cite{mao2017neural}&MDP & DQN with A3C&Client & Last segment quality, current buffer state, rebuffering time, and channel capacities&Bitrate selection for segment  & Video quality, rebuffering even, and buffer state& DASH system\tabularnewline \cline{2-9}
  &\cite{huang2018qarc}&MDP & DQN with CNN and RNN&Client & Predicted video quality, current buffer state, rebuffering time, and channel capacities&Bitrate selection for segment  & Video quality, rebuffering even, and buffer state& DASH system\tabularnewline \cline{2-9}
 &\cite{chinchali2018cellular}&MDP & DQN using A3C and LSTM&Base station & Congestion metric, current network connections, and cell efficiency&Traffic rate decisions for mobile users & HVFT traffic, traffic loss to existing applications, and the amount of served bytes& HVFT application\tabularnewline \cline{2-9}
  &\cite{ferreira2018multi} &MDP & DQN using FNN&Base station & Measurements of BER, throughput, spectral efficiency, power consumption, and transmit power efficiency& Symbol rate, energy per symbol, modulation mode, number of bits per symbol, and encoding rate& Same as the state& Space communication system \tabularnewline \cline{2-9}
\hline
\end{tabular}
\par\end{centering}
\end{table*}

%=====================
\section{Caching and Offloading}
%=====================
\label{sec:caching_offloading}

%Recent advances in the areas of content caching, networking, data offloading and computation will have profound impacts on the developments of mobile edge computing.

As one of the key features of information-centric networking, in-network caching can efficiently reduce duplicated content transmissions. The studies on wireless caching has shown that access delays, energy consumption, and the total amount of traffic can be reduced significantly by caching contents in wireless devices. Big data analytics~\cite{zhong2017deep} also demonstrate that with limited cache size, proactive caching at network edge nodes can achieve 100\% user satisfaction while offloading 98\% of the backhaul traffic. Joint content caching and offloading can address the gap between the mobile users' large data demands and the limited capacities in data storage and processing. This motivates the study on Mobile Edge Computing (MEC). By deploying both computational resources and caching capabilities close to end users, MEC significantly improves energy efficiency and QoS for applications that require intensive computations and low latency. A unified study on caching, offloading, networking, and transmission control in MEC scenarios involves very complicated system analysis because of strong couplings among mobile users with heterogeneities in application demand, QoS provisioning, mobility pattern, radio access interface, and wireless resources. A learning-based and model-free approach becomes a promising candidate to manage huge state space and optimization variables, especially by using DNNs. In this section, we review the modeling and optimization of caching and offloading policies in wireless networks by leveraging the DRL framework.

\subsection{Wireless Proactive Caching}
Wireless proactive caching has attracted great attentions from both academia and industry. Statistically, a few popular contents are usually requested by many users during a short time span, which accounts for most of the traffic load. Therefore, proactively caching popular contents can avoid the heavy traffic burden of the backhaul links. In particular, this technique aims at pre-caching the contents from the remote content servers at the edge devices or BSs that are close to the end users. If the requested contents are already cached locally, the BS can directly serve the end users with small delay. Otherwise, the BS requests these contents from the original content server and updates the local cache based on the caching policy, which is one of the main design problem for wireless proactive caching.

 %The policy design is a common problem in content caching. DRL is applied to determine which contents and how long each content will be stored in the cache.

\subsubsection{QoS-Aware Caching}
Content popularity is the key factor used to solve the content caching problem. With a large number of contents and their time-varying popularities, DQL is an attractive strategy to tackle this problem with high-dimensional state and action spaces. The authors in~\cite{zhong2017deep} present a DQL scheme to improve the caching performance. The system model consists of a single BS with a fixed cache size. For each request, the BS as an agent makes a decision on whether or not to store the currently requested content in the cache. If the new content is kept, the BS determines which local content will be replaced. The state is the feature space of the cached contents and the currently requested content. The feature space consists of the total number of requests for each content in a specific short-, medium-, and long-term. There are two types of actions: (i) to find a pair of contents and exchange the cache states of the two contents and (ii) to keep the cache states of the contents unchanged. The aim of the BS is to maximize the long-term cache hit rate, i.e., reward.

The DQL scheme in~\cite{zhong2017deep} trains the policy by using the DDPG method~\cite{grad-policy} and employs Wolpertinger architecture~\cite{wolpertinger} to reduce the size of the action space and avoid missing an optimal policy. The Wolpertinger architecture consists of three main parts: an actor network, K-Nearest Neighbors (K-NN), and a critic network. The actor network is to avoid a large action space. The critic network is to correct the decision made by the actor network. The DDPG method is applied to update both critic and actor networks. K-NN can help to explore a set of actions to avoid poor decisions. The actor and critic networks are then implemented by using FNNs. The simulation results show that the proposed DQL scheme outperforms the first-in first-out scheme in terms of long-term cache hit rate.

Maximizing the long-term cache hit rate in~\cite{zhong2017deep} implies that the cache stores the most popular contents. In a dynamic environment, contents stored in a cache have to be replaced according to the users' dynamic requests. {An optimization of the placement or replacement of cached contents is studied in~\cite{deepcache17} by a deep learning method. The optimization algorithm is trained by a DNN in advance and then used for real-time caching or scheduling with minimum delay.} The authors in~\cite{schaarschmidt2016learning} propose an optimal caching policy to learn the cache expiration times, i.e., Time-To-Live (TTL), for dynamically changing requests in content delivery networks. The system includes a cloud database server and multiple mobile devices that can issue queries and update entries in a single database. The query results can be cached for a specified time interval at server-controlled caches. All cached queries will become invalid if one of the cached records has been updated. A large TTL will strain cache capacities while a small TTL increases latencies significantly if the database server is physically remote.

Unlike the DDPG approach used in~\cite{zhong2017deep}, the authors in~\cite{schaarschmidt2016learning} propose to utilize Normalized Advantage Functions (NAFs) for continuous DQL scheme to learn optimal cache expiration duration. The key problem in continuous DQL is to select an action maximizing the Q-function, while avoiding performing a costly numerical optimization at each step. {The use of NAFs obviates a second actor network that needs to be trained separately. Instead, a single neural network is used to output both a value function and an advantage term. The DQL agent at the cloud database uses an encoding of a query itself and the query miss rates as the system states, which allows for an easier generalization. The system reward is linearly proportional to the current load, i.e., the number of cached queries divided by the total capacity. This reward function can encourage longer TTLs when fewer queries are cached, and shorter TTLs when the load is close to the system capacity. Considering incomplete measurements for rewards and next-states at run-time, the authors introduce the Delayed Experience Injection (DEI) approach that allows the DQL agent to keep track of incomplete transitions when measurements are not immediately available. The authors evaluate the learning algorithm by Yahoo! cloud serving benchmark with customized web workloads~\cite{YCSB}. The simulation results verify that the learning approach based on NAFs and DEI outperforms a statistical estimator.}

\subsubsection{Joint Caching and Transmission Control}
The caching policies determine where to store and retrieve the requested content efficiently, e.g., by learning the contents' popularities~\cite{zhong2017deep} and cache expiration time~\cite{schaarschmidt2016learning}. Another important aspect of caching design is the transmission control of the content delivery from caches to end users, especially for wireless systems with dynamic channel conditions. To avoid mutual interference in multi-user wireless networks, the transmission control decides which cached contents can be transmitted concurrently as well as the most appropriate control parameters, e.g., transmit power, precoding, data rate, and channel allocation. Hence, the joint design of caching and transmission control is required to enable efficient content delivery in multi-user wireless networks.

The authors in~\cite{he2017cache,he2017optimization,he2017deeptrans} propose a DQL framework to address the joint caching and interference alignment to tackle mutual interference in multi-user wireless networks. The authors consider an MIMO system with limited backhaul capacity and the caches at the transmitter. The precoding design for interference alignment requires the global CSI at each transmitter. A central scheduler is responsible for collecting CSI and cache status from each user via the backhaul, scheduling the users' transmission, and optimizing the resource allocation. {By enabling content caching at individual transmitters, we can decrease the demand for data transfer and thus save more backhaul capacity for real-time CSI update and sharing. Using the DQL-based approach at the central scheduler can reduce the explicit demand for CSI and the computational complexity in matrix optimization, especially with time-varying channel conditions. The DQL agent implements the DNN to approximate the Q-function with experience replay in training. To make the learning process more stable, the target Q-network parameter is updated by the Q-network for every a few time instants. The collected information is assembled into a system state and sent to the DQL agent, which feeds back an optimal action for the current time instant.} The action indicates which users to be active, and the resource allocation among active users. The system reward represents the total throughput of multiple users. {An extended work of~\cite{he2017cache} and~\cite{he2017optimization} with a similar DQL framework is presented in~\cite{he2017deeptrans}, in which a CNN-based DQN is adopted and evaluated in a more practical conditions with imperfect or delayed CSI. Simulation results show that the performance of the MIMO system is significantly improved in terms of the total throughput and energy efficiency.}

Interference management is an important requirement of wireless systems. The application-related QoS or user experience is also an essential metric. Different from~\cite{he2017cache,he2017deeptrans,he2017optimization}, the authors in~\cite{he2018green} propose a DQL approach to maximize Quality of Experience (QoE) of IoT devices by jointly optimizing the cache allocation and transmission rate in content-centric wireless networks. The system state is specified by the {nodes' caching conditions, e.g., the service information and cached contents}, as well as the transmission rates of the cached contents. The aim of the DQL agent is to minimize continuously the network cost or maximize the QoE. The proposed DQL framework is further enhanced with the use of PER and DDQN. PER replays important transitions more frequently so that DQN can learn from samples more efficiently. The use of DDQN can stabilize the learning by providing two value functions in separated neural networks. This avoids an overestimation of the DQN with the increasing number of actions. These two neural networks are not completely decoupled as the target network is a periodic copy of estimation network. {A discrete simulator ccnSim~\cite{sim} is used to model the caching behavior in various graph structures. The output data trace of the simulator is then imported to Matlab and used to evaluate the learning algorithm.} The simulation results show that the DQL framework by using PER and DDQN outperforms the standard penetration test scheme in terms of QoE.

{The QoE can be used to characterize the users' perception of Virtual Reality (VR) services. The authors in~\cite{chen2018echo} address the joint content caching and transmission strategy in a wireless VR network, where UAVs capture videos on live games and transmit them to small-cell BSs servicing the VR users.} Millimeter wave (mmWave) downlink backhaul links are used for VR content transmission from the UAVs to BSs. The BSs can also cache the popular contents that may be requested frequently by end users. The joint content caching and transmission problem is formulated as an optimization to maximize the users' reliability, i.e., the probability that the content transmission delay satisfies the instantaneous delay target. The maximization involves the control of transmission format, users' association, the set and format of cached contents. A DQL framework combining the Liquid State Machine (LSM) and Echo State Network (ESN) is proposed for each BS to find the optimal transmission and caching strategies. As a randomly generated spiking neural network~\cite{mlearning}, LSM can store information about the network environment over time and adjust the users' association policy, cached contents and formats according to the users' content requests. {It has been used in~\cite{lsm17} to predict the users' content request distribution while having only limited information regarding the network and different users.} Conventional LSM uses FNNs as the output function, which demands high complexity in training due to the computation of gradients for all of the neurons. Conversely, the proposed DQL framework uses an ESN as the output function, which uses historical information to find the relationship between the users' reliability, caching, and content transmission. It also has a lower complexity in training and a better memory for network information. {Simulation results show that the proposed DQL framework can yield 25.4\% gain in terms of users' reliability compared to the baseline Q-learning.}

\subsubsection{Joint Caching, Networking, and Computation} Caching and transmission control will become more involved in a HetNet that integrates different communication technologies, e.g., cellular system, device-to-device network, vehicular network, and networked UAVs, to support various application demands. The network heterogeneity raises the problem of complicated system design that needs to address challenging issues such as mutual interference, differentiated QoS provisioning, and resource allocation, hopefully in a unified framework. Obviously this demands a joint optimization far beyond the extent of joint caching and transmission control.

Accordingly, the authors in~\cite{he2017big} propose a DQL framework for energy-efficient resource allocation in green wireless networks, jointly considering the couplings among networking, in-network caching and computation. The system consists of a Software-Defined Network (SDN) with multiple virtual networks and mobile users requesting for video on-demand files that require a certain amount of computational resource at either the content server or at local devices. In each virtual network, an authorized user issues a request to download files from a set of available SBSs in its neighborhood area. The wireless channels between each mobile user and the SBSs are characterized as Finite-State Markov Channels (FSMC). The states are the available cache capacity at the SBSs, the channel conditions between mobile users and SBSs, the computational capability of the content servers and mobile users. The DQL agent at each SBS decides an association between each mobile user and SBS, where to perform the computational task, and how to schedule the transmissions of SBSs to deliver the required data. The objective is to minimize the total energy consumption of the system from data caching, wireless transmission, and computation.
%{\color{blue}Simulation results show that the total energy consumption in different testing scenarios is very high at the beginning of the learning process and gradually decreasing to a stable value when the learning converges. Moreover, the energy consumption of the unified DRL framework considering caching, networking, and computing is significantly lower than other DRL frameworks that only focus on part of the control variables.}

{The DQL scheme proposed in~\cite{he2017big} has been applied to improve the performance of Vehicular Ad doc NETworks (VANETs) in~\cite{he2017resource,he2017deep,he2018integrated}. The network model includes multiple BSs, Road Side Units (RSUs), MEC servers, and content servers. All devices are controlled by a mobile virtual network operator. The vehicles request for video contents that can be cached at the BSs or retrieved from remote content servers. The authors in~\cite{he2017resource} formulate the resource allocation problem as a joint optimization of caching, networking, and computing, e.g., compressing and encoding operations of the video contents. The system states include the CSI from each BS, the computational capability, and cache size of each MEC/content server. The network operator feeds the system state to the FNN-based DQN and gets the optimal policy that determines the resource allocation for each vehicle. To exploit spatial correlations in learning, the authors in~\cite{he2017deep} enhance Q-learning by using CNNs in DQN. This makes it possible to extract high-level features from raw input data. Two schemes have been introduced in~\cite{he2018integrated} to improve stability and performance of the ordinary DQN method. Firstly, DDQN is designed to avoid over-estimation of Q-value in ordinary DQN. Hence, the action can be decoupled from the target Q-value generation. This makes the training process faster and more reliable. Secondly, the dueling DQN approach is also integrated in the design with the intuition that it is not always necessary to estimate the reward by taking some action. The state-action Q-value in dueling DQN is decomposed into one value function representing the reward in the current state, and the advantage function that measures the relative importance of a certain action compared with other actions. The enhanced DQL agent combining these two schemes can achieve better performance and faster training speed.

Considering the huge action space and high complexity with the vehicle's mobility and service delay deadline $T_d$, a multi-time scale DQN framework is proposed in~\cite{mobilityaware} to minimize the system cost by the joint design of communication, caching and computing in VANET. The policy design accounts for limited storage capacities and computational resources at the vehicles and the RSUs. The small timescale DQN is for every time slot and aims to maximize the exact immediate reward. Additionally, the large timescale DQN is designed for every $T_d$ time slots within the service delay deadline, and used to estimate the reward considering the vehicle's mobility in a large timescale.}

The aforementioned DQL framework for VANETs, e.g., \cite{he2017deep,he2017resource,he2018integrated}, has also been generalized to smart city applications in~\cite{he2017software}, which necessitates dynamic orchestration of networking, caching, and computation to meet different servicing requirements. Through Network Function Virtualization (NFV)~\cite{han2015network}, the physical wireless network in smart cities can be divided logically into several virtual ones by the network operator, which is responsible for network slicing and resource scheduling, as well as allocation of caching and computing capacities. The use cases in smart cities are presented in~\cite{hesecure,trust-base}, which apply the generalized DQL framework to improve the security and efficiency for trust-based data exchange, sharing, and delivery in mobile social networks through the resource allocation and optimization of MEC allocation, caching, and D2D (Device-to-Device) networking.
%{\color{red}(Note by Shimin: \cite{he2017resource,he2017deep,he2018integrated,he2017software,hesecure} are very similar works by the same group of authors. They basically use the simplest DQN to address a joint optimization of caching, networking, and computing. There is no significant differences in modeling and solutions.)}

%{The system state of DRL framework includes channel state, computation capability, content indicator, version indicator, and the trust value for the content providers. The action includes which video provider is assigned to the subscribed user, whether or not the computation offloading (video transcoding) should be performed, and whether or not the video provider should cache the new video.}

\begin{figure}[t!]
\centering
\includegraphics[width=0.5\textwidth]{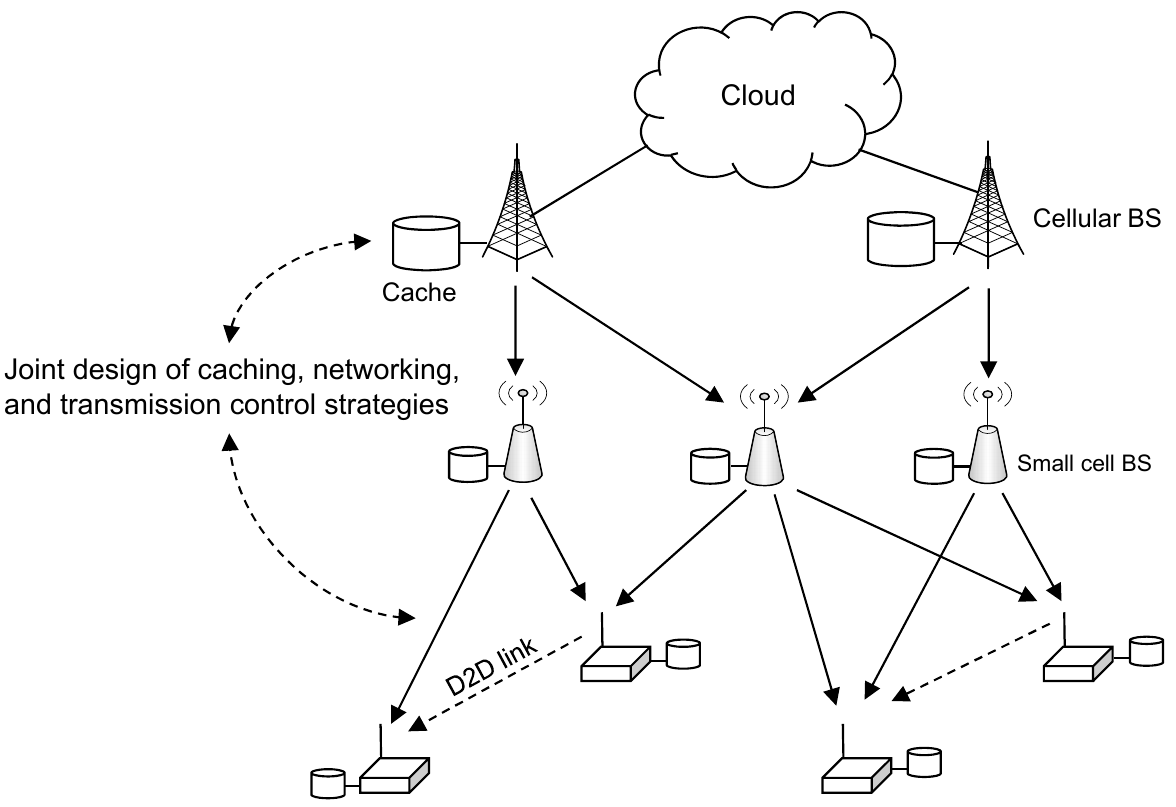}
\caption{\small Joint caching, networking, and transmission control to optimize cache hit rate~\cite{zhong2017deep},
cache expiration time~\cite{schaarschmidt2016learning},
interference alignment~\cite{he2017cache,he2017optimization,he2017deeptrans},
Quality of Experience~\cite{he2018green,chen2018echo},
energy efficiency~\cite{he2017big},
resource allocation~\cite{he2017resource,he2017deep,he2018integrated},
traffic latency, or redundancy~\cite{he2017software,hesecure}.}\label{fig_caching_networking}
\end{figure}

\subsection{Data and Computation Offloading}

With limited computation, memory and power supplies, IoT devices such as sensors, wearable devices, and handheld devices become the bottleneck to support advanced applications such as interactive online gaming and face recognition. To address such a challenge, IoT devices can offload the computational tasks to nearby MEC servers, integrated with the BSs, Access Points (APs), and even neighboring Mobile Users (MUs). As a result, data and computation offloading can potentially reduce the processing delay, save the battery energy, and even enhance security for computation-intensive IoT applications. However, the critical problem in the computation offloading is to determine the offloading rate, i.e., the amount of computational workload, and choose the MEC server from all available servers. If the chosen MEC server experiences heavy workloads and degraded channel conditions, it may take even longer time for the IoT devices to offload data and receive the results from the MEC server. Hence, the design of an offloading policy has to take into account the time-varying channel conditions, user mobility, energy supply, computation workload and the computational capabilities of different MEC servers.

The authors in~\cite{zhangdeep} focus on minimizing the mobile user's cost and energy consumption by offloading cellular traffic to WLAN. Each mobile user can either access the cellular network, or the complimentary WLAN as illustrated in Fig.~\ref{fig_caching_networking}(a), but with different monetary costs. The mobile user also has to pay a penalty if the data transmission does not finish before the deadline. The mobile user's data offloading decision can be modeled as an MDP. The system state includes the mobile user's location and the remaining file size of all data flows. The mobile user will choose to transmit data through either WLAN or cellular network, and decide how to allocate channel capacities to concurrent flows. Without knowing the mobility pattern in advance, the DQL is proposed for each mobile user to learn the optimal offloading policy from past experiences. CNNs are employed in the DQL to predict a continuous value of the mobile user's remaining data. Simulation results reveal that the DQN-based scheme generally outperforms the dynamic programming algorithm for the MDP. The reason is that the DQN can learn from experience while the dynamic programming algorithm cannot obtain the optimal policy with incorrect transition probability.

%The design of computation offloading policy is similar to data offloading proposed in~\cite{zhangdeep}.

The allocation of limited computational resources at the MEC server is critical for cost and energy minimization. The authors in~\cite{Ji2018deep} consider an MEC-enabled cellular system, in which multiple mobile users can offload their computational tasks via wireless channels to one MEC server, co-located with the cellular BS as shown in Fig.~\ref{fig_caching_networking}(b). {Each mobile user has a computational-intensive task, characterized by the required computational resources, CPU cycles, and the maximum tolerable delay.} The capacity of the MEC server is limited to accommodate all mobile users' task loads. The bandwidth sharing between different mobile users' offloading also affects the overall delay performance and energy consumptions. {The DQL is used to minimize the cost of delay and power consumptions for all mobile users, by jointly optimizing the offloading decision and computational resource allocation. The system states include the sum of cost of the entire system and the available computational capacity of the MEC server. The action of BS is to determine the resource allocation and offloading decision for each mobile user. To limit the size of action space, a pre-classification step is proposed to check the mobile users' feasible set of actions.}

In contrast to \cite{Ji2018deep}, multiple BSs in an ultra-dense network is considered in~\cite{chen2018performance} and~\cite{chen2018optimized}, as shown in Fig.~\ref{fig_caching_networking}(c), with the objective of minimizing the long-term cost of delay in computation offloading. All computational tasks are offloaded to the shared MEC server via different BSs. Besides the allocation of computational resources and transmission control, the offloading policy also has to optimize the association between mobile users and the BSs. With dynamic network conditions, the mobile users' decision-making can be formulated as an MDP. The system states are the channel conditions between the mobile user and the BSs, the states of energy and task queues. The cost function is defined as a weighted sum of the execution delay, the handover delay and the computational task dropping cost. The authors in~\cite{chen2018optimized} firstly propose a DDQN-based DQL algorithm to learn the optimal offloading policy without knowing the network dynamics. By leveraging the additive structure of the utility function, the Q-function decomposition combined with the DDQN further leads to a novel online SARSA-based DRL algorithm. Numerical experiments show that the new algorithm achieves a significant improvement in computation offloading performance compared with the baseline policies, e.g., the DQN-based DQL algorithm and some heuristic offloading strategies without learning. The high density of SBSs can relieve the data offloading pressure in peak traffic hours but consume a large amount of energy in off-peak time. Therefore, the authors in \cite{DRAG}, \cite{li2018deep}, and \cite{liu2018deepnap} propose a DQL-based strategy for controlling the (de)activation of different SBSs to minimize the energy consumption without compromising the quality of provisioning. In particular, in \cite{DRAG}, the on/off decision framework uses a DQL scheme to approximate both the policy and value functions in an actor-critic method. The reward of the DQL agent is defined as a cost function relating to energy consumption, QoS degradation, and the switching cost of SBSs. The DDPG approach is also employed together with an action refinement scheme to expedite the training process. Through extensive numerical simulations, the proposed scheme is shown to greatly outperform other baseline methods in terms of both energy and computational efficiency.

With a similar model to that in~\cite{chen2018optimized}, computation offloading finds a proper application for cloud-based malware detection in~\cite{wan2017reinforcement}. {A review of the threat models and the RL-based solutions for security and privacy protection in mobile offloading and caching are discussed in~\cite{securitymodel18}}. With limited energy supply, computational resources, and channel capacity, mobile users cannot always update the local malware database and process all application data in time and thus are vulnerable to zero-day attacks~\cite{zero}. By leveraging the remote MEC server, all mobile users can offload their application data and detection tasks via different BSs to the MEC/security server with larger and more sophisticated malware database, more computational capabilities, and powerful security services. {This can be modeled by a dynamic malware detection game in which multiple mobile users interact with each other in resource competition, e.g., the allocation of wireless channel capacities and the computational capabilities of the MEC/security server. A DQL scheme is proposed for each mobile user to learn its offloading data rate to the MEC/security server. The system states include the channel state and the size of application traces. The objective is to optimize the detection accuracy of the security server, which is defined as a concave function in the total amount of malware samples. The Q-value is estimated by using a CNN in the DQL framework. %Both convolutional layers use the rectified linear unit (ReLU) as the activation function.
The authors also propose the hotbooting Q-learning technique that provides a better initialization for Q-learning by exploiting the offloading experiences in similar scenarios. It can save exploration time at the initial stage and accelerate the learning speed compared with a standard Q-learning algorithm with all-zero initialization of the Q-value~\cite{allzeroq}. The proposed DQL scheme not only improves the detection speed and accuracy, but also increases the mobile users' battery life. The simulation results reveal that compared with the hotbooting Q-learning, the DQL-based malware detection has the faster learning rate, the higher accuracy, and the shorter detection delay.}

Multiple MEC servers have been considered in~\cite{min2017learning,twolayered18}, as illustrated in Fig.~\ref{fig_caching_networking}(d). The authors in~\cite{min2017learning} aim to design optimal offloading policy for IoT devices with energy harvesting capabilities. The system consists of multiple MEC servers, such as BSs and APs, with different capabilities in computation and communications. The IoT devices are equipped with energy storage and energy harvesters. They can execute computational tasks locally and offload the tasks to the MEC servers. The IoT device's offloading decision can be formulated as an MDP. The system states include the battery status, the channel capacity, and the predicted amount of harvested energy in the future. The IoT device evaluates the reward based on the overall delay, energy consumption, the task drop loss and the data sharing gains in each time slot. {Similar to~\cite{wan2017reinforcement}, the authors in~\cite{min2017learning} enhance Q-learning by the hotbooting technique to save the random exploration time at the beginning of learning. The authors also propose a fast DQL offloading scheme that uses hotbooting to initialize the CNN and accelerates the learning speed. } {The authors in~\cite{twolayered18} view the MEC-enabled BSs as different physical machines constituting a part of the cloud resources. The cloud optimizes the MUs' computation offloading to different virtual machines residing on the physical machines. A two-layered DQL algorithm is proposed for the offloading problem to maximize the utilization of cloud resources. The system state relates to the waiting time of each computational task and the number of virtual machines. The first layer is implemented by a CNN-based DQL framework to estimate an optimal cluster for each computational task. Different clusters of physical machines are generated based on the K-NN algorithm. The second layer determines the optimal serving physical machine within the cluster by Q-learning method. }
\begin{figure}[t!]
\centering
\includegraphics[width=0.5\textwidth]{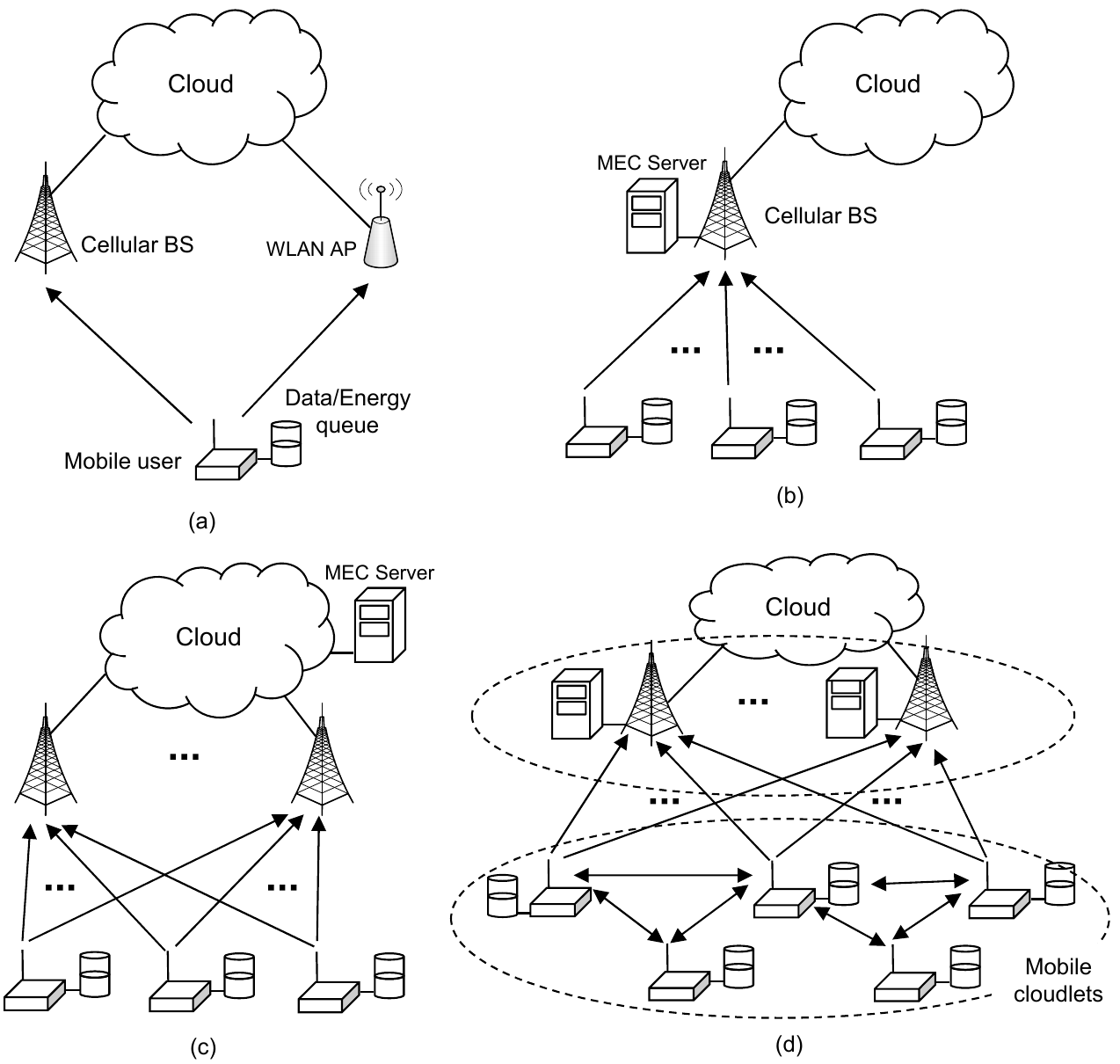}
\caption{\small Data/computation offloading models in cellular networks: (a) Offloading cellular traffic to WLAN~\cite{zhangdeep}, (b) Offloading to a single MEC-enabled BS~\cite{Ji2018deep}, (c) Offloading to one shared MEC server via multiple BSs~\cite{chen2018performance,chen2018optimized,wan2017reinforcement}, (d) Offloading to multiple MEC-enabled BSs~\cite{min2017learning,twolayered18} and mobile cloudlets~\cite{duc2018deep_jrl,duc2018deep}.}\label{fig_caching_networking}
\end{figure}

The aforementioned works all focus on data or computation offloading in cellular system via BSs to remote MEC servers, e.g., \cite{zhangdeep,Ji2018deep,chen2018performance,chen2018optimized,twolayered18,wan2017reinforcement,min2017learning}. In~\cite{duc2018deep} and~\cite{duc2018deep_jrl}, the authors study QoS-aware computation offloading in an ad-hoc mobile network. By making a certain payment, the mobile user can offload its computational tasks to nearby mobile users constituting a mobile cloudlet, as shown in Fig.~\ref{fig_caching_networking}(d). Each mobile user has a first-in-first-out queue with limited buffer size to store the arriving tasks arriving as a Poisson process. The mobile user selects nearby cloudlets within D2D communication range for task offloading. The offloading decision depends on the states including the number of remaining tasks, the quality of the links between mobile users and the cloudlet, and the availability of the cloudlet's resources. The objective is to maximize a composite utility function, subject to the mobile user's QoS requirements, e.g., energy consumption and processing delay. {The utility function is firstly an increasing function of the total number of tasks that have been processed either locally or remotely by the cloudlets. It is also related to the user's benefit such as energy efficiency and payment for task offloading.} This problem can be formulated as an MDP, which can be solved by linear programming and Q-learning approaches, depending on the availability of information about the state transition probabilities. This work is further enhanced by leveraging DNN or DQN to learn the decision strategy more efficiently. {A similar model is studied in~\cite{deepoffloading17}, where the computation offloading is formulated as an MDP to minimize the cost of computation offloading. The solution to the MDP can be used to train a DNN by supervised learning. The well-trained DNN is then applied to unseen network conditions for real-time decision-making. Simulation results show that the use of deep supervised learning achieves significant performance gain in offloading accuracy and cost saving.}

Data and computation offloading is also used in fog computing. The mobile application demanding a set of data and computational resources can be hosted in a container, e.g., virtual machine of a fog node. With user's mobility, the container has to be migrated or offloaded to other nodes and dynamically consolidated. With the container migration, some nodes with low resource utilization can be switched off to reduce power consumption. The authors in~\cite{tang2018migration} model the container migration as a multi-dimensional MDP, which is solved by the DQL. {The system states consist of the delay, the power consumption and the migration cost. The action includes the selection policy that selects the containers to be emigrated from each source node, and the allocation policy that determines the destination node of each container. The action space can be optimized for more efficient exploration by dividing fog nodes into under-utilization, normal-utilization, and over-utilization groups. By powering off under-utilization nodes, all their containers will be migrated to other nodes to reduce power consumption. The training process is also optimized by using DDQN and PER which assigns different priorities to the transitions in experience memory.
%A higher priority will be assigned if a transition does not fit well to the current estimate of Q function.
This helps the DQL agent at each fog node to perform better in terms of faster learning speed and more stability. Simulation results reveal that the DQL scheme achieves fast decision-making and outperforms the existing baseline approaches significantly in terms of delay, power consumption, and migration cost.}

\textbf{Summary:} This section reviews the applications of the DQL for wireless caching and data/computation offloading, which are inherently coupled with networking and allocation of channel capacity, computational resources, and caching capabilities, etc. We observe that the DQL framework for caching is typically centralized and mostly implemented at the network controller, e.g., the BS, service provider, and central scheduler, which is more powerful in information collection and cross-layer policy design. On the contrary, the end users have more control over their offloading decisions, and hence we observe more popular implementation of the DQL agent at local devices, e.g., mobile users, IoT devices, and fog nodes. Though an orchestration of networking, caching, data and computation offloading in one unified DQL framework is promising for network performance maximization, we face many challenges in designing highly-stable and fast-convergent learning algorithms, due to excessive delay and unsynchronized information collection from different network entities.

\begin{table*}
\caption{A summary of approaches using DQL for caching and offloading.} \label{table_summary_caching_offloading}
\begin{centering}
\begin{tabular}{|>{\centering\arraybackslash}m{0.7cm}|>{\centering}m{0.5cm}|>{\centering\arraybackslash}m{0.8cm}|>{\centering}m{1.8cm}|>{\centering}m{1.4cm}|>{\centering}m{3.5cm}|>{\centering}m{2.2cm}|>{\centering}m{2.5cm}|>{\centering}m{1.4cm}|}
\hline
 \cellcolor{mygray}  \textbf{\noun{Issues}} &  \cellcolor{mygray} \textbf{\noun{Ref.}}&  \cellcolor{mygray} \textbf{\noun{Model}}&  \cellcolor{mygray} \textbf{\noun{Learning algorithms}}&  \cellcolor{mygray} \textbf{\noun{Agent}}  &  \cellcolor{mygray} \textbf{\noun{States}} &  \cellcolor{mygray}\textbf{\noun{Actions}} & \cellcolor{mygray} \textbf{\noun{Rewards}} &  \cellcolor{mygray} \textbf{\noun{Networks}} \tabularnewline
\hline
\hline
\parbox[t]{2mm}{\multirow{9}{*}{\rotatebox[origin=c]{90}{ \hspace{-5 cm} Wireless proactive caching}}}
&\cite{zhong2017deep}& MDP & DQN using actor-critic, DDPG & Base station &Cached contents and requested content & Replace selected content or not & Cache hit rate \\
(score 1 or 0) & CRN \tabularnewline \cline{2-9}
&\cite{he2017big}& MDP &DQN using FNN & Base station & Channel states and computational capabilities & User association, computational unit, content delivery &  Energy consumption & CRN \tabularnewline \cline{2-9}
&\cite{schaarschmidt2016learning}& MDP & DQN using NAFs & Cloud database & Encoding of a query, query cache miss rate & Cache expiration times & Cache hit rates, CDN utilization & Cloud database \tabularnewline \cline{2-9}
&\cite{he2017cache}\cite{he2017optimization}& MDP & DQN using FNN & Central scheduler & Channel coefficients, cache state & Active users and resource allocation &  Network throughput & MU MIMO system \tabularnewline \cline{2-9}
& \cite{he2017deeptrans}& MDP &DQN using CNN & Central scheduler & Channel coefficients, cache state & Active users and resource allocation &  Network throughput & MU MIMO system \tabularnewline \cline{2-9}
%&\cite{he2017optimization}& MDP &DQN with FNN & Central scheduler & Channel coefficients, cache state & Active candidate users and resource allocation &  Network throughput & MU MIMO system \tabularnewline \cline{2-9}
&\cite{he2018green}& MDP &DDQN & Service provider & Conditions of cache nodes, transmission rates of content chunks & The content chunks to cache and to remove & Network cost, QoE & Content centric IoT \tabularnewline \cline{2-9}
&\cite{chen2018echo}& MDP & DQN using LSM and ESN & Base station & Historical content request & User association, cached contents and formats & Reliability & Cellular system \tabularnewline \cline{2-9}
&\cite{he2017deep}\cite{he2017software}& MDP &DQN using CNN & Service provider & Available BS, MEC, and cache & User association, caching, and offloading & Composite revenue & Vehicular ad hoc network \tabularnewline \cline{2-9}
&\cite{he2017resource}& MDP &DQN using FNN & Service provider & Available BS, MEC, and cache & User association, caching, and offloading & Composite revenue & Vehicular ad hoc network \tabularnewline \cline{2-9}
&\cite{he2018integrated}& MDP &DDQN and dueling DQN & Service provider & Available BS, MEC, and cache & User association, caching, and offloading & Composite revenue & Vehicular ad hoc network \tabularnewline \cline{2-9}
&\cite{hesecure}& MDP & DQN using CNN & Base station & Channel state, computational capability, content/version indicator, and the trust value & User association, caching, and offloading & Revenue & Mobile social network \tabularnewline \cline{2-9}
\hline
\parbox[t]{2mm}{\multirow{9}{*}{\rotatebox[origin=c]{90}{ \hspace{-3 cm} Data and computation offloading}}}
&\cite{zhangdeep}& MDP & DQN using CNN & Mobile user & User's location and remaining file size & Idle, transmit via WLAN or cellular network & Total data rate & Cellular system \tabularnewline \cline{2-9}
&\cite{Ji2018deep}& MDP & DQN using FNN & Base station & Sum of cost and computational capacity of the MEC server & Offloading decision and resource allocation & Sum of cost of delay and energy consumption & Cellular system  \tabularnewline \cline{2-9}
&\cite{chen2018performance} & MDP & DQN using FNN  & Mobile user & Channel qualities, states of energy and task queues &  Offloading and resource allocation & Long term cost function & Cellular system \tabularnewline \cline{2-9}
&\cite{chen2018optimized} & MDP & DDQN, SARSA & Mobile user & Channel qualities, states of energy and task queues &  Offloading decision and computational resource allocation & Long term cost function & Cellular system \tabularnewline \cline{2-9}
&\cite{wan2017reinforcement} & Game & DQN using CNN, hotbooting Q-learning & Mobile user & Channel states, size of App traces & Offloading rate &  Utility related to detection accuracy, response speed, and the transmission cost & Cellular system \tabularnewline \cline{2-9}
%&\cite{min2017learning} & MDP & DQN using CNN, hotbooting Q-learning & IoT device & Battery status, the channel capacity, and the predicted amount of harvested energy & Selection of MEC servers and the portion of data offloading &  Composite utility related to delay, energy consumption, task drop loss, and data sharing gains & Cellular system \tabularnewline \cline{2-9}
%&\cite{duc2018deep}\cite{duc2018deep_jrl} & MDP & Q-learning & Mobile user & The remaining tasks, link quality, and the availability of the cloudlet's resources & Tasks to be locally processed or offloaded &  Composite utility related to QoS and payment & Mobile ad hoc network \tabularnewline \cline{2-9}
& \cite{tang2018migration} & MDP & DDQN & Fog node & Delay, container's location and resource allocation & Container's next location &  Composite utility related to delay, power consumption, and migration cost & Fog computing \tabularnewline \cline{2-9}
\hline
\end{tabular}
\par\end{centering}
\end{table*}

%=====================
\section{Network Security and Connectivity Preservation}
%=====================
\label{sec:network_connectivity}
Future networks become more decentralized and ad-hoc in nature which are vulnerable to various attacks such as Denial-of-Service (DoS) and cyber-physical attack. Recently, the DQL has been used as an effective solution to avoid and prevent the attacks. In this section, we review the applications of DQL in addressing the following security issues:
\begin{itemize}
\item \textit{Jamming attack:} In the jamming attack, attackers as jammers transmit Radio Frequency (RF) jamming signals with high power to cause interference to the legitimate communication channels, thus reducing the SINR at legitimate receivers. Anti-jamming techniques such as the frequency hopping
\cite{popovski2006strategies} and user mobility, i.e., moving out from the heavy jamming area, have been commonly used. However, without being aware of the radio channel model and the jamming methods, it is challenging for the users to choose an appropriate frequency channel as well as to determine how to leave and avoid the attack. DQL enables the users to learn an optimal policy based on their past observations, and thus DQL can be used to address the above challenge.
\item \textit{Cyber-physical attack:} The cyber-physical attack is an integrity attack in which an attacker manipulates data to alter control signals in the system. This attack often happens in autonomous systems such as Intelligent Transportation Systems (ITSs) and increases the risk of accidents to Autonomous Vehicles (AVs). The DQL allows the AVs to learn optimal actions based on their time-varying observations of the attacker' activities. Thus, the DQL  can be used to achieve robust and dynamic control of the AV to the attacks.
\item \textit{Connectivity preserving:} This refers to maintaining the connectivity among the robots, e.g., UAVs, to support the communication and exchange of information among them. The system and network environment is generally dynamic and complex, and thus the DQL which allows each robot to make dynamic decisions based on its state can be effectively used to preserve the connectivity in the system.
\end{itemize}
%=====================
\subsection{Network Security}
%=====================
This section discusses the applications of DQL to address the jamming attack and the cyber-physical attack.
%=====================
\subsubsection{Jamming Attack}
%=====================
\begin{figure}[t!]
\centering
\includegraphics[width=\linewidth]{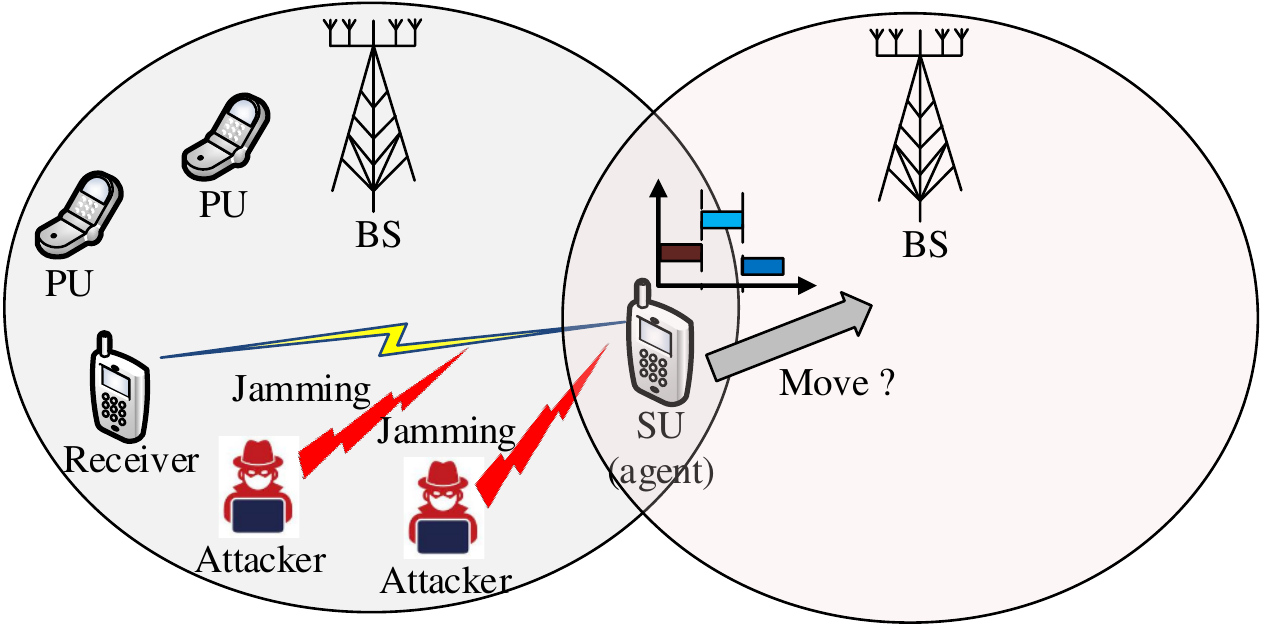}
 \caption{\small Jamming attack in cognitive radio network~\cite{han2017two}.}
 \label{jamming_attack_frequency_hopping}
\end{figure}
A pioneer work using the DQL for the anti-jamming is~\cite{han2017two}. The network model is a Cognitive Radio Network (CRN) as shown in Fig.~\ref{jamming_attack_frequency_hopping} which consists of one Secondary User (SU), multiple Primary Users (PUs), and multiple jammers. The network has a set of frequency channels for hopping. At each time slot, each jammer can arbitrarily select one of the channels to send its jamming signal, and the SU, i.e., the agent, needs to choose a proper action based on the SU's current state. The action is (i) selecting one of the channels to send its signals or (ii) leaving the area to connect to another BS. The jammers are assumed to avoid causing interference to the PUs, and thus the SU's current state consists of the number of PUs and the discretized SINR of the SU signal at the last time slot. The objective of the SU is to maximize its expected discounted utility over time slots. Note that when the SU chooses to leave the area to connect to another BS, it spends a mobility cost. Thus, the utility is defined as a function of the SINR of the SU signal and the mobility cost. Since the number of frequency channels may be large that results in a large action set, the CNN is used for the DQL to quickly learn the optimal policy. As shown in the simulation results, the proposed DQL has a faster convergence speed than that of the Q-learning algorithm. Moreover, considering the scenario with two jammers, the proposed DQL outperforms the frequency-hopping method in terms of the SINR and the mobility cost.

The model in~\cite{han2017two} is constrained to two jammers. As the number of jammers in the network increases, the proposed scheme may not be effective. The reason is that it becomes hard for the SU to find good actions when the number of jammed channels increases. An appropriate solution, as proposed in ~\cite{xiao2018anti}, allows the receiver of the SU to leave its current location. Since the leaving incurs the mobility cost, the receiver, i.e., the agent, needs an optimal policy, i.e., staying at or leaving the current location, to maximize its utility. In this scenario, the DQL based on CNN can be used for the receiver to find the optimal action to maximize its expected utility. Here, the utility and state of the receiver are essentially defined similarly to that of the agent in~\cite{han2017two}. In particular, the state includes the discretized SINR of the signal measured by the receiver at the last time slot.

The above approaches, i.e., in~\cite{han2017two} and \cite{xiao2018anti}, define states of the agents based on raw SINR values of the signals. In practical wireless environments, the number of SINR values may be large and even infinite. Moreover, the raw SINR can be inaccurate and noisy. To cope with the challenge of the infinite number of states, the DQL can use a recursive
Convolutional Neural Network (RCNN) as proposed in~\cite{liu2018anti}. By using the pre-processing layer and recursive convolution layers, the RCNN is able to remove noise from the network environment and extract useful features of the SINR, i.e., discrete spectrum sample values greater than a noise threshold, thus reducing the computational complexity. The network model and the problem formulation considered in~\cite{liu2018anti} are similar to those in~\cite{han2017two}. However, instead of directly using the raw SINR, the state of the SU is the extracted features of the SINR. Also, the action of the SU includes only frequency-hopping decision. The simulation results show that the proposed DQL based on the RCNN can converge in both fixed and dynamic jamming scenarios while the Q-learning cannot converge in the dynamic jamming one. Furthermore, the proposed DQL can achieve the average throughput close to that of the optimal scheme, i.e., an anti-jamming scheme with completely known jamming actions.
%For the future work, a general scenario with multiple SUs can be considered. In this case, the action of a particular SU needs to take account into the channel selections of other SUs.   The power control schemes estimate the channel conditions and adjust the transmit power to override the jamming signal and improve the communication quality. However, in modern network such as IoT, it is challenging to accurately estimate the channels due to the dynamic topology of the network. In such a network, the authors in \cite{chen2018dqn} proposed to use the DQN to achieve the an optimal power control policy without being aware of the channel variation information and the jamming pattern
%\cite{xiao2017two}: Anti-jamming mobile communication/general mobile network\\

Instead of finding the frequency-hopping decisions, the authors in~\cite{chen2018dqn} propose the use of DQL to find an optimal power control policy for the anti-jamming. The model is an IoT network including IoT devices and one jammer. The jammer can observe the communications of the transmitter and chooses a jamming strategy to reduce the SINR at the receiver. Thus, the transmitter chooses an action, i.e., transmit power level, to maximize its utility. Here, the utility is the difference between the SINR and the energy consumption cost due to the transmission. Note that choosing the transmit power impacts the future jamming strategy, and thus the interaction between the transmitter and the jammer can be formulated as an MDP. The transmitter is the agent, and the state is SINR measured at its receiver at the last time slot. The DQN using the CNN is then adopted to find an optimal power control policy for the transmitter to maximize its expected accumulated discounted reward, i.e., the utility, over time slots. The simulation results show that the proposed DQL can improve the utility of the transmitter up to 17.7\% compared with the Q-learning scheme. Also, the proposed DQL reduces the utility of the jammer around 18.1\% compared with the Q-learning scheme.

\begin{figure}[t!]
\centering
\includegraphics[width=\linewidth]{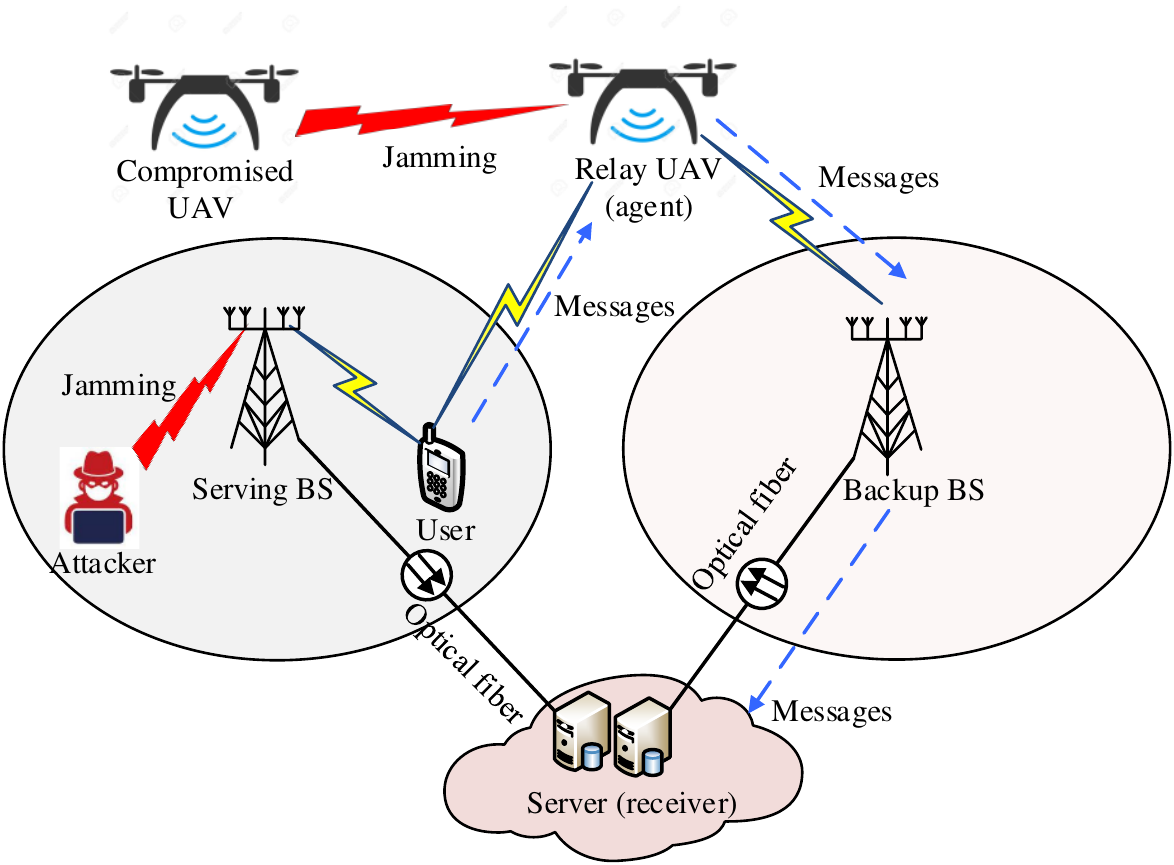}
 \caption{\small Anti-jamming scheme based on UAV~\cite{lu2018uav}.}
 \label{jamming_attack_power_allocation}
\end{figure}

%\cite{lu2018uav}: Anti-jamming/UAV-aided 5G\\ We can consider the UAV as an agent,
To prevent the jammer's observations of communications, the transmitter can change its communication strategy, e.g., by using relays that are far from the jamming area. The relays can be UAVs as proposed in~\cite{lu2018uav}. The model consists of one UAV, i.e., a relay, one jammer, one mobile user and its serving BS (see Fig.~\ref{jamming_attack_power_allocation}). The mobile user transmits messages to its server via the serving BS. In the case that the serving BS is heavily jammed, the UAV helps the mobile user to relay the messages to the server through a backup BS. In particular, depending on the SINR and Bit Error Rate (BER) values sent from the serving BS, the UAV as an agent decides the relay power level to maximize its utility, i.e., the difference between the SINR and the relay cost. The relay power level can be considered to be the UAV's actions, and the SINR and BER are its states. As such, the next state observed by the UAV is independent of all the past states and actions. The problem is formulated as an MDP. To quickly achieve the optimal relay policy for the UAV, the DQL based on CNN is then adopted. The simulation results in~\cite{lu2018uav} show that the proposed DQL scheme takes only 200 time slots to converge to the optimal policy, which is 83.3\% less than that of the relay scheme based on Q-learning~\cite{xiao2018uav}. Moreover, the proposed DQL scheme reduces the BER of the user by 46.6\% compared with the hill climbing-based UAV relay scheme~\cite{lv2017anti}.
%saves the energy consumption by 33.6\% compared with QL-based relay scheme.

%\cite{xiao2018user}: Smart attack prevention/UAV transmission\\
The scheme proposed in~\cite{lu2018uav} assumes that the relay UAV is sufficiently far from the jamming area. However, as illustrated in Fig.~\ref{jamming_attack_power_allocation}, the attacker can use a compromised UAV close to the relay UAV to launch the jamming attack to the relay UAV. In such a scenario, the authors in \cite{xiao2018user} show that the DQL can still be used to address the attack. The system model is based on physical layer security and consists of one UAV and one attacker. The attacker is assumed to be ``smarter'' than that in the model in~\cite{lu2018uav}. This means that the attacker can observe channels that the UAV uses to communicate with the BS in the past time slots and then chooses jamming power levels on the target channels. Therefore, the UAV needs to find a power allocation policy, i.e., transmit power levels on the channels, to maximize the secrecy capacity of the UAV-BS communication. Similar to~\cite{lu2018uav}, the DQL based on CNN is used which enables the UAV to choose its actions, i.e., transmit power levels on the channels, based on its state, i.e., the attacker's jamming power level in the last time slot. The reward is the difference between the secrecy capacity of the UAV and BS and the energy consumption cost.

The simulation results in \cite{xiao2018user} show that the proposed DQL can improve the UAV's utility up to 13\% compared with the baseline scheme~\cite{bowling2002multiagent} which uses the Win or Learn Faster-Policy Hill Climbing (WoLF-PHC) to prevent the attack. Also, the safe rate of the UAV, i.e., the probability that the UAV is attacked, obtained by the proposed DQL is 7\% higher than that of the baseline. However, the proposed DQL is applied only to a single-UAV system. For the  future work, scenarios with multiple UAVs need to be considered. In such a scenario, more computational overhead is expected and multi-agent DQL algorithms can be applied.

\subsubsection{Cyber-Physical Attack}
In autonomous systems such as ITSs, the attacker can seek to inject faulty data to information transmitted from the sensors to the AVs. The AVs which receive the injected information may inaccurately estimate the safe spacing among them. This increases the risk of AV accidents. Vehicular communication security algorithms, e.g.,~\cite{chen2018cyber}, can be used to minimize the spacing deviation. However, the attacker's actions in these algorithms are assumed to be stable which may not be applicable in practical systems. The DQL that enables the AVs to learn optimal actions based on their time-varying observations of the attacker' actions can be thus used.
%manipulate the spacing and deviate the spacing from the optimal safe state. In contrast, the AV tries to minimize the spacing deviation. weight of measurement The interaction between the attacker and the AV can be thus modeled as a non-cooperative game.
\begin{figure}[t!]
\centering
\includegraphics[width=7.9 cm, height=4.4 cm]{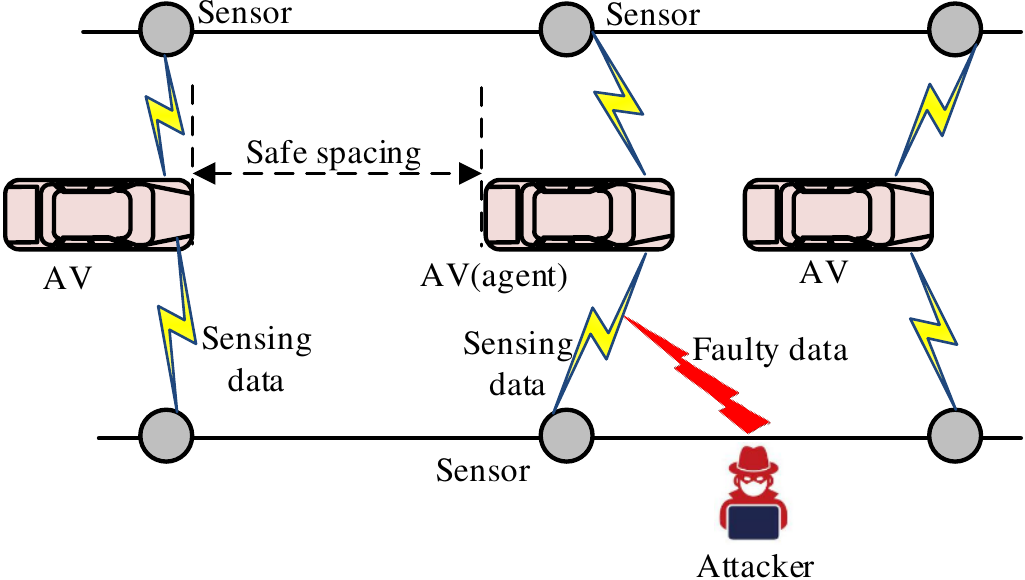}
 \caption{\small Car-following model with cyber-physical attack.}
 \label{cyber_physical_attack_ITS}
\end{figure}

The first work using the DQL for the cyber-physical attack in an ITS can be found in~\cite{ferdowsi2018robust}. The system is a \textit{car-following} model~\cite{brackstone1999car} of the General Motors as shown in Fig.~\ref{cyber_physical_attack_ITS}. In the model, each AV updates its speed based on measurement information received from the closest road smart sensors. The attacker attempts to inject faulty data to the measurement information. However, the attacker cannot inject the measurements of different sensors equally due to its resource constraint. Thus, the AV can choose less-faulty measurements by selecting a vector of measurement weights. The objective of the attacker is to maximize the deviation, i.e., the utility, from the safe spacing between the AV and its nearby AV while that of the AV is to minimize the deviation. The interaction between the attacker and the AV can be modeled as a zero-sum game. The authors in~\cite{ferdowsi2018robust} show that the DQL can be used to find the equilibrium strategies. In particular, the action of the AV is to choose a weight vector. Its state includes the past actions, i.e., the weight vectors, and the past deviation values. Since the actions and deviations have continuous values, the state space is infinite. Thus, LSTM units that are able to extract useful features are adopted for the DQL to reduce the state space. The simulation results show that by using the past actions and deviations for learning the attacker's action, the proposed DQL scheme can guarantee a lower steady-state deviation than the Kalmar filter-based scheme~\cite{chen2018cyber}. Moreover, by using the LSTM units, the results show that the proposed DQL scheme can converge much faster than the baseline scheme.

\begin{figure}[t!]
\centering
\includegraphics[width=7.6 cm, height=5.3 cm]{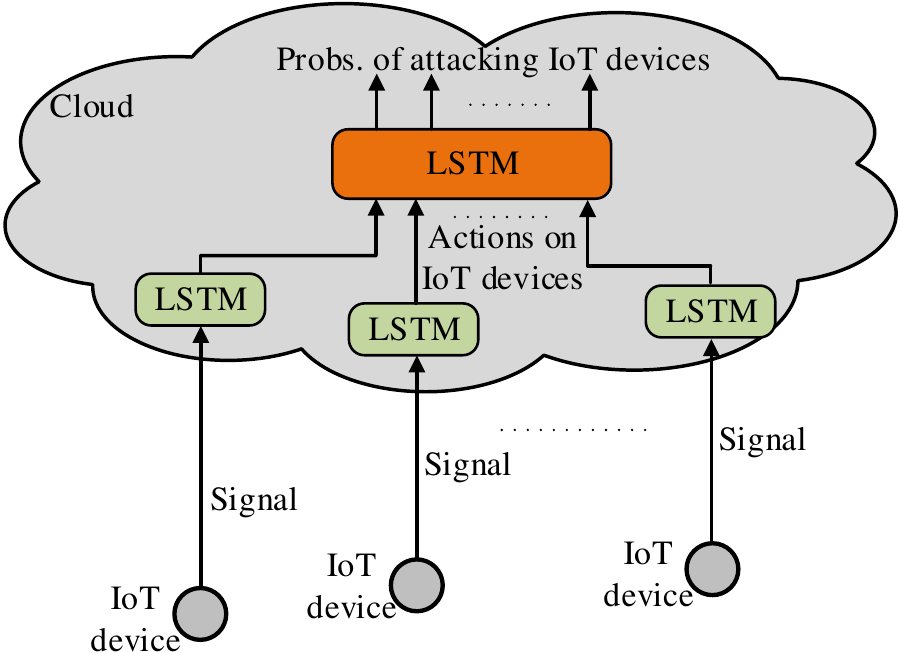}
 \caption{\small Cyber-physical detection in IoT systems using DQL.}
 \label{cyber_physical_attack_IoT}
\end{figure}

%======================
%\cite{ferdowsi2018deep}: signal authentication and security/IoT\\
Another work that uses the LSTM to extract useful features from the measurement information to detect the cyber-physical attack is proposed in~\cite{ferdowsi2018deepICC}. The model is an IoT system including a cloud and a set of IoT devices. The IoT devices generate signals and transmit the signals to the cloud (see Fig.~\ref{cyber_physical_attack_IoT}). The cloud uses the received signals for estimation and control of the IoT devices' operation. An attacker can launch the cyber-physical attack by manipulating the IoT devices' output signals that causes control errors at the cloud and degrades the performance of the IoT system. To detect the attack, the cloud uses LSTM units to extract stochastic features or fingerprints such as flatness, skewness, and kurtosis, of the IoT devices' signals. The cloud sends the fingerprints back to the IoT devices, and the IoT devices embed, i.e., watermark, the fingerprints inside the signals. The cloud uses the fingerprints to authenticate the IoT devices' signals to detect the attack.

The algorithm proposed in \cite{ferdowsi2018deepICC} is also called \textit{dynamic watermarking} \cite{satchidanandan2017dynamic} which is able to detect the cyber-physical attack and to prevent eavesdropping attacks. However, the algorithm requires large computational resources at the cloud for the IoT device signal authentication. Consequently, the cloud can only authenticate a limited number of vulnerable IoT devices. The cloud can choose the vulnerable IoT devices by observing their security status. However, this can be impractical since the IoT devices may not report their security status. Thus, the authors in \cite{ferdowsi2018deep} propose to use the DQL that enables the cloud to decide which IoT devices to authenticate with the incomplete information. Since IoT devices with more valuable data are likely to be attacked, the reward is defined as a function of data values of IoT devices. The cloud's state includes attack actions of the attacker on the IoT devices in the past time slots. The actions of the attacker on the IoT devices can be obtained by using the dynamic watermarking algorithm in~\cite{ferdowsi2018deepICC} (see Fig.~\ref{cyber_physical_attack_IoT}). The DQL then uses an LSTM unit to find the optimal policy. The input of the LSTM unit is the state of the cloud, and the output includes probabilities of attacking the IoT devices. By using a real dataset from the accelerometers, the simulation results show that the proposed DQL can improve the cloud's utility up to 30\% compared with the case in which the cloud chooses the IoT devices with equal probability.
%\cite{akazaki2018falsification}: cyber-physical systems\\

%\cite{pattanaik2017robust}: Adversarial attacks/general network\\

%\cite{zhangresearch}: optimal attack strategy/UAVs\\

%\cite{feng2017deep}: cyber-physical system/general network\\
%\cite{tsagkaris2018early}: early warning/wireless broadband network\\

%\cite{liu2017deep}: Decision-making/Unmanned Combat Aerial Vehicle
%(UCAV) air combat\\
%\cite{atallah2017deep}: Road-side units scheduling/Vehicular ad-hoc networks\\

%=====================
\subsection{Connectivity Preservation}
\begin{figure}[t!]
\centering
\includegraphics[width=7.7 cm, height=5.3 cm]{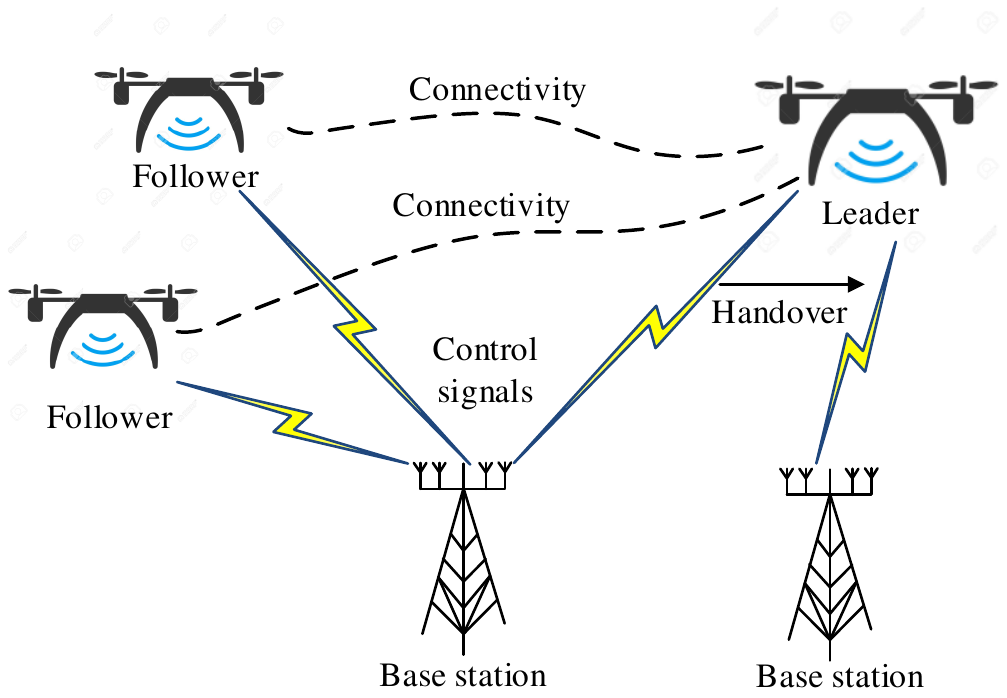}
 \caption{\small Connectivity preservation of a multi-UAV network.}
 \label{connectivity_preservation}
\end{figure}

%=====================
%=====================
% \cite{huang2017deep}: connectivity preserving/multi-robot system\\
Multi-robot systems such as multi-UAV cooperative networks have been widely applied in many fields such as military, e.g., enemy detecting. In the cooperative multi-robot system, the connectivity among the robots, e.g., UAVs in Fig~\ref{connectivity_preservation}, is required to enable the communication and exchange of information. To tackle the connectivity preservation problem, the Artificial Potential Field (APF) algorithm~\cite{vadakkepat2000evolutionary} has been used. However, the algorithm cannot be directly adopted when the robots are undertaking missions in dynamic and complex environments. The DQL which allows each robot to make dynamic decisions based on its own state can be effectively applied to preserve the connectivity in the multi-robot system. Such an approach is proposed in~\cite{huang2017deeppreserve}.

The model in~\cite{huang2017deeppreserve} consists of two robots or UAVs, i.e., one leader robot and one follower robot. In the model, a central control, i.e., a ground BS, adjusts the velocity of the follower such that the follower stays in the communication range of the leader at all time (see Fig~\ref{connectivity_preservation}). The connectivity preservation problem can be thus formulated as an MDP. The agent is the BS, and the states are the relative position and the velocity of the leader with respect to the follower. The action space consists of possible velocity values of the follower. Taking an action returns a reward which is +1 if the follower is in the range of the leader, and -1 otherwise. A DQN using FNN is used which enables the BS to learn an optimal policy to maximize the expected discounted cumulative reward. The input of the DQN includes the states of the two robots, and the output is the action space of the follower. The simulation results show that the proposed scheme can achieve better connectivity between the two robots than that of the APF method. However, a general scenario with more than one leader and one follower needs to be investigated.

%==================================================
%\cite{huang2017deep}: Preserve connectivity/multi-robot system\\
Considering the general scenario, the authors in~\cite{huang2017deep} address the connectivity preservation between multiple leaders and multiple followers. The robot system is definitely connected if any two robots are connected via a direct link or multi-hop link. To express the connectivity in such a robot system, the authors introduce the concept of \textit{algebraic connectivity}\cite{poonawala2015collision} which is the second smallest eigenvalue of a Laplacian matrix. The robot system is connected if the algebraic connectivity of the system is positive. Thus, the problem is to adjust the velocity of the followers such that the algebraic connectivity is positive over time slots. This problem can be formulated as an MDP in which the agent is the ground BS, the state is a combination of the states of all robots, the action is a set of possible velocity values for the followers. The reward is +1 if the algebraic connectivity of the system increases or holds, and becomes a penalty of -1 if the algebraic connectivity decreases. Similar to \cite{huang2017deeppreserve}, a DQN is adopted. Due to the large action space of the followers, the actor-critic neural network~\cite{mnih2016asynchronous} is used. The simulation results show that the followers always follow the motion of the leaders even if the leaders' trajectory dynamically changes. However, the proposed DQN requires more time to converge than that in \cite{huang2017deeppreserve} because of the presence of more followers.

The proposed schemes in \cite{huang2017deeppreserve} and \cite{huang2017deep} do not consider a minimum distance between the leaders and followers. The leaders and followers can collide with each other if the distance between them is too short. Thus, the BS needs to guarantee the minimum distance between them. One solution is to have the minimum distance in the reward as proposed in~\cite{wang2017autonomous}. In particular, if the leader is too close to its follower, the reward of the system is penalized regarding the minimum distance. The DQL algorithm proposed in~\cite{huang2017deep} is then used such that the BS learns proper actions, e.g., turning left and right, to maximize the cumulative reward.

%============================================
%\cite{wang2018handover}: Handover control/ultra-dense network\\ The central controller clusters the users into clusters. Users in a cluster have the same mobility patterns, i.e., locations and speed
%The robots discussed in~\cite{huang2017deeppreserve},~\cite{huang2017deep}, and~\cite{wang2017autonomous}, are UAVs which are typically controlled by base stations.

When BSs are densely deployed, the UAVs or mobile users need to trigger a frequent handover to preserve the connectivity. The frequent handover increases communication overhead and energy consumption of the mobile users, and interrupts data flows. Thus, it is essential to maintain an appropriate handover rate. The authors in~\cite{wang2018handover} address the handover decision problem in an ultra-density network. The network model consists of multiple mobile users, SBSs, and one central controller. At each time slot, the user needs to decide its serving SBS. The handover decision process can be modeled as an MDP, and the DQL is adopted to find an optimal handover policy for each user to minimize the number of handover occurrences while ensuring certain throughput. The state of the user, i.e., the agent, includes reference signal quality received from candidate SBSs and the last action of the user. The reward is defined as the difference between the data rate of the user and its energy consumption for the handover process. Given a high density of users, the DQL using A3C and LSTM is adopted to find the optimal policy in short training time. The simulation results show that the proposed DQL can achieve better throughput and lower handover rate than those of the upper confidence bandit algorithm~\cite{shen2016non} with similar training time.

%============================================
%\cite{long2017towards}: Collision avoidance/multi-robot systems\\
%\cite{garratt2017supervised}: target tracking/UAVs\\ Self-Organization Networks (SONs) which can automatically handle network faults as self-healing by adjusting network parameters are an effective solution
%===========================================
To enhance the reliability of the communication between the SBSs and the mobile users, the SBSs should be able to handle network faults and failure automatically as self-healing. The DQL can be applied as proposed in~\cite{faris2018deep} to make optimal parameter adjustments based on the observation of the network performance. The model is the 5G network including one MBS. The MBS as an agent needs to handle network faults such as transmit diversity faults and antenna azimuth change, e.g., because of wind. These faults are represented as the MBS's state that is the number of active alarms. Based on the alarms, the MBS can take actions including (i) enabling the transmit diversity and (ii) setting the antenna azimuth to default value. The reward that the MBS receives is the scores, e.g., -1, 0, and +1, depending on the number of faults happening. The DQL is used to learn the optimal policy. The simulation results show that the proposed DQL can achieve network throughput close to that of the oracle-based self-healing, i.e., the upper-performance bound, but incurs less fault message passing overhead.
\begin{table*}
\caption{\label{tab: }A summary of approaches using DQL for network security and connectivity preservation.}
\label{DQN_network_security_connectivity_preservation_summary_table}
\begin{centering}
\begin{tabular}{|>{\centering\arraybackslash}m{0.7cm}|>{\centering}m{0.5cm}|>{\centering\arraybackslash}m{0.8cm}|>{\centering}m{1.8cm}|>{\centering}m{1.4cm}|>{\centering}m{3.5cm}|>{\centering}m{2.2cm}|>{\centering}m{2.5cm}|>{\centering}m{1.4cm}|}
\hline
 \cellcolor{mygray}  \textbf{\noun{Issues}} &  \cellcolor{mygray} \textbf{\noun{Ref.}}&  \cellcolor{mygray} \textbf{\noun{Model}}&  \cellcolor{mygray} \textbf{\noun{Learning algorithms}}&  \cellcolor{mygray} \textbf{\noun{Agent}}  &  \cellcolor{mygray} \textbf{\noun{States}} &  \cellcolor{mygray}\textbf{\noun{Actions}} & \cellcolor{mygray} \textbf{\noun{Rewards}} &  \cellcolor{mygray} \textbf{\noun{Networks}} \tabularnewline
\hline
\hline
\parbox[t]{2mm}{\multirow{9}{*}{\rotatebox[origin=c]{90}{\hspace{-2cm} Network security}}}

 &\cite{han2017two}&Game& DQN using CNN &Secondary user &Number of PUs and signal SINR&Channel selection and leaving decision &SINR and mobility cost& CRN\tabularnewline \cline{2-9}

  &\cite{xiao2018anti}&Game& DQN using CNN &Receiving
transducer &Signal SINR& Staying and leaving decisions&SINR and mobility cost& Underwater acoustic network\tabularnewline \cline{2-9}

  &\cite{liu2018anti}&MDP& DQN using RCNN &SU &Signal SINR& Channel selection&SINR and mobility cost& CRN\tabularnewline \cline{2-9}

  &\cite{chen2018dqn}&MDP& DQN using CNN &Transmit IoT device &Signal SINR& Channel selection&SINR and energy consumption cost& IoT\tabularnewline \cline{2-9}

    &\cite{lu2018uav}&MDP& DQN using CNN &Relay UAV&Signal SINR and BER& Relay power& SINR and relay cost& UAV\tabularnewline \cline{2-9}

 &\cite{xiao2018user}&MDP& DQN using CNN &Transmit UAV&Jamming power& Transmit power& Secrecy capacity and energy consumption cost& UAV\tabularnewline \cline{2-9}

  &\cite{ferdowsi2018robust}&Game& DQN using LSTM units &Autonomous vehicle&Deviation values & Measurement weight selection&Safe spacing deviation& ITS\tabularnewline \cline{2-9}

   &\cite{ferdowsi2018deep}&Game& DQN using LSTM units &Cloud&Attack actions on IoT devices & IoT device set selection&IoT devices' data values& IoT\tabularnewline \cline{2-9}

\hline
\parbox[t]{2mm}{\multirow{9}{*}{\rotatebox[origin=c]{90}{\hspace{-1 cm} Connectivity preservation}}}
%\multirow{1}{*}{Rate control} &&  & & & & \tabularnewline
 &\cite{huang2017deeppreserve}&MDP & DQN using FNN &Ground base station & Relative positions and the velocity of robots&Velocity decision &Sore +1 and -1& Robot system\tabularnewline \cline{2-9}
  &\cite{huang2017deep}&MDP & DQN using A3C&Ground base station & Relative positions and the velocity of robots&Velocity decision&Sore +1 and -1& Robot system\tabularnewline \cline{2-9}
  &\cite{wang2017autonomous}&POMDP & DQN using A3C&Ground base station &Information of distances among robots&Turning left and turning right decisions &Sore +1 and -1& Robot system\tabularnewline \cline{2-9}
    &\cite{wang2018handover}&MDP &  DQN using A3C and LSTM&Mobile users& Reference signal received quality and the last action&Serving SBS selection  & Data rate and energy consumption&Ultra-dense network\tabularnewline \cline{2-9}
     &\cite{faris2018deep}&MDP & DQN using CNN&MBS & The number of active alarms&Enabling transmit diversity and changing antenna azimuth & Score -1, 0, +1, and +5& Self-organization network\tabularnewline \cline{2-9}
\hline
\end{tabular}
\par\end{centering}
\end{table*}

\textbf{Summary:} This section reviews applications of DQL for the network security and connectivity preservation. The reviewed approaches are summarized along with the references in Table~\ref{DQN_network_security_connectivity_preservation_summary_table}. We observe that the CNN is mostly used for the DQL to enhance the network security. Moreover, DQL approaches for the anonymous system such as robot systems and ITS receive more attentions than other networks. However, the applications of DQL for the cyber-physical security are relatively few and need to be investigated.
%=====================
\section{Miscellaneous Issues}
%=====================
\label{sec:misc_issues}
This section reviews applications of DRL to solve some other issues in communications and networking. The issues include (i) traffic engineering and routing, (ii) resource sharing and scheduling, and (iii) data collection.
\subsection{Traffic Engineering and Routing}
% =============================================
%\cite{xu2018experience}(r25) routing problem/general communication network:

Traffic Engineering (TE) in communication networks refers to Network Utility Maximization (NUM) by optimizing a path to forward the data traffic, given a set of network flows from source to destination nodes. Traditional NUM problems are mostly model-based. However, with the advances of wireless communication technologies, the network environment becomes more complicated and dynamic, which makes it hard to model, predict, and control. The recent development of DQL methods provides a feasible and efficient way to design experience-driven and model-free schemes that can learn and adapt to the dynamic wireless network from past observations.

{Routing optimization is one of the major control problems in traffic engineering. The authors in~\cite{stampa2017deep} present the first attempt to use the DQL for the routing optimization. Through the interaction with the network environment, the DQL agent at the network controller determines the paths for all source-destination pairs. The system state is represented by the bandwidth request between each source-destination pair, and the reward is a function of the mean network delay. The DQL agent leverages the actor-critic method for solving the routing problem that minimizes the network delay, by adapting routing configurations automatically to current traffic conditions. The DQL agent is trained using the traffic information generated by a gravity model~\cite{gravity}. The routing solution is then evaluated by OMNet+ discrete event simulator~\cite{omnet}. The well-trained DQL agent can produce a near-optimal routing configuration in a single step and thus the agent is agile for real-time network control. The proposed approach is attractive as the traditional optimization-based techniques require a large number of steps to produce a new configuration. The authors in~\cite{xu2018experience} consider a similar network model with multiple end-to-end communication sessions. Each source-destination pair has a set of candidate paths that can transport the traffic load. Experimental results show that the conventional DDPG method does not work well for the continuous control problem in~\cite{xu2018experience}. One possible explanation is that DDPG utilizes uniform sampling for experience replay, which ignores different significance of the transition samples.

The authors in~\cite{xu2018experience} also combine two new techniques to optimize DDPG particularly for traffic engineering problems, i.e., TE-aware exploration and actor-critic-based PER methods. The TE-aware exploration leverages the shortest path algorithm and NUM-based solution as the baseline during exploration. The PER method is conventionally used in DQL, e.g.,~\cite{he2018green} and~\cite{tang2018migration}, while the authors in~\cite{xu2018experience} integrate the PER method with the actor-critic framework for the first time. The proposed scheme assigns different priorities to transitions in the experience replay. Based on the priority, the proposed scheme samples the transitions in each epoch. The system state consists of throughput and delay performance of each communication session. The action specifies the amount of traffic load going through each of the paths. By learning the dynamics of network environment, the DQL agent aims to maximize the total utility of all the communication sessions, which is defined based on end-to-end throughput and delay~\cite{alpha_func}. Packet-level simulations using NS-3~\cite{ns3}, tested on well-known network topologies as well as random topologies generated by BRITE~\cite{brite}, reveal that the proposed DQL scheme significantly reduces the end-to-end delay and improves the network utility, compared with the baseline schemes including DDPG and the NUM-based solutions.}

The networking and routing optimization become more complicated in the UAV-based wireless communications. The authors in~\cite{wang2017autonomous} model autonomous navigation of one single UAV in a large-scale unknown complex environment as a POMDP, which can be solved by actor-critic-based DRL method. {The system state includes its distances and orientation angles to nearby obstacles, the distance and angle between its present position and the destination. The UAV's action is to turn left or right or keep ahead. The reward is composed of four parts: an exponential penalty term if it is too close to any obstacles, a linear penalty term to encourage minimum time delay, the transition and direction rewards if the UAV is getting close to the target position in a proper direction. Instead of using conventional DDPG for continuous control, the Recurrent Deterministic Policy Gradient (RDPG) is proposed for the POMDP by approximating the actor and critic using RNNs. Considering that RDPG is not suitable for learning using memory replay, the authors in~\cite{wang2017autonomous} propose the fast-RDPG method by utilizing the actor-critic framework with function approximation~\cite{func_appx}. The proposed method derives policy update for POMDP by directly maximizing the expected long-term accumulated discounted reward. }

%which is designed to reduce correlation in the observation sequence,

{Path planning for multiple UAVs connected via cellular systems is studied in~\cite{challita2018deep} and~\cite{challita2018cellular}. Each UAV aims to achieve a tradeoff between maximizing energy efficiency and minimizing both latency and interference caused to the ground network along its path. The network state observable by each UAV includes its distances and orientation angles to cellular BSs, the orientation angle to its destination, and the horizontal coordinates of all UAVs. The action of each UAV includes an optimal path, transmit power, and cell association along its path. The interaction among UAVs is cast as a dynamic game and solved by a multi-agent DRL framework.
%A major challenge in this game is the need for full knowledge of the ground network topology, service requirements, and other UAVs' locations.
The use of ESN in the DRL framework allows each UAV to retain previous memory states and make a decision for unseen network states, based on the reward obtained from previous states. ESN is a new type of RNNs with feedback connections, consisting of the input, recurrent, and output weight matrices. ESN training is typically quick and computationally efficient compared with other RNNs. Deep ESNs can exploit the advantages of a hierarchical temporal feature representation at different levels of abstraction, hence disentangling the difficulties in modeling complex tasks. Simulation results show that the proposed scheme improves the tradeoff between energy efficiency, wireless latency, and the interference caused to the ground network. Results also show that each UAV's altitude is a function of the ground network density and the UAV's objective function is an important factor in achieving the UAV's target.}

%that belong to the family of reservoir computing (RC)~\cite{mlearning}

Besides networked UAVs, vehicle-to-infrastructure also constitutes an important part and provides rich application implications in 5G ecosystem. The authors in~\cite{zhu2017communication} adopt the DQL to achieve an optimal control policy in communication-based train control system, which is supported by bidirectional train-ground communications. The control problem aims to optimize the handoff decision and train control policy, i.e., accelerate or decelerate, based on the states of stochastic channel conditions and real-time information including train position, speed, measured SNR from APs, and handoff indicator. The objective of the DQL agent is to minimize a weighted combination of operation profile tracking error and energy consumption.
%{\color{red}(Note by Shimin: I saw some papers in the paper pool focusing on V2V communications. Maybe these two papers regarding vehicle-to-infrastructure can be put together with V2V communications.)}

%=====================
\subsection{Resource Sharing and Scheduling}
%=====================
System capacity is one of the most important performance metrics in wireless communication networks. System capacity enhancements can be based on the optimization of resource sharing and scheduling among multiple wireless nodes. The integration of DRL into 5G systems would revolutionize the resource sharing and scheduling schemes from model-based to model-free approaches and meet various application demands by learning from the network environment.

The authors in~\cite{yang2018decco} study the user scheduling in a multi-user massive MIMO system. User scheduling is responsible for allocating resource blocks to BSs and mobile users, taking into account the channel conditions and QoS requirements. Based on this user scheduling strategy, {a DRL-based coverage and capacity optimization is proposed to obtain dynamically the scheduling parameters and a unified threshold of QoS metric. The performance indicators are calculated as the average spectrum efficiency of all the users. The system state is an indicator of the average spectrum efficiency. The action of the scheduler is a set of scheduling parameters to maximize the reward as a function of the average spectrum efficiency. The DRL scheme uses policy gradient method to learn a policy function (instead of a Q-function) directly from trajectories generated by the current policy. The policy network is trained with a variant of the REINFORCE algorithm~\cite{func_appx}.} The simulation results in~\cite{yang2018decco} show that compared with the optimization-based algorithms that suffer from incomplete network information, the policy gradient method achieves much better performance in terms of network coverage and capacity.

% {Nonetheless, the success of DQL depends on the availability of a large amount of real or simulated data for training. To deal with limited data in wireless system, the authors in~\cite{klautau5g} present a traffic simulator to generate channel realizations representing 5G mmWave MIMO scenarios with mobility in both transceivers and objects.

%As one application, the generated data is used to train the DRL agent to predict the optimal beam for mmWave MIMO communications~\cite{klautau5g}.

In~\cite{xu2017deep}, the authors focus on dynamic resource allocation in a cloud radio access network and present a DQL-based framework to minimize the total power consumption while fulfilling mobile users' QoS requirements. The system model contains multiple Remote Radio Heads (RRHs) connected to a cloud BaseBand Unit (BBU). The information of RRHs can be shared in a centralized manner. The system state contains information about the mobile users' demands and the RRHs' working states, e.g., active or sleep. According to the system state and the result of last execution, the DQL agent at the BBU decides whether to turn on or off certain RRH(s), and how to allocate beamforming weight for each active RRH. The objective is to minimize the total expected power consumption. The authors propose a two-step decision framework to reduce the size of action space. In the first step, the DQL agent determines the set of active RRHs by Q-learning and DNNs. In the second step, the BBU derives the optimal resource allocation for the active RRHs by solving a convex optimization problem. Through the combination of DQL and optimization techniques, the proposed framework results in a relatively small action space and low online computational complexity. Simulation results show that the framework achieves significant power savings while satisfying user demands and is robust in highly dynamic network environment. The aforementioned works mostly focus on simulations and numerical comparisons. With one step further, the authors in~\cite{hackett2017implementation} implement a multi-objective DQL framework as the radio-resource-allocation controller for space communications. The implementation uses modular software architecture to encourage re-use and easy modification for different algorithms, which is integrated into the real space-ground system developed by NASA Glenn Research Center.

In emerging and future wireless networks, BSs are deployed with a high density, and thus the interference among the BSs must be considered. The authors in \cite{nasir2018deep} propose to use a DQL scheme which allows the BSs to learn their optimal power control policy. In the proposed scheme, each BS is an agent, the action is choosing power levels, and the state includes interference that the BS caused to its neighbors in the last time slot. The objective is to maximize the BS's data rate. The DQN using FNN is then adopted to implement the DQL algorithm. For the future work, a joint power control and channel selection can be considered.

Network slicing \cite{foukas2017network} and NFV~\cite{han2015network} are two emerging concepts for resource allocation in the 5G ecosystem to provide cost-effective services with better performance. The network infrastructure, e.g., cache, computation, and radio resources, is comparatively static while the upper-layer Virtualized Network Functions (VNFs) are dynamic to support time-varying application-specific service requests. The concept of network slicing is to divide the network resources into multi-layer slices, managed by different service renderers independently with minimal conflicts. The concept of Service Function Chaining (SFC) is to orchestrate different VNFs to provide required functionalities and QoS provisioning.

%, which is subject to users' different understanding and preferences of various experience metrics

The authors in~\cite{chen2018reinforcement} propose a DQL scheme for QoS/QoE-aware SFC in NFV-enabled 5G systems. {Typical QoS metrics are bandwidth, delay, throughput, etc. The evaluation of QoE normally involves the end-user's participation in rating the service based on direct user perception. The authors quantify QoE by measurable QoS metrics without end-user involvements, according to the Weber-Fechner Law (WFL)~\cite{wfl} and exponential interdependency of QoE and QoS hypothesis~\cite{iqx}. These two principles actually define nonlinear relationship between QoE and QoS. The system state represents the network environment including network topology, QoS/QoE status of the VNF instances, and the QoS requirements of the SFC request. The DQL agent selects a certain direct successive VNF instance as an action. The reward is a composite function of the QoE gain, the QoS constraint penalty, and the OPEX penalty. A DQL based on CNNs is implemented to approximate the action-value function. The authors in~\cite{zhao2018deep} review the application of a DQL framework in two typical resource management scenarios using network slicing. For radio resource slicing, the authors simulate a scenario containing one single BS with different types of services. The reward can be defined as a weighted sum of spectrum efficiency and QoE. For priority-based core network slicing, the authors simulate a scenario with 3 SFCs demanding different computational resources and waiting time. The reward is the sum of waiting time in different SFCs. Simulation results in both scenarios show that the DQL framework could exploit more implicit relationship between user activities and resource allocation in resource constrained scenarios, and enhance the effectiveness and agility for network slicing.}

%For radio resource slicing, the authors simulate a scenario containing one single BS with different types of services. The reward can be defined as a weighted sum of spectrum efficiency and QoE. The state is the number of arrived packets in each slice within a specific time window and the action of the DRL agent at the BS is to adjust the bandwidth allocate of each slice, by updating its Q network every second. For priority-based core network slicing, the authors simulate a scenario with 3 SFCs demanding different computation resources and waiting time. The state can be set as the priority and time-stamp of last arrived flows in each SFC and the action is to allocate SFC for the current flow. The reward is the sum of waiting time in different SFCs.

Resource allocation and scheduling problems are also important for computer clusters or database systems. This usually leads to an online decision-making problem depending on the information of workload and environment. The authors in~\cite{mao2016resource} propose a DRL-based solution, DeepRM, {by employing policy gradient methods~\cite{func_appx} to manage resources in computer systems directly from experience. The same policy gradient method is also used in~\cite{yang2018decco} for user scheduling and resource management in wireless systems. DeepRM is a multi-resource cluster scheduler that learns to optimize various objectives such as minimizing average job slowdown or completion time. The system state is the current allocation of cluster resources and the resource profiles of jobs in the queue. The action of the scheduler is to decide how to schedule the pending jobs.} By simulations with synthetic dataset, DeepRM is shown to perform comparably or better than state-of-the-art heuristics, e.g., Shortest-Job-First (SJF). It adapts to different conditions and converges quickly, without any prior knowledge of system behavior. In~\cite{li2018model}, the authors use the actor-critic method to address the scheduling problem in a general-purpose distributed data stream processing systems, which deal with processing of continuous data flow in real time or near-real-time. {The system model contains multiple threads, processes, and machines. The system state consists of the current scheduling decision and the workload of each data source. The scheduling problem is to assign
each thread to a process of a machine. The agent at the scheduler determines the assignment of each thread, with the objective of minimizing the average processing time. The DRL framework includes three components, i.e., an actor network, an optimizer producing a K-NN set of the actor network's output action, and the critic network predicting the Q-value for each action in the set. The action is selected from the K-NN set with the maximum Q-value. The use of optimizer may avoid unstable learning and divergence problems in conventional actor-critic methods~\cite{unstable}.}
%=====================
%\subsection{Crowdsensing and Social Networking}
%=====================
%=====================
\subsection{Power Control and Data Collection}
%=====================
With the prevalence of IoT and smart mobile devices, mobile crowdsensing becomes a cost-effective solution for network information collection to support more intelligent operations of wireless systems. The authors in~\cite{li2017intelligent} consider spectrum sensing and power control in non-cooperative cognitive radio networks. There is no information exchange between PUs and SUs. As such, the SU outsources the sensing task to a set of spatially distributed sensing devices to collect information about the PU's power control strategy. {The SU's power control can be formulated as an MDP. The system state is determined by the Received Signal Strength (RSS) at individual sensing devices. The SU chooses its transmit power from the set of pre-specified power levels based on the current state. A reward is obtained if both primary and SUs can fulfill their SNR requirements. Considering the randomness in RSS measurements, the authors propose a DQL scheme for the SU to learn and adjust its transmit power. The DQL is then implemented by a DQN by using FNN. The simulation results show that the proposed DQL scheme is able to  converge to a close-to-optimal solution.}

The authors in \cite{oda2017design} leverage the DQL framework for sensing and control problems in a Wireless Sensor and Actor Network (WSAN), which is a group of wireless devices with the ability to sense events and to perform actions based on the sensed data shared by all sensors. {The system state includes processing power, mobility abilities, and functionalities of the actors and sensors. The mobile actor can choose its moving direction, networking, sensing and actuation policies to maximize the number of connected actor nodes and the number of sensing events.}

The authors in~\cite{wang2018cell} focus on mobile crowdsensing paradigm, where data inference is incorporated to reduce sensing costs while maintaining the quality of sensing. The target sensing area is split into a set of cells. The objective of a sensing task is to collect data (e.g., temperature, air quality) in all the cells. A DQL-based cell selection mechanism is proposed for the mobile sensors to decide which cell is a better choice to perform sensing tasks. {The system state includes the selection matrices for a few past decision epochs. The reward function is determined by the sensing quality and cost in the chosen cells. To extract temporal correlations in learning, the authors propose the DRQN that uses LSTM layers in DQL to capture the hidden patterns in state transitions. Considering inter-data correlations, the authors use the transfer learning method to reduce the amount of data in training. That is, the cell selection strategy learned for one task can benefit another correlated task. Hence, the parameters of DRQN can be initialized by another DRQN with rich training data. Simulations are conducted based on two real-life datasets collected from sensor networks, i.e., the Sensor-Scope dataset~\cite{epfl} in the EPFL campus and the U-Air dataset of air quality readings in Beijing~\cite{beijing}. The experiments verify that DRQN reduces up to 15\% of the sensed cells with the same inference quality guarantee.} The authors in~\cite{zhang2018learning} combine UAV and unmanned vehicle in mobile crowdsensing for smart city applications. The UAV cruises in the above of the target region for city-level data collection. {Meanwhile, the unmanned vehicle carrying mobile charging stations moves on the ground and can charge the UAV at a preset charging point.
%Both UAV and unmanned vehicle are equipped with high-precision sensors and have limited energy supply.

The target region is divided into multiple subregions and each subregion has a different sample priority. The authors in~\cite{zhang2018learning} propose a DQL-based control framework for the unmanned vehicle to schedule its data collection, constrained by limited energy supply. The system state includes information about the sample priority of each subregion, the location of charging point, and the moving trace of the UAV and unmanned vehicle. The UAV and unmanned vehicle can choose the moving direction. The DQL framework utilizes CNNs for extracting the correlation of adjacent subregions, which can increase the convergence speed in training. The DQL algorithm can be enhanced by using a feasible control solution as the baseline during exploration. The PER method is also used in DQL to assign higher priorities to important transitions so that the DQL agent can learn from samples more efficiently. The proposed scheme is evaluated by using real dataset of taxi traces in Rome~\cite{taxi}. Simulation results reveal that the proposed DQL algorithm can obtain the highest data collection rate compared with the MDP and other heuristic baselines.}

{Mobile crowdsensing is vulnerable to faked sensing attacks, as selfish users may report faked sensing results to save their sensing costs and avoid compromising their privacy. The authors in~\cite{securemcs18} formulate the interactions between the server and a number of crowdsensing users as a Stackelberg game. The server is the leader that sets and broadcasts its payment policy for different sensing accuracy. In particular, the higher payment is set for more sensing accuracy. Based on the server's sensing policy, each user as a follower then chooses its sensing effort and thus the sensing accuracy to receive the payment. The payment motivates the users to put in sensing efforts and thus the payment decision process can be modeled as an MDP. In a dynamic network, the server uses the DQL to derive the optimal payment to maximize its utility, based on the system state consisting of the previous sensing quality and the payment policy. The DQL uses a deep CNN to accelerate the learning process and improve the crowdsensing performance against selfish users. Simulation results show that the DQL-based scheme produces a higher sensing quality, lower attack rate, and higher utility of the server, exceeding those of both the Q-learning and the random payment strategies.}

Social networking is an important component of smart city applications. The authors in~\cite{zhang2017social} aim to extract useful information by observing and analyzing the users' behaviors in social networking. One of the main difficulties is that the social behaviors are usually fuzzy and divergent. The authors model pervasive social networking as a monopolistically competitive market, which contains different users as data providers selling information at a certain price. Given the market model, the DQL can be used to estimate the users' behavior patterns and find the market equilibrium. Considering the costly deep learning structure, the authors in~\cite{zhang2017social} propose a Decentralized DRL (DDRL) framework that decomposes the costly deep component from the RL algorithms at individual users. The deep component can be a feature extractor integrated with the network infrastructure and provide mutual knowledge for all individuals. Multiple RL agents can purchase the most desirable data from the mutual knowledge. The authors combine well-known RL algorithms, i.e., Q-learning and learning automata, to estimate users' patterns which are described by vectors of probabilities representing the users' preferences or altitudes to different information. In social networking and smart city applications with human involvement, there can be both labeled and unlabeled data and hence a semi-supervised DRL framework can be designed, by combining the strengths of DNNs and statistical modeling to improve the performance and accuracy in learning. Then, the authors in~\cite{mohammadi2018semisupervised} introduce the semi-supervised DRL framework that utilizes variational auto-encoders~\cite{kingma2014semi} as an inference engine to infer the classification of unlabeled data. As a case study, the proposed DRL framework is customized to provide indoor localization based on the RSS from Bluetooth devices. The positioning environment contains a set of positions. Each position is associated with the set of RSS values from the set of anchor devices with known positions. The system state includes a vector of RSS values, the current location, and the distance to the target. The DQL agent, i.e., the positioning algorithm itself, chooses a moving direction to minimize the error distance to the target point. Simulations tested on real-world dataset show an improvement of 23\% in terms of the error distance to the target compared with the supervised DRL scheme.

{\textbf{Summary:} In this section, we review miscellaneous uses of DRL in wireless and networked systems. DRL provides a flexible tool in rich and diversified applications, conventionally involving dynamic system modeling and multi-agent interactions. All these imply a huge space of state transitions and actions. These approaches are summarized along with the references in Table~\ref{table_summary_misc}. We observe that the NUM problems in 5G ecosystem for traffic engineering and resource allocation face very diversified control variables, including discrete indicators, e.g., for BS (de)activation, user/cell association, and path selection, as well as continuous variables such as bandwidth allocation, transmit power, and beamforming optimization. Hence, both DQL and policy gradient methods are used extensively for discrete and continuous control problems, respectively. }

\begin{table*}[t]
\caption{\label{tab: }A summary of applications of DQL for traffic engineering, resource scheduling, and data collection.}
\label{table_summary_misc}
\begin{centering}
\begin{tabular}{|>{\centering\arraybackslash}m{0.7cm}|>{\centering}m{0.5cm}|>{\centering\arraybackslash}m{0.8cm}|>{\centering}m{1.8cm}|>{\centering}m{1.4cm}|>{\centering}m{3.5cm}|>{\centering}m{2.2cm}|>{\centering}m{2.5cm}|>{\centering}m{1.4cm}|}
\hline
 \cellcolor{mygray}  \textbf{\noun{Issues}} &  \cellcolor{mygray} \textbf{\noun{Ref.}}&  \cellcolor{mygray} \textbf{\noun{Model}}&  \cellcolor{mygray} \textbf{\noun{Learning algorithms}}&  \cellcolor{mygray} \textbf{\noun{Agent}}  &  \cellcolor{mygray} \textbf{\noun{States}} &  \cellcolor{mygray}\textbf{\noun{Actions}} & \cellcolor{mygray} \textbf{\noun{Rewards}} &  \cellcolor{mygray} \textbf{\noun{Scenarios}} \tabularnewline
\hline
\hline
\parbox[t]{2mm}{\multirow{9}{*}{\rotatebox[origin=c]{90}{ \hspace{- 1.5 cm} Traffic engineering and routing}}}
&\cite{stampa2017deep} & MDP & DQN using actor-critic networks  & Network controller & Bandwidth request of each node pair & Traffic load split on different paths & Mean network delay  & 5G network  \tabularnewline \cline{2-9}
&\cite{xu2018experience} & NUM &DQN using actor-critic networks  & Network controller & Throughput and delay performance &  Traffic load split on different paths & $\alpha$-fairness utility  & 5G network   \tabularnewline \cline{2-9}
&\cite{wang2017autonomous}& POMDP & DQN using actor-critic networks  & UAV & Local sensory information, e.g., distances and angles & Turn left or right &  Composite reward & UAV navigation \tabularnewline \cline{2-9}
&\cite{challita2018deep}  \cite{challita2018cellular}& Game & DQN using ESN & UAV & Coordinates, distances, and orientation angles  & Path, transmit power, and cell association & Weighted sum of energy efficiency, latency, and interference & Cellular-connected UAVs \tabularnewline \cline{2-9}
&\cite{zhu2017communication}& MDP & DQN using FNN & Train scheduler & Channel conditions, train position, speed, SNR, and handoff indicator & Making handoff of connection, or accelerate or decelerate the train & Tracking error and energy consumption & Vehicle-to-infrastructure system\tabularnewline \cline{2-9}
\hline
\parbox[t]{2mm}{\multirow{9}{*}{\rotatebox[origin=c]{90}{ \hspace{-1 cm} Resource sharing and scheduling}}}
%&\cite{yang2018decco}& MDP & Policy gradient & User scheduler & Continuous values of CASE and CESE & Scheduling parameters & Weighted function of CASE and CESE & Massive MIMO \tabularnewline \cline{2-9}
&\cite{xu2017deep}& MDP & DQN using FNN& Cloud baseband unit & MUs' demands and the RRHs' working states & Turn on or off certain RRH(s), and beamforming allocation & Expected power consumption & Cloud RAN \tabularnewline \cline{2-9}
&\cite{chen2018reinforcement}& MDP & DQN with CNN & Network controller &  Network topology, QoS/QoE status, and the QoS requirements & Successive VNF instance & Composite function of QoE gain, QoS constraints penalty, and OPEX penalty & Cellular system  \tabularnewline \cline{2-9}
&\cite{zhao2018deep} & MDP & DQN using FNN  & Network controller & The number of arrived packets/the priority and time-stamp of flows  &  Bandwidth/SFC allocation  & Weighted sum of spectrum efficiency and QoE/waiting time in SFCs & 5G network \tabularnewline \cline{2-9}
%&\cite{mao2016resource} & MDP & Policy gradient, REINFORCE  & Cluster scheduler & Current allocation of cluster resources, resource profiles of jobs in queue & When to schedule the pending jobs & Average job slowdown or completion time & Computer system \tabularnewline \cline{2-9}
&\cite{li2018model} & MDP & DQN using actor-critic networks & Central scheduler & Current scheduling decision and the workload & Assignment of each thread & Average processing time  & Distributed stream data processing  \tabularnewline \cline{2-9}
\hline
\parbox[t]{2mm}{\multirow{9}{*}{\rotatebox[origin=c]{90}{ \hspace{-1.5 cm} Data collection}}}
&\cite{li2017intelligent}  & MDP & DQN using FNN & Secondary user & Received signal strength at individual sensors &  Transmit power & Fixed reward if QoS satisfied  & CRN \tabularnewline \cline{2-9}
&\cite{wang2018cell} & MDP & DRQN, LSTM, transfer learning & Mobile sensors & Cell selection matrices & Next cell for sensing &  A function of the sensing quality and cost & WSN \tabularnewline \cline{2-9}
& \cite{zhang2018learning} & MDP & DQN using CNN & UAV and unmanned vehicle & Subregions' sample priority, charging point's location, and trace of the UAV and unmanned vehicle & Moving direction of the UAV and unmanned vehicle & Fixed reward related to subregions' sample priority & UAV and vehicle \tabularnewline \cline{2-9}
& \cite{securemcs18} & Game & DQN using CNN & Crowdsensing server & Previous sensing quality and payment policy & Current payment policy & Utility & Mobile crowdsensing    \tabularnewline \cline{2-9}
&\cite{zhang2017social}  & Game & DDQN& Mobile users & Current preferences & Positive or negative altitude &  Reward or profit  & Mobile social network \tabularnewline \cline{2-9}
%&\cite{mohammadi2018semisupervised} & MDP & Semisupervised DRL & IoT devices & RSSI values, current location, and distance to the target & One of the eight moving directions & The reciprocal of the distance error to the target point & Indoor positioning \tabularnewline \cline{2-9}
\hline
\end{tabular}
\par\end{centering}
\end{table*}

%
%DQN with LSTM: \cite{wang2018cell}
%
%Double Q-learning:  \cite{he2018green}\cite{chen2018optimized}\cite{hesecure}
%
%DQN SARSA: \cite{chen2018optimized}
%
%DQN using CNN: \cite{zhangdeep}\cite{he2017deep}\cite{min2017learning}\cite{wan2017reinforcement}\cite{hesecure}\cite{zhang2018learning}\cite{chen2018reinforcement}
%
%DQN using hotbooting-Q: \cite{min2017learning}\cite{wan2017reinforcement}
%
%DQN using RNN: \cite{wang2017autonomous}\cite{wang2018cell}
%
%DQN using A3C: \cite{wang2018handover}
%
%%DQN using DDPG: \cite{zhong2017deep}
%
%DQN using LSM (liquid state machine) or ESN (echo state network): \cite{chen2018echo}\cite{chen2018echo}\cite{challita2018deep}
%
%DQN using normalized advantage functions (NAFs): \cite{schaarschmidt2016learning}
%
%DQN using actor-critic based prioritized experience replay (PER): \cite{xu2018experience}
%
%DQN using actor-critic based deterministic policy gradient algorithm: \cite{zhong2017deep}\cite{stampa2017deep}\cite{li2018model}
%
%DQN using actor-critic based recurrent deterministic policy gradient (RDPG):\cite{wang2017autonomous}

%=====================
\section{Challenges, Open Issues, and Future Research Directions}
%=====================
\label{sec:challenge_issue_future}
 %HetNet_Spectrum_allocation

Different approaches reviewed in this survey evidently show
that DRL can effectively address various emerging issues in communications and networking. There are existing challenges, open issues, and new research directions which are discussed as follows.

\subsection{Challenges}
\subsubsection{State Determination in Density Networks} The DRL approaches, e.g., \cite{nan2018deep}, allow the users to find an optimal access policy without having complete and/or accurate network information. However, the DRL approaches often require the users to report their local states at every time slot. To observe the local state, the user needs to monitor Received Signal Strength Indicators (RSSIs) from its neighboring BSs, and then it temporarily connects to the BS with the maximum RSSI. However, the future networks will deploy a high density of the BSs, and the RSSIs from different BSs may not be different. Thus, it is challenging for the users to determine the temporary BS~\cite{cao2018aif}.
\subsubsection{Knowledge of Jammers' Channel Information} The DRL approach for wireless security as proposed in \cite{xiao2018user} enables the UAV to find optimal transmit power levels to maximize the security capacity of the UAV and the BS. However, to formulate the reward of the UAV, a perfect knowledge of channel information of the jammers is required. This is challenging and even impossible in practice.

\subsubsection{Multi-agent DRL in Dynamic HetNets} Most of the existing works focus on the customizations of DRL framework for individual network entities, based on locally observed or exchanged network information. Hopefully, the network environment is relatively static to ensure convergent learning results and stable policies. This requirement may be challenged in a dynamic heterogenous 5G network, which consists of hierarchically nested IoT devices/networks with fast changing service requirements and networking conditions. In such a situation, the DQL agents for individual entities have to be light-weighted and agile to the change of network conditions. This implies a reduce to the state and action spaces in learning, which however may compromise the performance of the convergent policy. The interactions among multiple agents also complicate the network environment and cause a considerable increase to the state space, which inevitably slows down the learning algorithms.

\subsubsection{Training and Performance Evaluation of DRL Framework} The DRL framework requires large amounts of data for both training and performance evaluation. In wireless systems, such data is not easily accessible as we rarely have referential data pools as other deep learning scenarios, e.g., computer vision. Most of the existing works rely on simulated dataset, which undermines the confidence of the DRL framework in practical system. The simulated data set is usually generated by a specific stochastic model, which is a simplification of the real system and may overlook the hidden patterns. Hence, a more effective way for generating simulation data is required to ensure that the training and performance evaluation of the DRL framework are more consistent with practical system.

% The CNN and RNN can be adopted to extract potential spatial features, e.g., location information of the BSs, to facilitate the state determination of users.

%These features can be utilized to facilitate the state determination of users. optimal access policy for the user while maximizing the user’s throughput.

\subsection{Open Issues}
\subsubsection{Distributed DRL Framework in Wireless Networks} The DRL framework requires large amounts of training for DNNs. This may be implemented at a centralized network controller, which has sufficient computational capacity and the capability for information collection. However, for massive end users with limited capabilities, it becomes a meaningful task to design distributed implementation for the DRL framework that decomposes resource-demanding basic functionalities, e.g., information collection, sharing, and DNN training, from reinforcement learning algorithms at individual devices. The basic functionalities can be integrated with the network controller. It remains an open issue for the design of network infrastructure that supports these common functionalities for distributed DRL. The overhead of information exchange between end users and network controller also has to be well controlled.
\subsubsection{ Balance between Information Quality and Learning Performance} The majority of the existing works consider the orchestration of networking, transmission control, offloading, and caching decisions in one DRL framework to derive the optimal policy, e.g.,~\cite{he2017big,he2017deep,he2017resource,he2018integrated,mobilityaware,he2017software,hesecure,trust-base}. However, from a practical viewpoint, the network system will have to pay substantially increasing cost for information gathering. The cost is incurred from large delay, pre-processing of asynchronous information, excessive energy consumption, reduced learning speed, etc. Hence, an open issue is to find the optimal balance between information quality and learning performance so that the DQL agent does not consume too much resources only to achieve insignificantly marginal increase in the learning performance.
% One potential resolution for this issue can be a correlation study (e.g., principle component analysis) to determine the most important network information that will affect the learning performance significantly.

\subsection{Future Research Directions}
\subsubsection{DRL for Channel Estimation in Wireless Systems}
Massive MIMO will be deployed for 5G to achieve high-speed communications at Gbps. For this, the channel estimation is the prerequisite for realizing massive MIMO. However, in a large-scale heterogeneous cellular network foreseen for 5G or beyond, the required channel estimation is very challenging. Thus, DRL will play an important role in acquiring the channel estimates with regard to dynamic time-varying wireless channels.

Also, we expect that the combination of Wireless Power Transfer (WPT) and Mobile Crowd Sensing (MCS), namely Wireless-Powered Crowd Sensing (WPCS) will be a promising technique for the emerging IoT services. To this end, a higher power transfer efficiency of WPT is very critical to enable the deployment of WPCS in low-power wide area network. A "large-scale array antenna based WPT" will achieve this goal of higher WPT efficiency, but the channel estimation should be performed with minimal power consumption at a sensor node. This is because of that the sensor must operate with self-powering via WPT from the dedicated energy source, e.g., power beacon, Wi-Fi or small-cell access point, and/or ambient RF sources, e.g., TV tower, Wi-Fi AP and cellular BS. In this regard, the channel estimation based on the receive power measurements at the sensor node is one viable solution, because the receive power can be measured by the passive-circuit power meter with negligible power consumption. DRL can be used for the time-varying wireless channels with temporal correlations over time by taking the receive power measurements from the sensor node as the input for DRL, which will enable the channel estimation for WPT efficiently.

\subsubsection{DRL for Crowdsensing Service Optimization} In MCS, mobile users contribute sensing data to a crowdsensing service provider and receive an incentive in return. However, due to limited resources, e.g., bandwidth and energy, the mobile user has to decide on whether and how much data to be uploaded to the provider. Likewise, the provider aiming to maximize its profit has to determine the amount of incentive to be given. The provider's decision depends on the actions of the mobile users. For example, with many mobile users contributing data to the crowdsensing service provider, the provider can lower the incentive. Due to a large state space of a large number of users and dynamic environment, DRL can be applied to obtain an optimal crowdsensing policy similar to~\cite{zhancrowdsensingsubmit}.

\subsubsection{DRL for Cryptocurrency Management in Wireless Networks} Pricing and economic models have been widely applied to wireless networks~\cite{luong2017resource},~\cite{luong2016data}. For example, wireless users pay money to access radio resources or mobile services. Alternatively, the users can receive money if they contribute to the networks, e.g., offering a relay or cache function. However, using real money and cash in such scenarios faces many issues related to accounting, security, and privacy. Recently, the concept of cryptocurrency based on the blockchain technology has been introduced and adopted in wireless networks, e.g.,~\cite{shi2017oppay}, which has been shown to be a secure and efficient solution. However, the value of cryptocurrency, i.e., token or coin, can be highly dynamic depending on many market factors. The wireless users possessing the tokens can decide to keep or spend the tokens, e.g., for radio resource access and service usage or exchange into real money. In the random cryptocurrency market environment, DRL can be applied to achieve the maximum long-term reward of the cryptocurrency management for wireless users as in~\cite{jiang2017cryptocurrency}.

\subsubsection{DRL for Auction}An auction has been effectively used for radio resource management, e.g., spectrum allocation~\cite{luong2017applications}. However, obtaining the solution of the auction, e.g., a winner determination problem, can be complicated and intractable when the number of participants, i.e., bidders and sellers, become very large. Such a scenario is typical in next-generation wireless networks such as 5G highly-dense heterogeneous networks. DRL appears to be an efficient approach for solving different types of auctions such as in~\cite{zhao2018deepbidding}.

%==================
\section{Conclusions}
\label{sec:conclusion}
This paper has presented a comprehensive survey of the
applications of deep reinforcement learning to communications and networking. First, we have presented an overview of reinforcement learning, deep learning, and deep reinforcement learning. Then, we have introduced various deep reinforcement learning techniques and their extensions. Afterwards, we have provided detailed reviews, analyses, and comparisons of the deep reinforcement learning to solve different issues in communications and networking. The issues include dynamic network access, data rate control, wireless caching, data offloading, network security, connectivity preservation, traffic routing, and data collection. Finally, we have outlined important challenges, open issues as well as future research directions.

%\cite{omidshafiei2017deep}
%\cite{dwivedi2018oqpmsav}
%\cite{tan2018sim}:

%\begin{thebibliography}{100}
\bibliographystyle{IEEEtran}
\bibliography{DeepRLdatabase}{}

%\end{thebibliography}

%\]
\end{document}